\documentclass[final,3p,times]{elsarticle}

\usepackage{graphicx} % Required for inserting images
\usepackage[utf8]{inputenc}
\usepackage{caption}
\usepackage{subcaption}
\usepackage{siunitx}
\usepackage{float} % For forcing figure placement
\usepackage{amsmath}
\usepackage{cite}
\begin{document}

% \begin{frontmatter}
\title{Efficient FGM optimization with a novel design space and DeepONet}

\author{Piyush Agrawal\textsuperscript{a}\textsuperscript{,*}}
\author{Ihina Mahajan\textsuperscript{a}\textsuperscript{,*}}
\author{Shivam Choubey\textsuperscript{a}}
\author{Manish Agrawal\textsuperscript{a}\textsuperscript{,**}}
\ead{manish.agrawal@iitrpr.ac.in}
\cortext[equal]{Piyush Agrawal and Ihina Mahajan have contributed equally to this work.}
\cortext[cor1]{Corresponding author.}

%\author[a,*]{Piyush Agrawa\corref{equal}}
%\author[a,*]{Ihina Mahajan\corref{equal}}
%\author[a]{Shivam Choubey}
%\author[a,**]{Manish Agrawal\corref{cor1}}
%\ead{manish.agrawal@iitrpr.ac.in}

%\cortext[cor1]{Corresponding author.}
%\cortext[equal]{Piyush Agrawal and Ihina Mahajan contributed equally to this work.}

\address[1]{Department of Mechanical Engineering, Indian Institute of Technology Ropar, Rupnagar-140001, Punjab, India}

\begin{abstract}
This manuscript proposes an optimization framework to find the tailor-made functionally graded material (FGM) profiles for thermoelastic applications. This optimization framework consists of (1) a random profile generation scheme, (2) deep learning (DL) based surrogate models for the prediction of thermal and structural quantities, and (3) a genetic algorithm (GA). From the proposed random profile generation scheme, we strive for a generic design space that does not contain impractical designs, i.e., profiles with sharp gradations. We also show that the power law is a strict subset of the proposed design space. We use a dense neural network-based surrogate model for the prediction of maximum stress, while the deep neural operator DeepONet is used for the prediction of the thermal field. The point-wise effective prediction of the thermal field enables us to implement the constraint that the metallic content of the FGM remains within a specified limit. The integration of the profile generation scheme and DL-based surrogate models with GA provides us with an efficient optimization scheme. The efficacy of the proposed framework is demonstrated through various numerical examples.

\end{abstract}
\maketitle

\section{Introduction}

Functionally Graded Materials (FGMs) such as Ni-Al$_{2}$O$_{3}$, Al-ZrO$_{2}$ are advanced heterogeneous materials that are specifically tailor-made to have a spatial variation of the material properties depending on the functional requirements \citep{miyamoto2013functionally}. The ability to tailor volume gradation profiles is of significant value in making them invaluable in the aerospace, biomedical, automotive, and electronics industries \citep{koizumi1997fgm, birman2007modeling}. Some of the specific applications of the FGM in extreme environments\citep{finot1996small, librescu2007material} are spacecraft heat shielding, heat exchanger tubes, plasma facings for fusion reactors, and advanced gas turbine engine nozzles. In this manuscript, our focus is on obtaining the tailored volume gradation profile of FGM for a given thermoelastic problem. The optimum FGM profile needs to satisfy the thermal and structural criterion, as well as generic criteria such as minimum weight. A stress measure such as maximum effective stress is used for structural criterion. For thermal criterion, the metallic contents of the FGM should not be subjected to more than a specified temperature threshold is adopted. The optimization framework of FGM in these thermoelastic settings involves profile generation, thermoelastic finite element analysis, DL based surrogate modeling and the deployment of a genetic algorithm optimization technique.

%%%%%%%%%%%%%%%%%%%%%%%%%%%%%%%%%%%%%%%5
The first step of the FGM optimization framework involves formulation of the design space consisting of various gradation profiles. In that regard, power-law distribution has been widely used for its simplicity in creating smooth transitions between materials. For instance, Jha et al., 2013 \citep{jha2013critical}; Kieback et al., 2003 \citep{kieback2003processing} utilized the power law to optimize thermal resistance and structural safety in FGM plates under thermal loads, achieving lower thermal stresses. Wattanasakulpong and Ungbhakorn (2012) \citep{wattanasakulpong2014linear}, have applied the power-law distribution to analyze vibration characteristics in functionally graded beams, demonstrating its versatility. Beyond the power law approach, exponential \citep{reddy2014three}  and sigmoidal \citep{yang2008free} distributions have also been explored. However, the representation capacity of these methods is limited, potentially leading to sub-optimal FGM profiles \citep{miyamoto2013functionally}.

On the other end, having a very large design space can lead to an inefficient optimization strategy.  In~\citep{do2019material}, first, random profiles are generated without any constraints, and then, to restrict the design space, the volume fraction gradient values are compared against a specified gradient value. This results in a moderate-sized design space, efficiently be used in optimization.  In this work, we propose a new profile generation scheme, which is generic but at the same time do not generate the profiles with high gradient. The proposed profile generation scheme is simple to implement and implicitly satisfies the requirement of having only smooth profiles. In the proposed scheme, the ratio of volume fraction values at successive grid points should be lower than bounded random number. The bounded nature ensures smoothness, while the randomness results in varied profiles. The proposed scheme eliminates the need for post-checking gradient values to ensure only smooth profiles in the design space. We also demonstrate that the design space obtained by the power law is a strict subset of our proposed design space. 

The next step involves finding the optimal profile from the design space of FGM. To achieve this, gradient-based \citep{golbahar2011three, huang2003thermal} and non-gradient-based methods are deployed in literature to find the optimum FGM profile. The non-gradient-based methods are more straightforward in implementation with various design spaces and constraints. The genetic algorithm is mostly used non-gradient-based method for the FGM optimization \citep{baykasouglu2017multiple, nguyen2017optimal, jamshidi2017optimal, venkataraman2003elasticity, ootao1999optimization} and here we employ this technique for the optimization purpose only. For instance, Goupee et al. \citep{goupee2006two} applied GA in optimizing FGMs for thermoelastic applications. Chiba et al. \citep{chiba2012optimisation} utilized GA to minimize thermal stress in an infinite FGM plate. Some of the other non-gradient-based methods that are used for the optimization of FGM include modified symbiotic organisms search \citep{do2019material}, imperialist competitive algorithm \citep{yas2014application}, Firefly Algorithm (FA) \citep{kamarian2014application}, and others \citep{roque2015differential, he2018multi}. In non-gradient-based optimization methods, one of the key steps is to calculate the fitness score, which determines the rank of each design within the generation. To evaluate the fitness scores, FEM is the most widely used numerical technique {\citep{maleki2015evolutionary,nguyen2017optimal, correia2019multiobjective,cho2002optimal, goupee2006two}. For example, Ngyuen et al. \citep{nguyen2017optimal} applied FEM to calculate the buckling load in thin-walled FGM beams. Correia et al. \citep{correia2019multiobjective} optimized a 3D FGM plate subjected to thermo-mechanical conditions by minimizing mass, stress, and cost. Cho et al. \citep{cho2002optimal} utilized FEA to determine stress values in their optimization under thermal loading conditions.

FEM although providing a reliable numerical methodology to compute the accurate stress/thermal fiel is computationally expensive in nature. The computationally intensive nature becomes one of the main hindrances to deploying the non-gradient optimization scheme in practical problems. DL-based surrogate computational models offer an alternate paradigm that, once trained can provide predictions with minimal computational effort. These DL-based surrogate models have gained popularity in both scientific and industrial applications. Even in the FGM optimization, they have been integrated into the optimization framework to evaluate the fitness score values. For example, Do et al. \citep{do2019material} integrated Deep neural networks (DNNs) with a modified symbiotic organism search (MSOS) algorithm to solve eigenvalue problems. Similarly, Ly et al. \citep{ly2022multi} used  Neural Networks (NNs) in the optimization to determine the natural frequency of FG beams. Baykasoglu et al. \citep{baykasouglu2017multiple} employed NN to calculate the peak force value during optimization using GA. Additionally, Yas et al. \citep{yas2014application} utilized DNN combined with the imperialistic competitive algorithm to predict the first natural frequency of FG beams. Ootao et al. \citep{ootao1998neural} employed NN to assess the thermal stress in optimizing volume fraction for the plate of FGM. From the literature, we find that the DL-based surrogate models, although widely deployed for FGM problems, are restricted to the prediction of specified real-valued metrics such as maximum stress, peak force, or eigenvalue and have not yet attempted to include point-wise thermal constraint in the optimization process.

In this manuscript, deep learning-based surrogate models have been deployed to predict the maximum effective stress and the thermal field. We have used DNN-based surrogate models to predict the maximum effective stress. While for the prediction of the temperature field in the domain, the operator-based deep learning networks, i.e., DeepONet, have been used. While DNN provides an effective strategy for predicting real-valued metrics such as maximum stress/maximum temperature, they are less effective in predicting the overall field. In this context, this operator learning \citep{li2021neural} represents a paradigm shift from traditional approaches like DNNs. In literature, attempts have been made to apply DeepONet for various structural and thermal problems \citep{shukla2024deep, garg2022assessment, liu2022causality, he2024sequential, liu2023operator}. However, to the best of our knowledge, till now, no attempts have been made to apply the operator technique for FGM optimization. The training data for these DL-based surrogate models has been generated through thermoelastic FEM analysis. The proposed profile generation scheme, thermoelastic FEM analysis, and deep learning-based surrogate models have been integrated into the genetic algorithm. We found that the DL-based surrogate models generally provide good accuracy; however, in limited cases, the accuracy of these DL models is low. This accuracy issue is due to the insufficient representation of the limiting cases in the training data. In view of this, we propose a hybrid strategy based on the DL-based surrogate models and FEM models for the fitness score calculation. The proposed hybrid strategy to calculate the fitness score ensures the twin requirement of the accuracy and computational efficiency of the fitness evaluations. 
 
In summary, this manuscript contributes to the field of designing the functionally graded materials in the following aspects: 
\begin{itemize}
    \item A novel profile generation scheme for FGM has been developed. This proposed profile generation scheme is generic and can be applied to various FGM applications.
    \item Deep learning-based surrogate models have been deployed for the fitness evaluation in the optimization. DNN is used to predict maximum effective stress, while DeepONet is used to predict the temperature field as a function of the volume gradation profile.
    \item To the best of our knowledge, this is the first attempt where point-wise thermal constraints are incorporated with deep-learning surrogate models.
    \item The proposed hybrid approach of using the FEM as well as DL-based surrogate models ensures both the accuracy and efficiency of the fitness function. 
    \item The overall efficacy of the proposed framework has been shown through various numerical simulations.
\end{itemize}

 The paper is organized as follows. Section 2 provides a detailed explanation of the proposed profile generation framework used to establish the design space in the optimization. Section 3 describes the finite element strategy for thermoelasticity problems in FGMs. The implementation and effectiveness of deep learning-based surrogate models are discussed in section 4. Section 5 provides details about the deployment of genetic algorithms for FGM optimization. Section 6 demonstrates the performance of the proposed framework on various numerical examples.

\section{Generation of the FGM design space}

In this work, our aim is to find the tailor made profiles of the Functionally Graded Material (FGM) for a specified thermoelastic application. To develop an effective optimization framework, we need first to formulate the design space for FGM profiles. The design space of the FGM should exhibit two main requirements for the optimal design: (1) The design space should be generic and consist of varied profiles. This requirement is essential to achieve high performance with the optimum solution. (2) The design space should not consist of unnecessary designs. Unnecessary profiles in the design space can lead to an increase in the computational cost of the optimization algorithm. Further, with a larger design space, training of deep learning-based surrogate models also becomes difficult.

Here, we propose a novel framework for generating the design space of FGM profiles. The proposed design algorithm is generic and, at the same time, has some basic restrictions to remove the impractical designs. Cases with abrupt or sharp changes in the volume fraction result in stress concentration and are thus considered as not desirable in nature. The proposed framework generates the design space having only smooth profiles.

\subsection{Generation of 1D profile}

In our framework, the first step towards generating FGM profiles involves creating one-dimensional profiles. We begin by representing the FGM profile through a volume fraction parameter denoted as $\phi$, where $\phi$ signifies the volume fraction of one of the distinct materials, while $1-\phi$ represents the volume fraction of the other material. In the proposed methodology, first, we consider the 1D FGM case, where \(\phi=0\) at one end and \(\phi=1\) at the other end. Note that in thermoelastic applications, generally, the end facing the high temperature is ceramic-rich, while the other end subjected to lower temperature is metal-rich. Thus, this setting is considered consistent with the practical applications of the FGM. Further, the proposed methodology is generic, and with only minimal modifications, the constraints of the \(\phi\) at the boundaries can be relaxed.

\begin{figure}[htbp]
  \centering
  \includegraphics[width=0.450\textwidth]{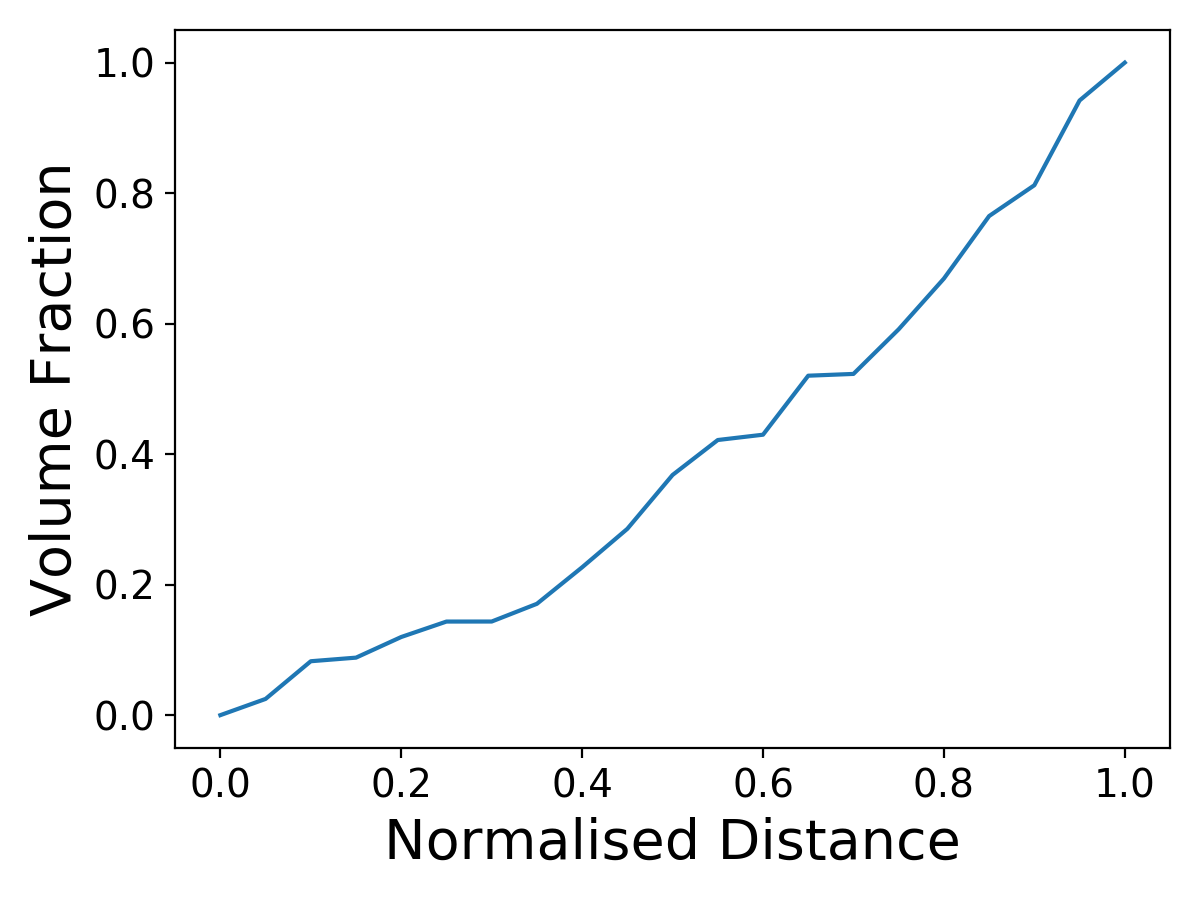}
  \caption{FGM profile with piece-wise linear interpolation.}
  \label{1d}
\end{figure}

In the proposed formulation, we consider that the variation of the volume fraction \(\phi\) is a piecewise linear function, as shown in Fig. \ref{1d}. To characterize this piecewise linear function, we discretize the domain into $(n)$ number of 1-D elements and  $(n+1)$ number of nodes. The function \(\phi\) can be uniquely determined by its values at the nodal points. The value of the volume fraction at the nodes is given by the vector $ \boldsymbol{\tilde{\phi}}$. In particular, the value of the volume fraction at any point in the domain can be obtained from the nodal volume fraction values using the equation given below:
\begin{equation}
\phi(x) = \boldsymbol{N \tilde{\phi_{e}}}    
\end{equation}
Where, \(x\in[x_{i},x_{i+1}]\). $\boldsymbol{\tilde{\phi_{e}}}$ is given by  $[\phi_{i},\;\phi_{i+1}]$ and linear interpolation functions $\boldsymbol{N}$ is given by:
\begin{equation}
\boldsymbol{N} = \left[\frac{1-\xi}{2},\quad \frac{1+\xi}{2} \right]   \nonumber
\end{equation}
where $\xi \in [-1, 1]$ is a parametric coordinate within an element.

We now describe the algorithm to generate the volume fraction values at the nodes. Since we are considering the case where the volume fraction value is always zero at one end, the nodal value at the zeroth node \(\phi_0\) is taken as zero. The value of the volume fraction at the first node is generated randomly using a uniform distribution within a specified range. Some specific considerations for this random value generation are required to have higher variance in the 1D profiles, which are discussed later in this section. After obtaining the volume fraction at the first node, the volume fraction at the successive nodes can be determined by the following recursive equation:
\begin{equation}
 \phi_{i+1} = \text{min}\;(1,\alpha_{i}\phi_{i}).   
\end{equation}

Where $\alpha_{i}$ is a random parameter independently generated from uniform distribution at all nodes. Further, $\alpha_{i} \in [\alpha_{l}, \alpha_{u}]$,  where, $\alpha_{l}$ and $\alpha_{u}$ are the lower limit and upper limit respectively.  In this manuscript, the $\alpha_{u}$ is generated randomly for each profile within a specified range, while the value of the $\alpha_{l}$ is taken as one for each profile. The lower bound $\alpha_{l}$ as one ensures the monotonicity of the generated FGM profiles. The lower bound can be generated randomly within [$a$, 1] where $a$ is less than one if monotonicity in the profiles is not desired.

\begin{figure}[htbp]
  \centering
  \includegraphics[scale=0.35]{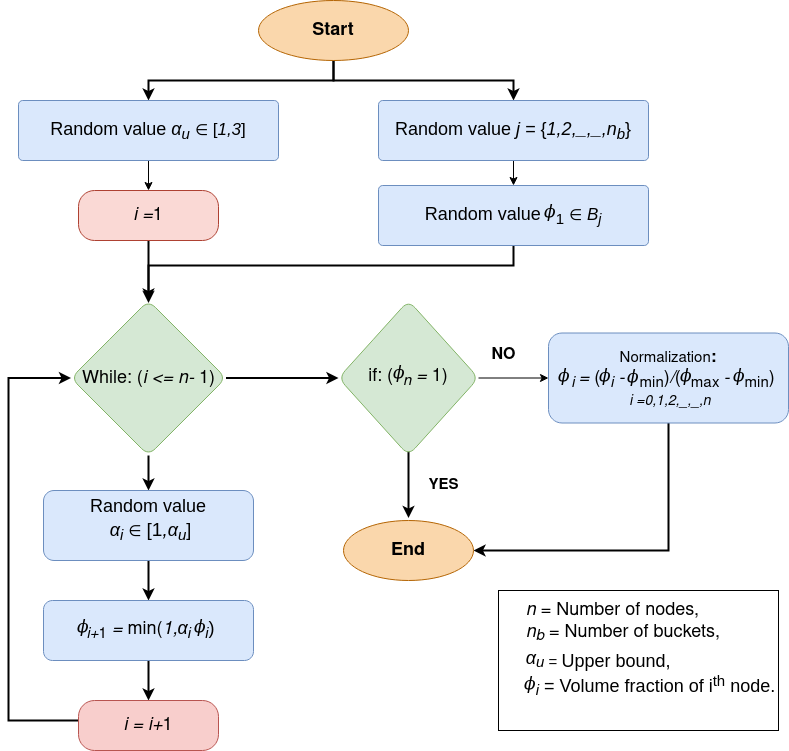}
  \caption{Flowchart of proposed FGM profile generation scheme.}
  \label{fig:profile generation}
\end{figure}

The overall proposed formulation for generating  FGM profiles is given in Fig. \ref{fig:profile generation}, and some of the obtained profiles from the formulation are shown in Fig. \ref{proposed_profile}. As shown in Fig. \ref{fig:profile generation}, we start by initializing the value of $\alpha_{u}$ and the volume fraction of the first node $\phi_{1}$. $\alpha_{u}$ is a random number generated from the range \([1, b]\), where \(b\) is always greater than 1. For example, in our problems, we have prescribed the value of \(b\) to be 3. Note that it is critical to take $\alpha_{u}$ as a random value, if we keep $\alpha_{u}$ as a fixed value, the overall variation in the design space is significantly limited. The range of the first nodal values is split into the number of buckets. For example in our case, \(\phi_1\) is split into two buckets $B = B_{1} \cup B_{2}$, where $B_{1}$=[0.001, 0.01] and $B_{2}$=[0.01, 0.1]. This split in buckets ensures that the profiles with low values (bucket $B_{1}$) as well as with a slightly higher range of values (bucket $B_{2}$) of \(\phi_1\) with equal probabilities are there in the design space. In the absence of multiple buckets, only the designs with the higher volume fraction range are likely to be there. The subsequent nodal volume fraction is obtained by multiplying the previous nodal volume fraction by the random parameter $\alpha_{i}$. In case the value of the volume fraction at the last node (\(\phi_n\)) is less than one,  we apply normalization to the vector $\boldsymbol{\phi}$, as mentioned in the flowchart. Note if the value of the \(\phi_n\)  is not desired to be always equal to one, this normalization step can be omitted.

\begin{figure}[htbp]
  \centering
  \begin{subfigure}[b]{0.45\textwidth}
    \centering
    \includegraphics[width=\textwidth]{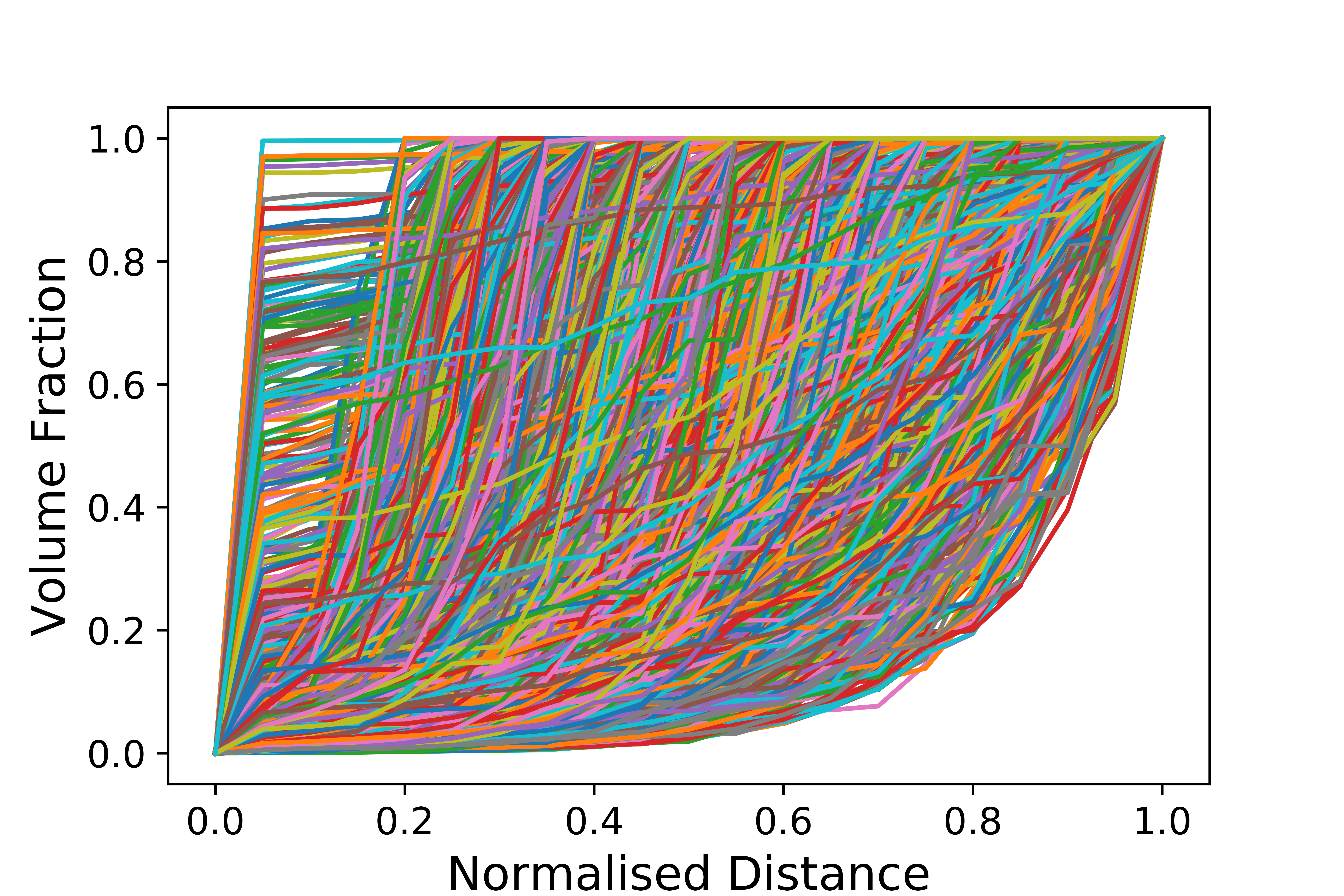}
    \caption{}
    \label{proposed_profile}
  \end{subfigure}
  \hfill
  \begin{subfigure}[b]{0.45\textwidth}
    \centering
    \includegraphics[width=\textwidth]{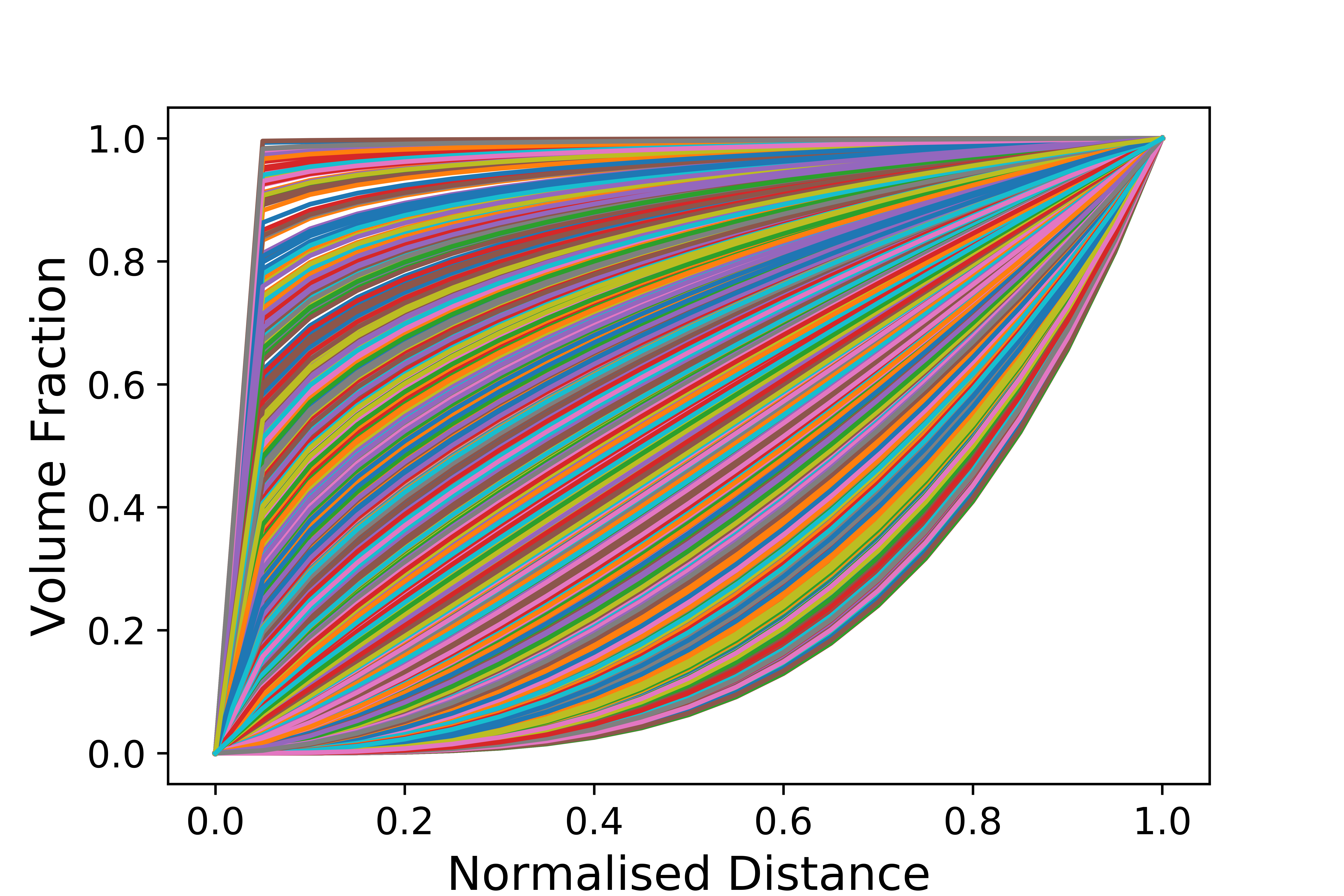}
    \caption{}
    \label{}
  \end{subfigure}
  \caption{Comparison of 1D FGM profiles from (a) proposed scheme, (b) power law.}
  \label{fig:2}
\end{figure}

It can be shown that the above-proposed profile generation scheme is more generic than the power law scheme. In particular, here, we show that the design space generated by the power law scheme is a subset of the proposed methodology. In the case of power law, the volume fraction at a particular location \(x\) is given by the following expression: 
\begin{equation}
    \phi^{(p)}(x)=\left(\frac{x}{L}\right)^m
    \label{power_law}
\end{equation}
Here, \(m\) is the power law index, whose typical range is between [0-4]. To show mathematically that the profiles generated by the power law can also be generated by our design scheme, we calculate the volume fraction ratio obtained from power law at successive locations:
\begin{align}
%    \phi_{1} &= a\left(\frac{\Delta x}{L} \right)^m,\\
    \phi^{(p)}_{i} &= \left(\frac{x_{i}}{L} \right)^m,\\
    \phi^{(p)}_{i+1} &= \left(\frac{x_{i}+\Delta x}{L} \right)^m,\\
    \frac{\phi^{(p)}_{i+1}}{\phi^{(p)}_{i}} &= \left(1 + \frac{\Delta x}{x_{i}}\right)^m,
    \label{6}
\end{align}

assuming that the ${\Delta x}/{x_{i}} \ll 1$, then by using binomial expansion in Eqn \ref{6}, we get:
\begin{align}
    \frac{\phi^{(p)}_{i+1}}{\phi^{(p)}_{i}} &\approx  1 + m\frac{\Delta x}{x_{i}} = 1 + \left(\frac{m}{n\beta_i}\right)
    \label{factor power law}
\end{align}
where, $n$+1= number of nodes used for profile discretization, $\beta_i$ = $x_{i}/L$, $\Delta x$ = $L/n$, $\beta_{i} \in [0,1]$. Further from power law, the volume fraction value at the first node \(\phi^{(p)}_1\) is equal to the $(1/n)^m$. So if we take the value of the \(\alpha_i\) equal to the right-hand side of the Eqn. \eqref{factor power law} and \(\phi_1\) as equal to \(\phi^{(p)}_1\), then the profile obtained from the proposed scheme will be almost identical to the power law scheme. Further, the proposed scheme is more generic since, as shown in Fig. \ref{diffrent profiles}, numerous profiles can be generated by the proposed scheme, which the power law scheme can not generate. 

\begin{figure}[htbp]
  \centering
  \includegraphics[width=0.5\textwidth]{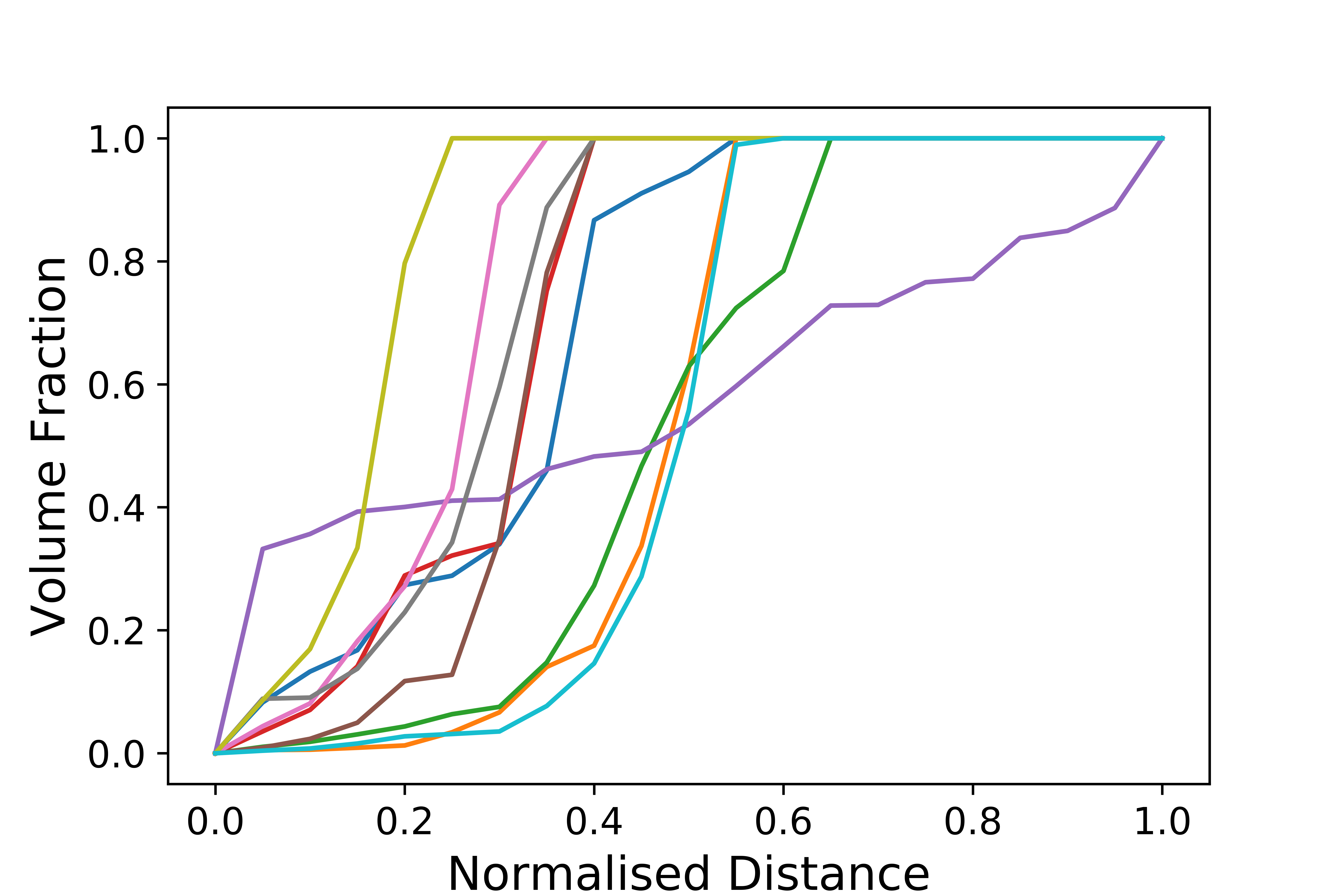}
  \caption{Sample profiles from the proposed profile generation scheme, that can not be represented by power law.}
  \label{diffrent profiles}
\end{figure}

\subsection{Generation of 2D profile}
\label{profile_generation}
In this subsection, we extend our 1D profile generation approach to the two-dimensional domain. We generate 2D profiles by combining two 
independent profiles, $\boldsymbol{\phi^x}$ and $\boldsymbol{\phi^y}$, obtained along the $x$-axis and $y$-axis, respectively. The 2D profile, 
denoted as $\boldsymbol{\phi^{xy}}$, is derived through the tensorial product of $\boldsymbol{\phi^x}$ and $\boldsymbol{\phi^y}$, and is expressed as:

\begin{equation}
\boldsymbol{\phi^{xy}} = \boldsymbol{\phi^x} \otimes \boldsymbol{\phi^y} 
 \end{equation}
 
This operation yields a 2D representation of the material distribution, encapsulating variations in the $x$ and $y$ directions. The generation of $\boldsymbol{\phi^x}$ and $\boldsymbol{\phi^y}$ is entirely independent
of each other. The lower bound values, $\alpha^{x}_{l}$ and $\alpha^{y}_{l}$, both equal to one, produce monotonic independent profiles $\boldsymbol{\phi^x}$ and $\boldsymbol{\phi^y}$ along the x-axis and y-axis, respectively. Consequently, the resultant profile
$\boldsymbol{\phi^{xy}}$, generated by the tensorial product of $\boldsymbol{\phi^x}$ and $\boldsymbol{\phi^y}$, is also monotonic in nature along both axes. The upper bounds $\alpha^{x}_{u}$ and $\alpha^{y}_{u}$ ensure smooth gradation of volume fraction along both axes. 

Several typical 2D profiles resulting from the combination of different $\boldsymbol{\phi^{x}}$ and $\boldsymbol{\phi^{y}}$ profiles are illustrated in Fig. \ref{2D FGM profiles}. These profiles are generated using $\alpha^{x}_{u}$ and $\alpha^{y}_{u}$ equal to three and $\alpha^{x}_{l}$ and $\alpha^{y}_{l}$ equal to one. The range of first node value ($\phi^{x}_{1}$) for the profile $\boldsymbol{\phi^{x}}$ is obtained using single bucket $B^{x} \in [0.001,1]$. For the range of first node value ($\phi^{y}_{1}$) for the profile $\boldsymbol{\phi^{y}}$, the range is split into two buckets: $B^{y} = B^{y}_{1} \cup B^{y}_{2}, $ where $B^{y}_{1} \in [0.001,0.01] $ and $ B^{y}_{2} \in [0.01,0.1]$. The value of the volume fraction at any point in the domain can be obtained from the nodal volume fraction values using the equation given below:
\begin{equation}
\phi(x,y) = \boldsymbol{N \tilde{\phi}^{xy}_{e}}  
\label{interpolation_eq}
\end{equation}
Where, \(x\in[x_{i},x_{i+1}]\).   \(y\in[y_{j},y_{j+1}]\), $\boldsymbol{\tilde{\phi}^{xy}_{e}}$ is given by  $[\phi^{xy}_{i,j},\;\phi^{xy}_{i+1,j},\; \phi^{xy}_{i+1,j+1},\; \phi^{xy}_{i,j+1}]$ and linear interpolation functions $\boldsymbol{N}$ is given by:
\begin{equation}
\boldsymbol{N} = \left[\frac{(1-\xi)(1-\eta)}{4},\quad \frac{(1+\xi)(1-\eta)}{4}, \quad \frac{(1+\xi)(1+\eta)}{4}, \quad \frac{(1_\xi)(1+\eta)}{4} \right]   \nonumber
\end{equation}
where $\xi, \eta \in [-1, 1]$ are parametric coordinates within an element. 

These 2D profiles demonstrate the diverse material compositions and spatial distributions achievable through our generation framework, laying the groundwork for comprehensive analysis and optimization of FGM in two dimensions.

\begin{figure}[htbp]
  \centering
  \includegraphics[width=0.8\textwidth]{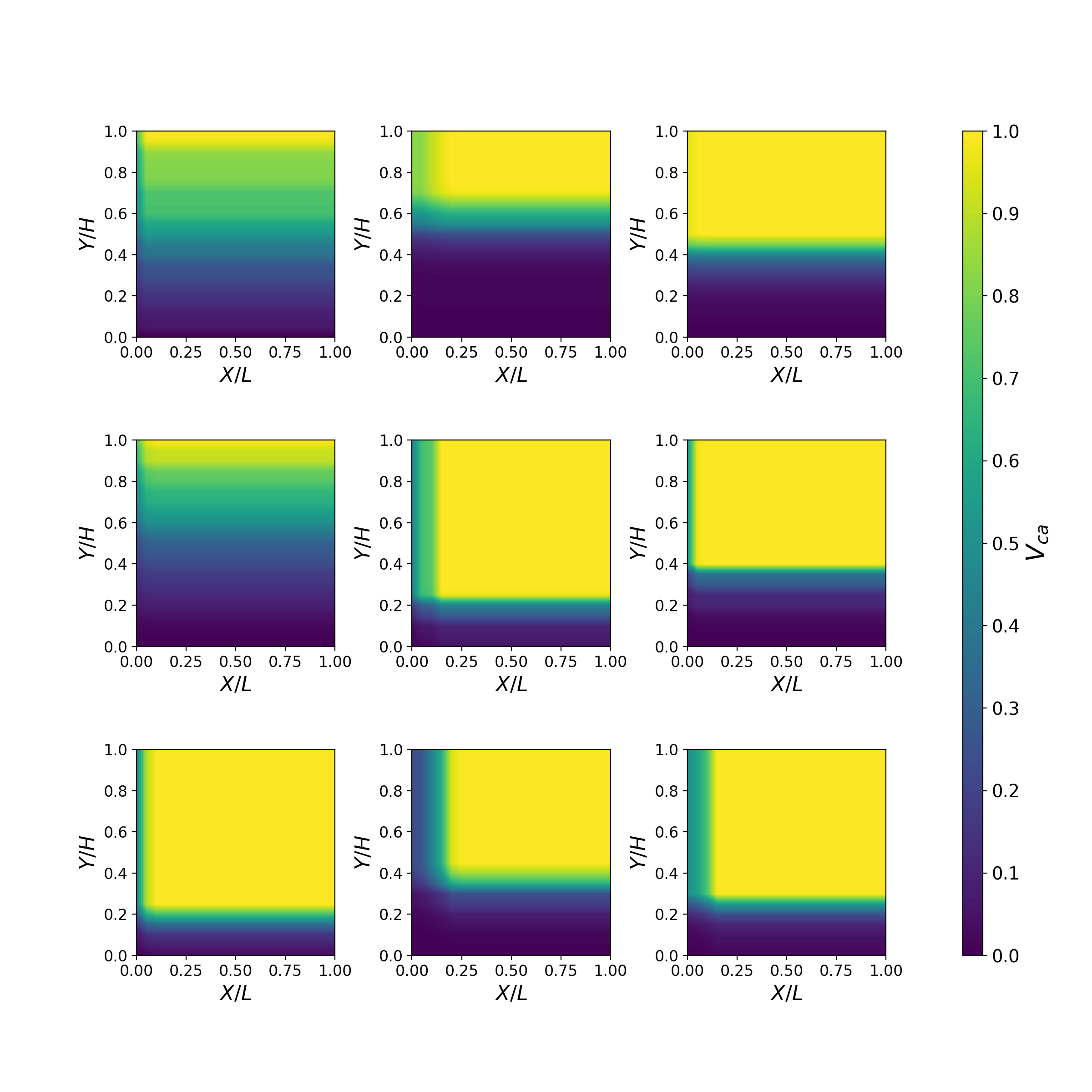}
  \caption{Sample two-dimensional FGM profiles obtained using profile generation scheme.}
  \label{2D FGM profiles}
\end{figure}

\section{Finite element analysis of thermoelasticity problems}
\label{FEM}
We consider the infinitesimal static thermoelastic deformation of the body occupying domain \(\mathit{\Omega}\). For this thermoelastic analysis, we conduct a linear thermal analysis followed by a linear elastic analysis using the finite element technique.

\subsubsection{Governing equation}
The temperature within the domain is governed by the balance of heat equation followed by Fourier's law of heat conduction:
\begin{equation}
 \boldsymbol{\nabla} \cdot \boldsymbol{q} + Q = 0 \quad \text{on } \mathit{\Omega},
 \label{feth1}
 \end{equation}
\begin{equation}
  \boldsymbol{q} = -k\boldsymbol{\nabla}\theta,
\end{equation}
 where $\boldsymbol{q}$ is the heat flux, $Q$ is the internal heat source, $k$ is the thermal conductivity, $\theta$ is the change in temperature from the reference point.\\

 Thermal boundary conditions are imposed in the following forms: specified heat flux $\hat{q}$ in the direction normal to $\mathit{\Gamma}_{q}$, convective heat transfer specified on the boundary $\mathit{\Gamma}_{c}$, and temperature $\hat{\theta}$ prescribed on the boundary $\mathit{\Gamma}_{\theta}$.\\

\begin{equation}
  \theta = \hat{\theta} \quad \text { on } \mathit{\Gamma}_{\theta},
 \end{equation}
 \begin{equation}
-\boldsymbol{q\cdot n} = \hat{q} \quad\text { on } \mathit{\Gamma}_{q},
 \end{equation}
 \begin{equation}
 -\boldsymbol{q \cdot n} +h(\theta - \theta_{\infty})= 0 \quad\text { on } \mathit{\Gamma}_{c},
 \end{equation}
Where, $\boldsymbol{n}$ is the unit vector normal to the boundary, $h$ is the coefficient of convection heat transfer and $\theta_{\infty}$ is the surrounding temperature.\\

After performing the thermal analysis, linear structural analysis is carried out with the help of the following governing and constitutive equations:
 \begin{equation}
\boldsymbol{\nabla\cdot\sigma} + \boldsymbol{b} = 0  \quad\text { on } \mathit{\Omega}, 
\label{fe1}
 \end{equation}
\begin{equation}
\boldsymbol{\sigma} = \boldsymbol{\sigma_{m}} - \beta\theta \boldsymbol{I},
\label{fe2}
 \end{equation}  
 \begin{equation}
\boldsymbol{\sigma_{m}} = \lambda(tr\boldsymbol{\epsilon})\boldsymbol{I} + 2\mu\boldsymbol{\epsilon},
\label{fe3}
 \end{equation} 
Where, $\boldsymbol{\sigma}$ is the Cauchy stress tensor, $\boldsymbol{\sigma_{m}}$ is the isothermal stress tensor, $\boldsymbol{b}$ is the body force, $\alpha$ is the thermal expansion coefficient, $\lambda$ and $\mu$ are the Lame constants, $\boldsymbol{I}$ is identity tensor. The infinitesimal strain tensor $\boldsymbol{\epsilon}$ is given in the terms of displacement vector by:
\begin{equation}
\boldsymbol{\epsilon} = \frac{1}{2}((\nabla\boldsymbol{u}) + ({\nabla}\boldsymbol{u})^T),
\label{fe4}
 \end{equation}

Mechanical boundary conditions are applied in the form of prescribed traction $\overline{\mathbf{t}}$ on the boundary $\mathit{\Gamma}_{t}$ and displacement $\boldsymbol{u}$ on the boundary $\mathit{\Gamma}_{u}$.
 \begin{equation}
  \boldsymbol{t}=\overline{\boldsymbol{t}} \quad \text { on } \mathit{\Gamma}_{t},
  \label{fe5}
 \end{equation}
 \begin{equation}
 \boldsymbol{u}=\boldsymbol{u_{0}} \quad \text { on } \mathit{\Gamma}_{u}, 
 \label{fe6}
\end{equation}

\subsubsection{Finite element formulation}
In FEM analysis, we have considered one-way coupling between thermal analysis (to determine the temperature distribution) and structural analysis (to determine the stress distribution). For the thermal analysis, the weak form is obtained by multiplying the variation $\theta_{\delta}$ and subsequently integrating by parts in Eqn. \eqref{feth1}:
\begin{equation}
 \int_{\mathit{\Omega}} \boldsymbol{\nabla} \theta_{\delta}.(k\boldsymbol{\nabla}\theta)\; d \mathit{\Omega} - \int_{\mathit{\Omega}} \theta_{\delta}Q \; d \mathit{\Omega}  -\int_{\mathit{\Gamma}_{q}} \theta_{\delta} \hat{q} \; d \mathit{\Gamma} + \int_{\mathit{\Gamma}_{c}} \theta_{\delta}h(\theta - \theta_{\infty}) \; d \mathit{\Gamma} = 0 \quad \forall \boldsymbol{\theta}_\delta,
\end{equation}
By using the variational form, we obtain discretized finite element equation for the linear thermal analysis.\\
\begin{equation}
  \boldsymbol{K}_{\theta\theta}\overline{\boldsymbol{\theta}} = \boldsymbol{f}_{\theta},
\end{equation}
where the matrix $\boldsymbol{K}_{\theta\theta}$ and $\boldsymbol{f}_{\theta}$ have the following meanings:\\
 \begin{equation}
\boldsymbol{K}_{\theta\theta} = \int_{\mathit{\Omega}} \boldsymbol{B}_{\theta}^T k\boldsymbol{B}_{\theta}\;d\mathit{\Omega} + \int_{\mathit{\Gamma}_{c}}h\boldsymbol{N}_{\theta}^T\boldsymbol{N}_{\theta}\;d\mathit{\Gamma},
\end{equation}
 \begin{equation}
\boldsymbol{f}_{\theta} = \int_{\mathit{\Omega}} Q\boldsymbol{N}_{\theta}^T\;d\mathit{\Omega} + \int_{\mathit{\Gamma}_{q}}\boldsymbol{N}_{\theta}^T\;d\mathit{\Gamma} + \int_{\mathit{\Gamma}_{c}}h\theta_{\infty}\boldsymbol{N}_{\theta}^T\;d\mathit{\Gamma},
\end{equation}

Next, we perform an elastic analysis to determine the stress distribution in the domain of the body. These stresses develop due to mechanical and thermal loading. The weak form for the elastic analysis is derived by multiplying $\boldsymbol{u}_{\delta}$ and subsequently integrating by parts in Eqn. \eqref{fe1}:
\begin{equation}
\int_{\mathit{\Omega}} \boldsymbol{\sigma_{m}}:\boldsymbol{\nabla} \boldsymbol{u}_{\delta}\;d \mathit{\Omega} - \int_{\mathit{\Omega}}\boldsymbol{u}_{\delta}.\boldsymbol{b}\; d \mathit{\Omega} + \int_{\mathit{\Gamma}_{t}}\boldsymbol{u}_{\delta}.\boldsymbol{t} \;d\mathit{\Gamma}_{t} + \int_{\mathit{\Omega}} \beta\theta \boldsymbol{\nabla}.\boldsymbol{u}_{\delta}\;d \mathit{\Omega} = 0 \quad \forall \boldsymbol{u}_\delta,
\end{equation}

In the FEM analysis, we use nine-noded quadrilateral elements. We apply a 3$\times$3 points Gauss quadrature rule in 2D for integration. The volume fraction values are linearly interpolated as given by the Eqn. \eqref{interpolation_eq}. This linear interpolation of the volume fraction ensures that the properties remain non-negative throughout the FEA. Properties such as Young's modulus, thermal conductivity, and thermal expansion coefficient are calculated at the Gauss points using the rule of mixtures given by the Eqn. (\ref{rule of mixture}).
\begin{equation}
    P(x,y) = P_{m}\phi_{m}(x,y) + P_{c}(1-\phi_{m}(x,y)),
    \label{rule of mixture}
\end{equation}
where, $P(x,y)$ = Property at the Gauss point, $P_{m}$ = Property of the metallic phase, $P_{c}$ = Property of the ceramic phase, $\phi_{m}$ = Volume fraction of metallic phase at the gauss point.

In our study, effective stress is defined as:
\begin{equation}
\sigma^e = \sqrt{\frac{3}{2}\tilde{\sigma}_{ij} \cdot \tilde{\sigma}_{ji}}\;,\quad \  \text{where:}\;  \tilde{\sigma} = \boldsymbol{\sigma_{m}} - \frac{1}{3}(\text{tr}\boldsymbol{\sigma_{m}})\boldsymbol{I}.     
\end{equation}

\section{Deep learning based surrogate models}
In this section, we describe the deep learning-based surrogate models utilized in FGM optimization. These surrogate models include a deep neural network (DNN) for predicting the maximum effective stress and a deep operator network (DeepONet) for predicting the temperature field.

\subsection{Deep neural network implementation}

Here, we aim to predict maximum effective stress ($\sigma^{e}_{max}$) values from FGM profiles using a deep neural network (DNN). The DNN is designed to establish a relationship between the input, consisting of FGM profiles ($\boldsymbol{\phi^x}$, $\boldsymbol{\phi^y}$), and the output, representing the $\sigma^{e}_{max}$.

The DNN is built using a sequential model, which includes an input layer, several hidden layers, and an output layer. The architecture incorporates multiple hidden layers with different numbers of neurons to efficiently model complex relationships within the dataset. This structure allows the network to process input data and generate the desired output, adapting to the unique characteristics of the problem at hand.

The DNN's configuration, as shown in Fig. \ref{DNN_diagram}, demonstrates the layer-by-layer structure and the corresponding neuron counts. This design provides flexibility and robustness, enabling the network to capture patterns and relationships essential for accurate predictions.

\begin{figure}[htbp]
%  \captionsetup{font=large}
  \centering
  \includegraphics[width=0.7\textwidth]{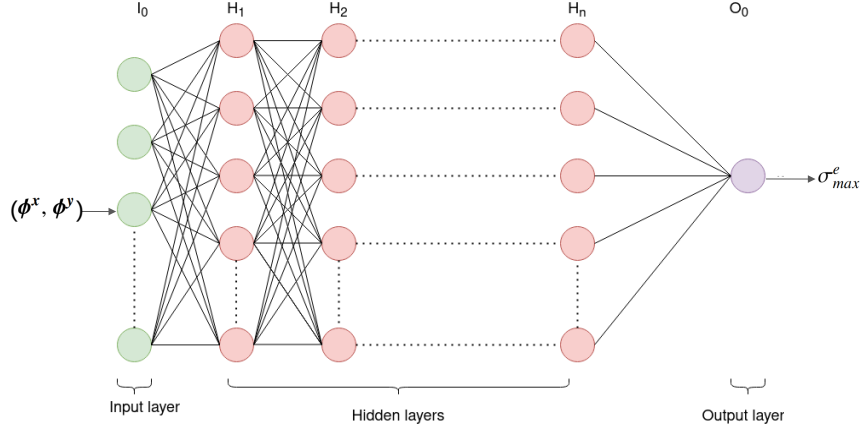}
  \caption{Schematic representation of a typical DNN.}
  \label{DNN_diagram}
\end{figure}

The neurons in the hidden layers operate based on a weighted sum of inputs defined by Eqn. \ref{eq:neuron_output}. The output of each neuron is determined by applying an activation function to the weighted sum of the inputs, with \(\omega_{kj}\) representing the weights, \(x_j\) denoting the inputs, \(b_k\) as the bias term, and \(\chi()\) as the activation function. The output calculation for each neuron is given by:

\begin{equation}
y_k = \chi\left( \sum_{j=1}^{m} \omega_{kj} x_j + b_k \right).
\label{eq:neuron_output}
\end{equation}

The weights and bias of the networks are obtained by minimizing the mean square error on a training dataset. We use R-squared (R²) as a metric to measure the correlation between predicted and actual stress values. As demonstrated in the numerical section, this DNN implementation effectively predicted $\sigma^{e}_{max}$ values based on the given input features.

\subsection{DeepONet implementation}
\label{Deeponet}
In this section, we describe the application of DeepONet for predicting temperature profiles from given FGM profiles. DeepONet is utilized to predict the temperature at various points within the domain by learning the dependence of temperature on the volume fraction of the FGM profile ($\boldsymbol{\phi^{x}}$,  $\boldsymbol{\phi^{y}}$) and the spatial location $(x,y)$ within the domain.\\
A typical DeepONet architecture, as shown in Fig. \ref{fig:5_d}, consists of two independent sub-neural networks called branch and trunk networks. The branch network learns the dependence of the input field ($\boldsymbol{\phi^{x}}$, $\boldsymbol{\phi^{y}}$) on the output field ($\boldsymbol{T(x,y)}$) and the trunk network learns the spatial dependence of the temperature field. The output layer dimension of both these sub-networks is kept the same, and then their dot product is taken to obtain the temperature field through split network architecture. 

%This way DeepONet is able to capture the dependence of $\boldsymbol{T}(x,y)$ on $\boldsymbol{\phi^{x}}$,  $\boldsymbol{\phi^{y}}$ and $\boldsymbol{(x,y)}$ effectively.
\begin{figure}[htbp]
%  \caAdilptionsetup{font=normalsize}
  \centering
  \includegraphics[width=0.7\textwidth]{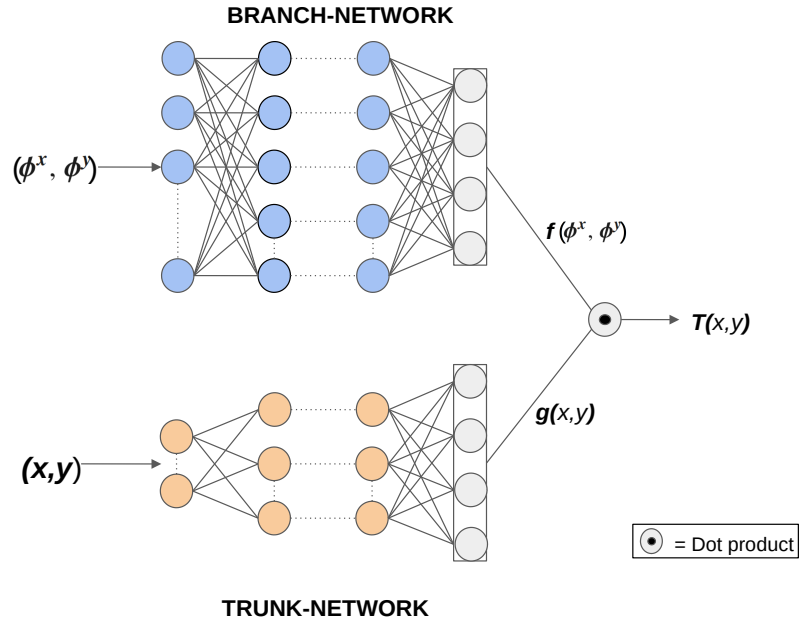}
  \caption{Schematic representation of a  DeepONet.}
  \label{fig:5_d}
\end{figure}
If we assume the two independent vector fields  \(\boldsymbol{f}\) and  \(\boldsymbol{g}\) of identical dimension \(c\). Wherein, \(\boldsymbol{f}(\boldsymbol{\phi^{x}},\boldsymbol{\phi^{y}})\) is only dependent on the source field \(\boldsymbol{\phi^{x}}\) and \(\boldsymbol{\phi^{y}}\), while the \(\boldsymbol{g}(x,y)\) is only dependent on the spatial coordinates. With these function choices, the output from the DeepONet network  can be  written in the following functional form:
\begin{equation}
    T(\boldsymbol{\phi^{x}}, \boldsymbol{\phi^{y}})(x,y) = \sum_{i=1}^{c}f_i(\boldsymbol{\phi^{x}}, \boldsymbol{\phi^{y}})\cdot g_i(x,y) 
    \label{e2}
\end{equation}
The term $T(\boldsymbol{\phi^{x}}, \boldsymbol{\phi^{y}})(x,y)$ represents the output field, while the vector fields  \(\boldsymbol{f}\) and  \(\boldsymbol{g}\) are approximated by branch and trunk network respectively. Through the split network, DeepONet is able to learn the operator transforming the input field ($\boldsymbol{\phi^x}, \boldsymbol{\phi^y}$) to the output field $T(\boldsymbol{\phi^{x}}, \boldsymbol{\phi^{y}})(x,y)$ effectively.

\begin{figure}[htbp]
  \centering
  \includegraphics[width=.60\textwidth]{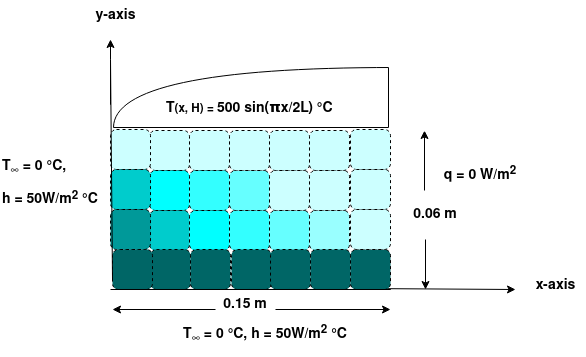}
  \caption{$\mathrm{Al}/\mathrm{ZrO_2}$ FGM plate, the thermal analysis of this plate is used demonstrate the efficacy of DeepONet.}
  \label{deeponet_diagram}
\end{figure}

To demonstrate the DeepONet capacity to predict the thermal field, we present a thermal analysis of a 2D FGM plate. The schematic diagram of one half of the plate is shown in Fig. \ref{deeponet_diagram}. The properties of material aluminum and zirconia are given in Table \ref{prop_2}. For the mentioned problem, the architecture of DeepONet consists of one hidden layers in the branch net and three hidden layers in the trunk net. Each hidden layer consists of 200 neurons. The output of the branch and trunk net consists of 250 neurons. A decaying learning rate is used for the training, with parameters: initial learning rate = 0.001, epochs = 10, batch size = 1024, reducing to 0.0001 for 20 epochs and batch size = 1024, and finally, learning rate becomes constant 0.0001 with 20 epochs and batch size 256. This model is compiled using Adam optimizer with mean square error (MSE) as a loss function and. The total dataset is 8000, with an 80-20 split into training and testing datasets. The model achieved an $\text{R}^2$ score of  0.9998 on the training data and 0.9994 on the testing data. Further, the accuracy of the model towards the temperature field prediction for the different FGM profiles is demonstrated in Fig. \ref{deep_prediction}.

%\blindtext

\begin{figure}[htbp]
    \includegraphics[width=.24\textwidth]{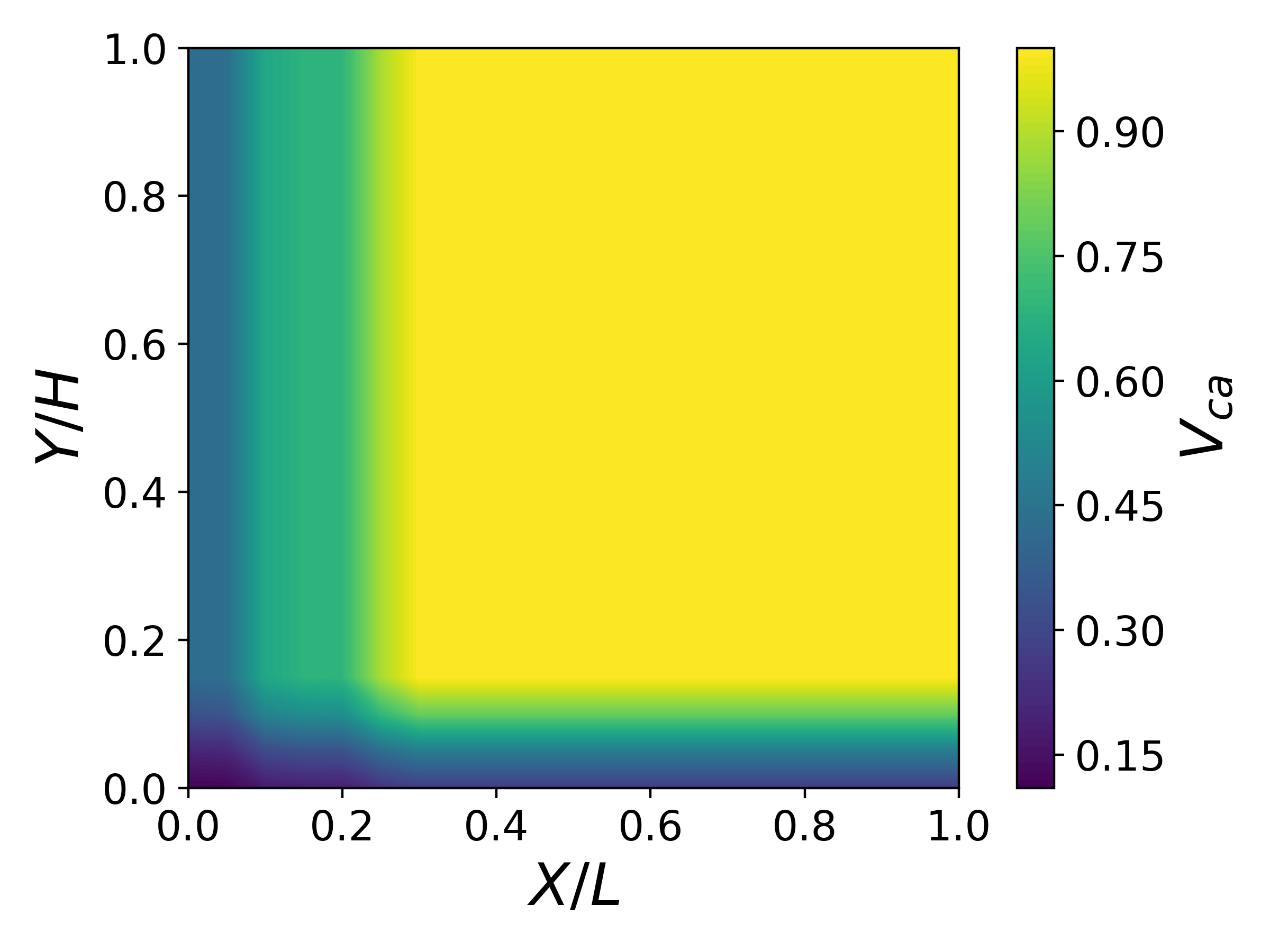}\hfill
    \includegraphics[width=.24\textwidth]{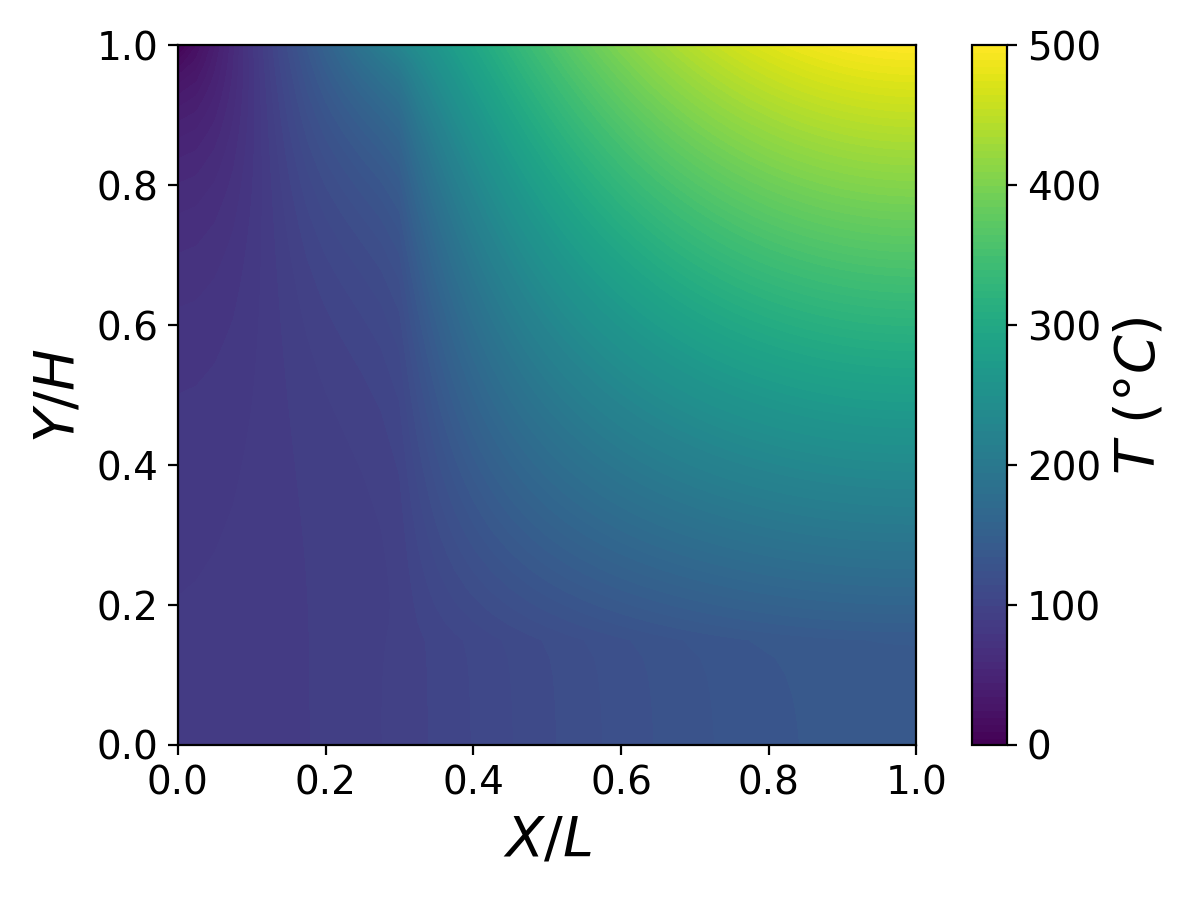}\hfill
    \includegraphics[width=.24\textwidth]{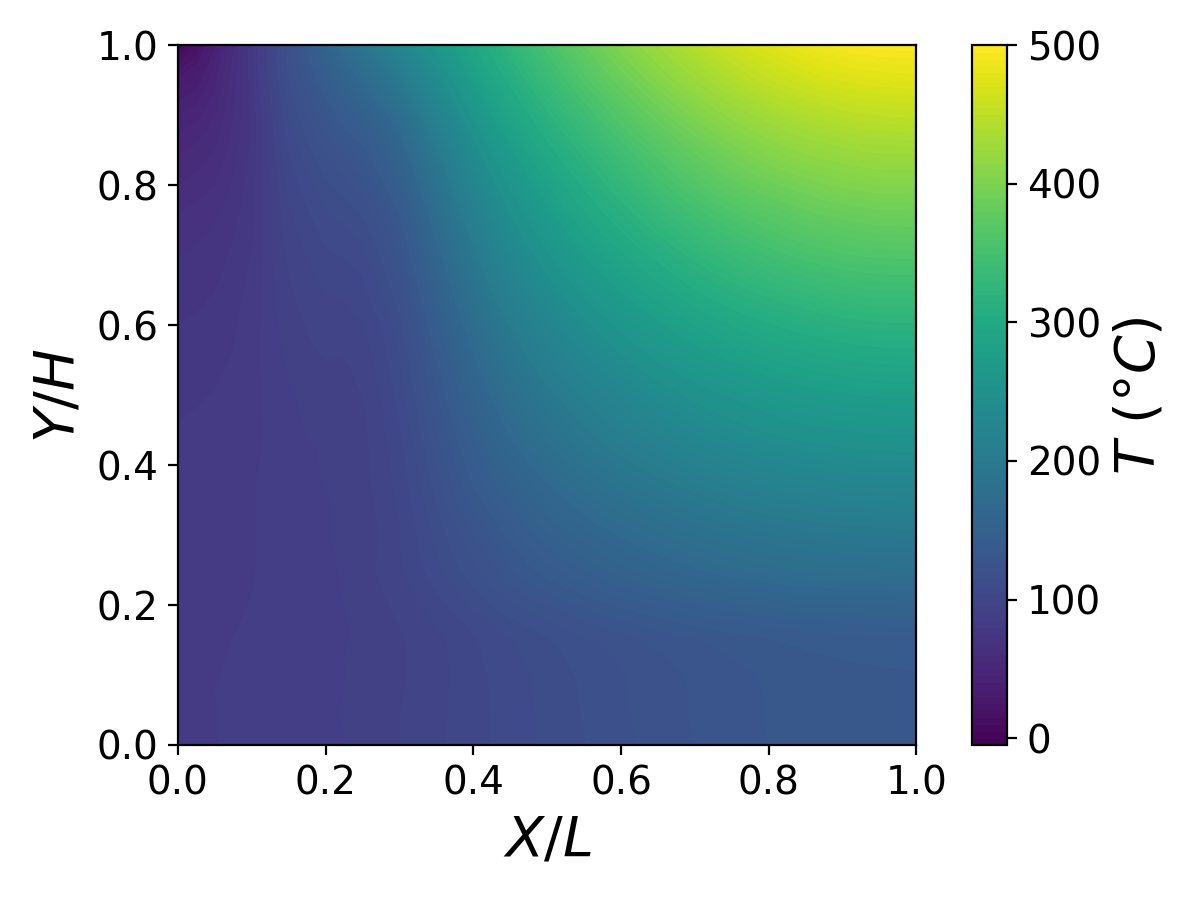}\hfill
    \includegraphics[width=.24\textwidth]{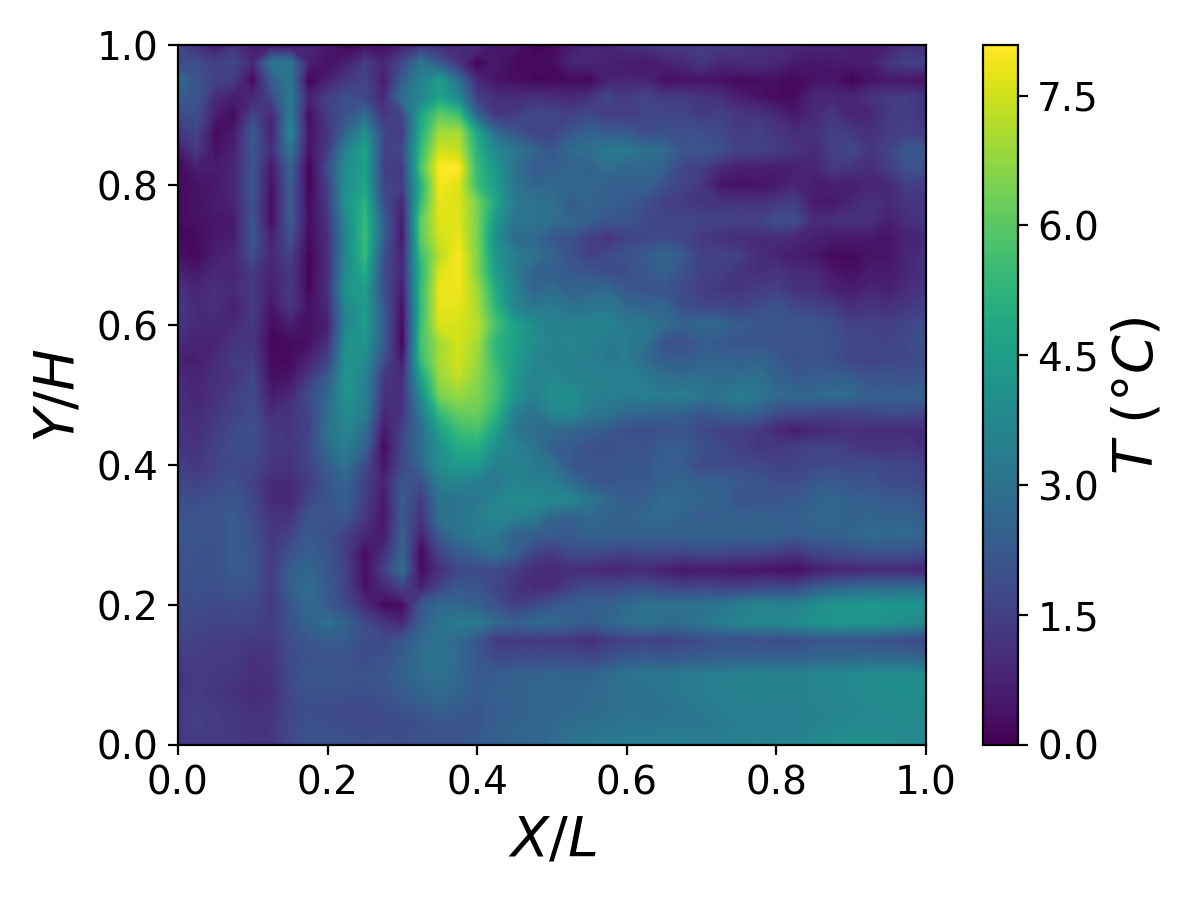}
    \\[\smallskipamount]
    \includegraphics[width=.24\textwidth]{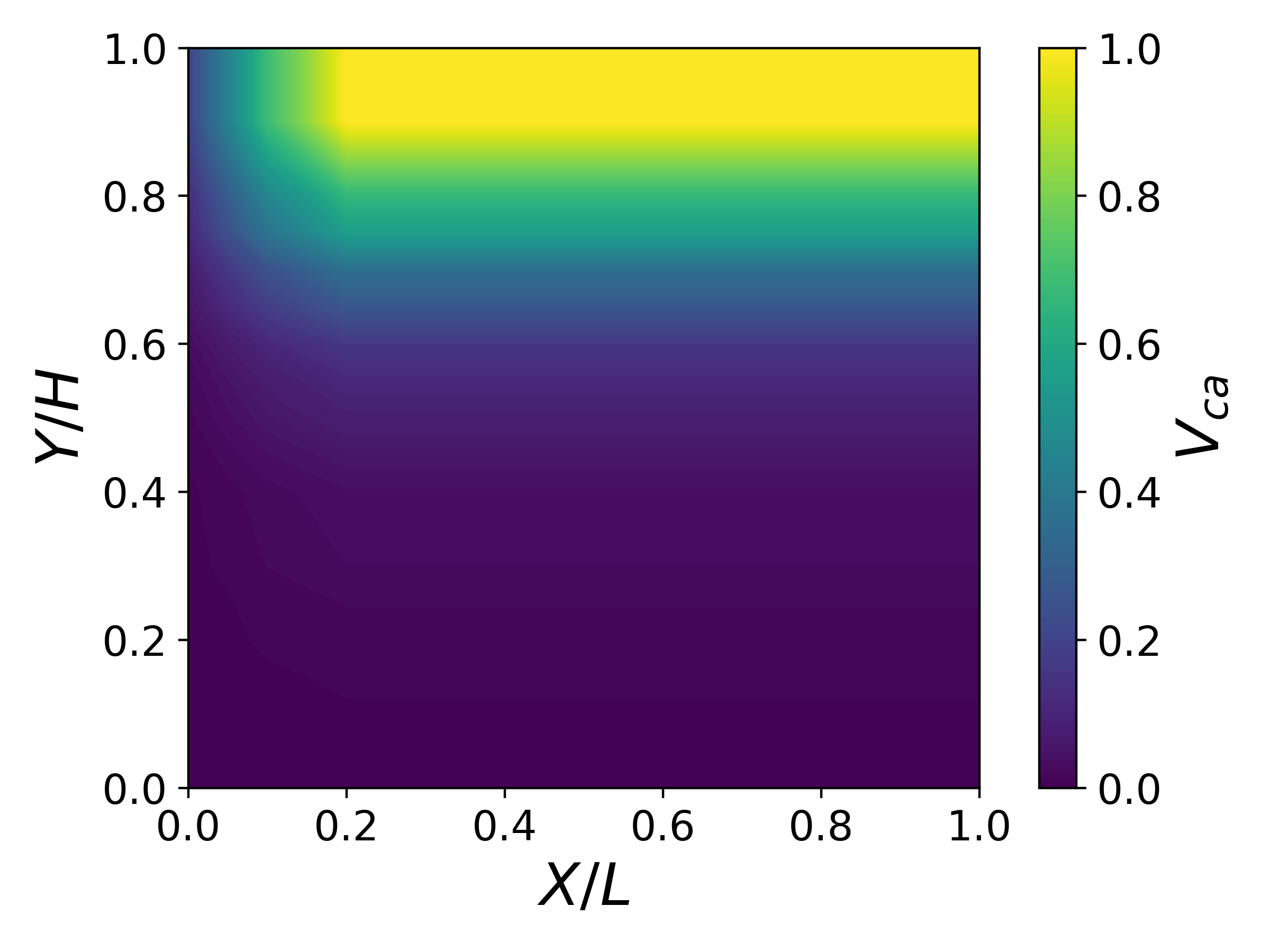}\hfill
    \includegraphics[width=.24\textwidth]{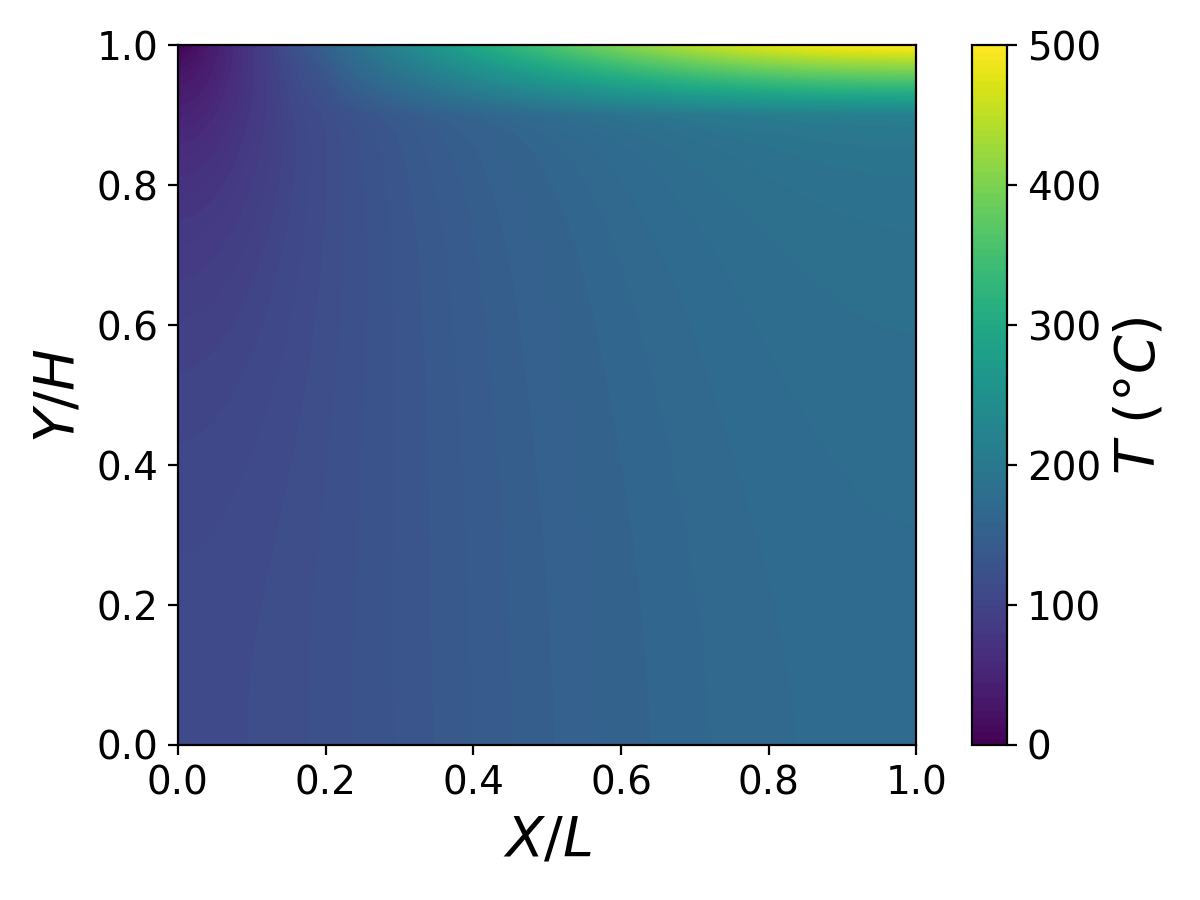}\hfill
    \includegraphics[width=.24\textwidth]{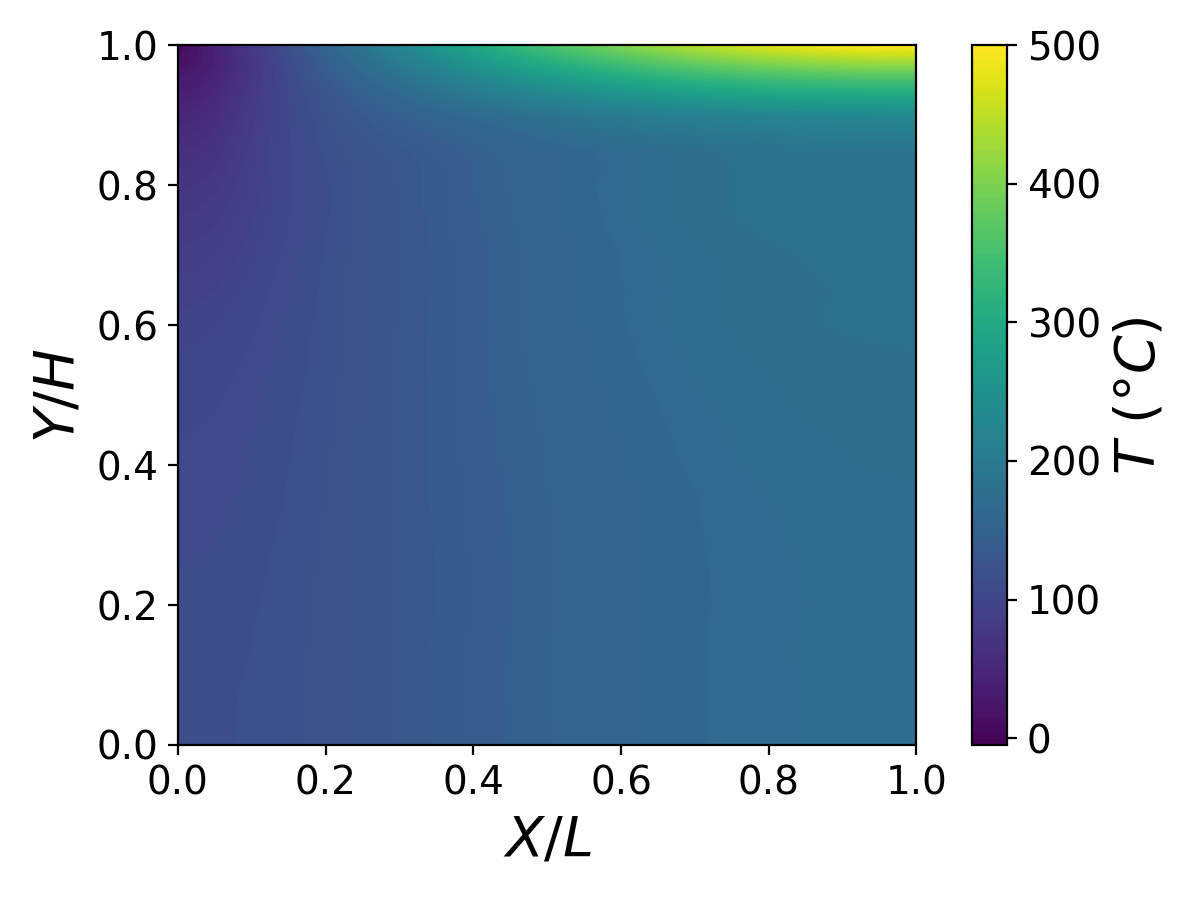}\hfill
    \includegraphics[width=.24\textwidth]{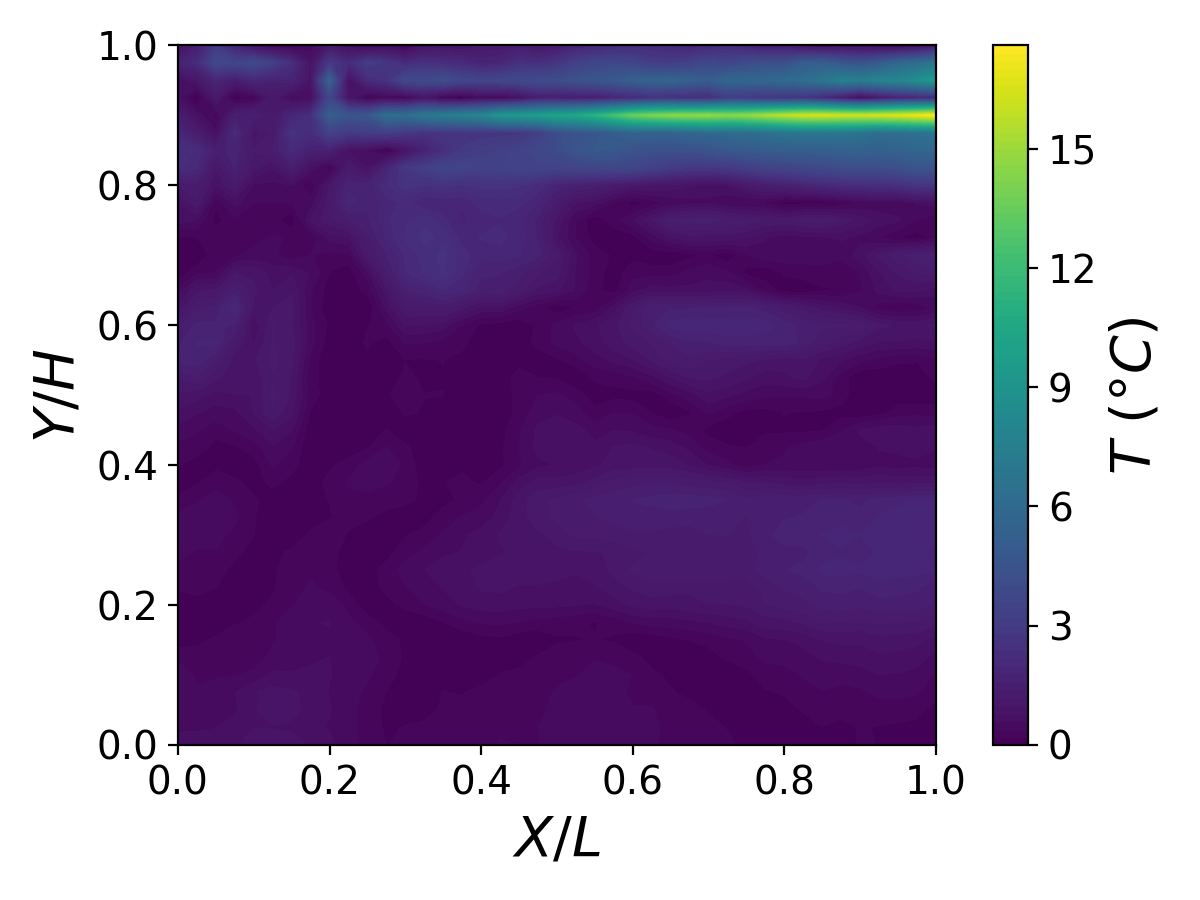}
    \\[\smallskipamount]
    \includegraphics[width=.24\textwidth]{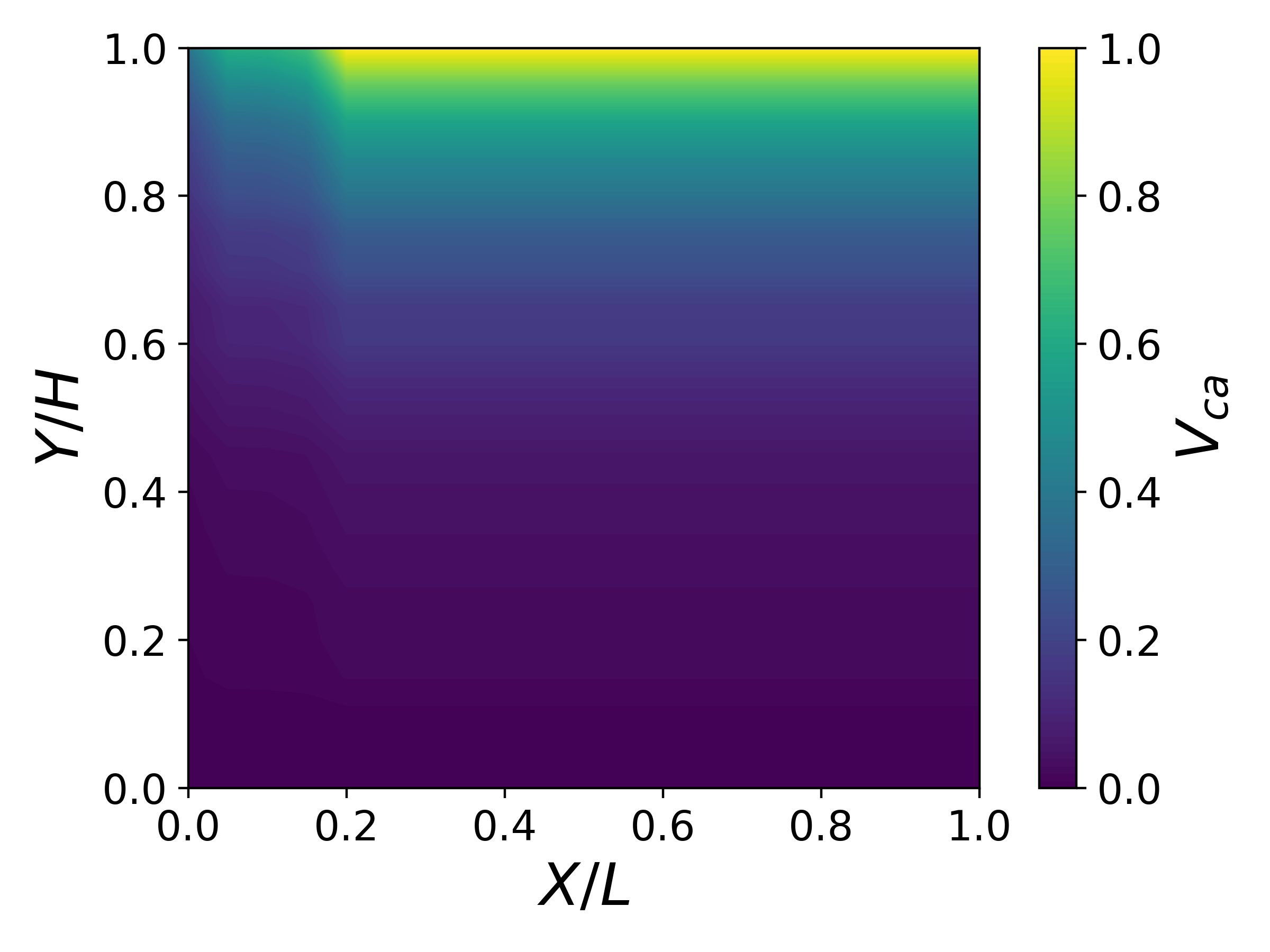}\hfill
    \includegraphics[width=.24\textwidth]{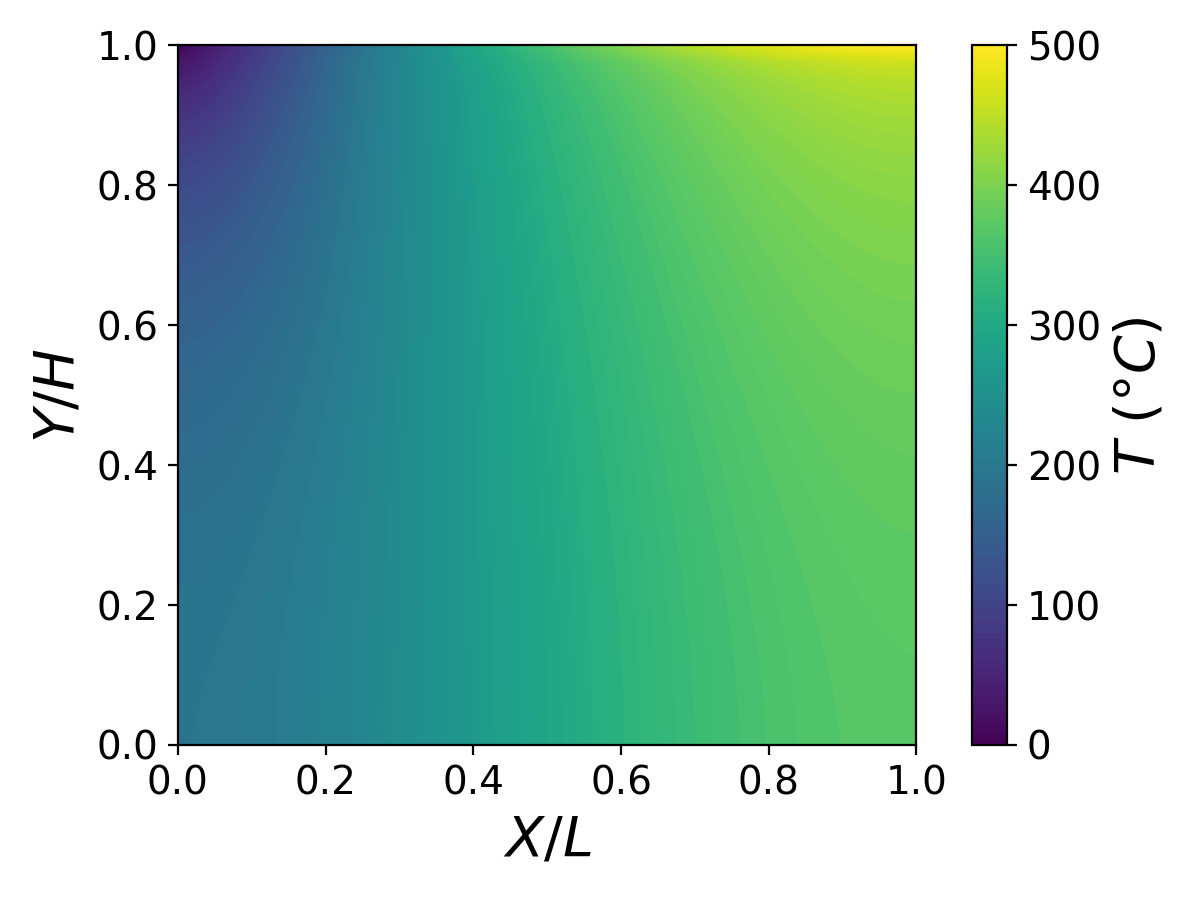}\hfill
    \includegraphics[width=.24\textwidth]{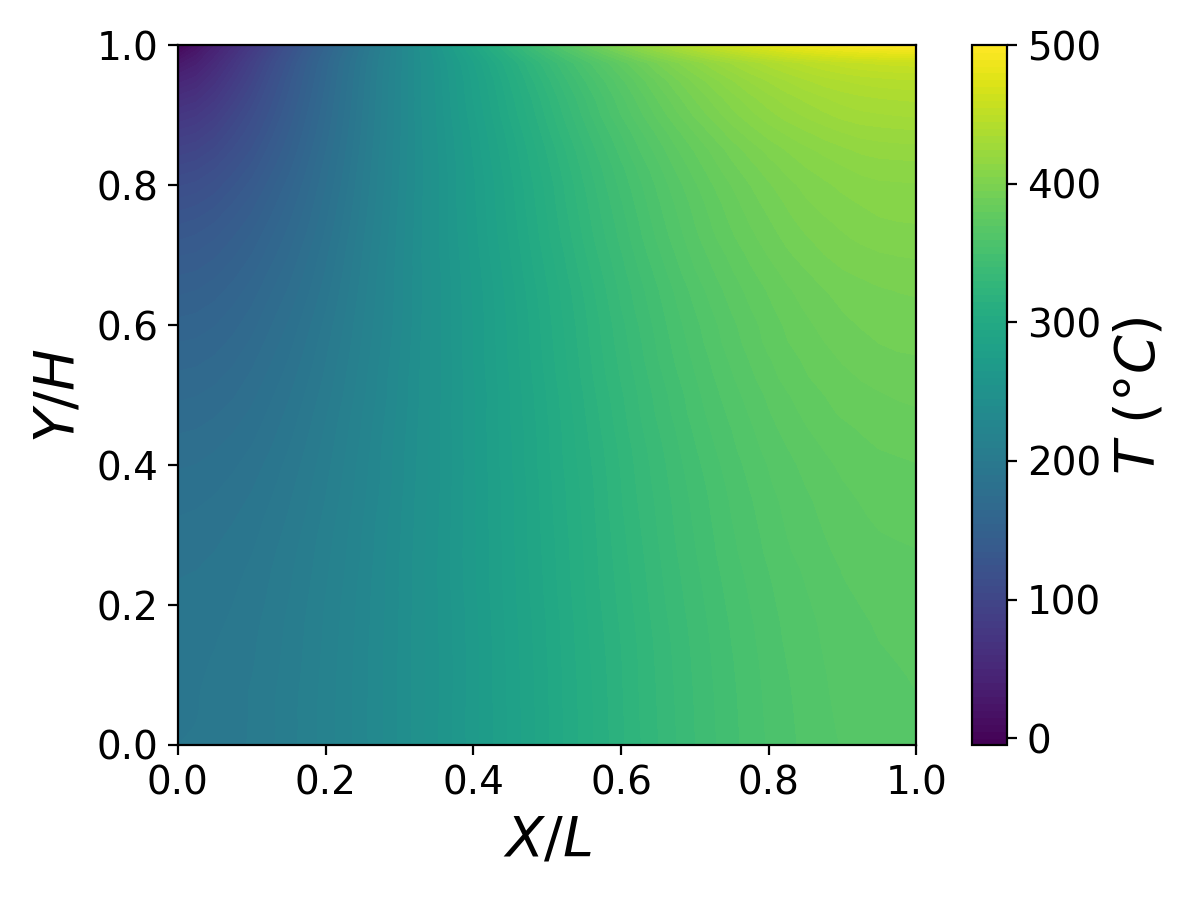}\hfill
    \includegraphics[width=.24\textwidth]{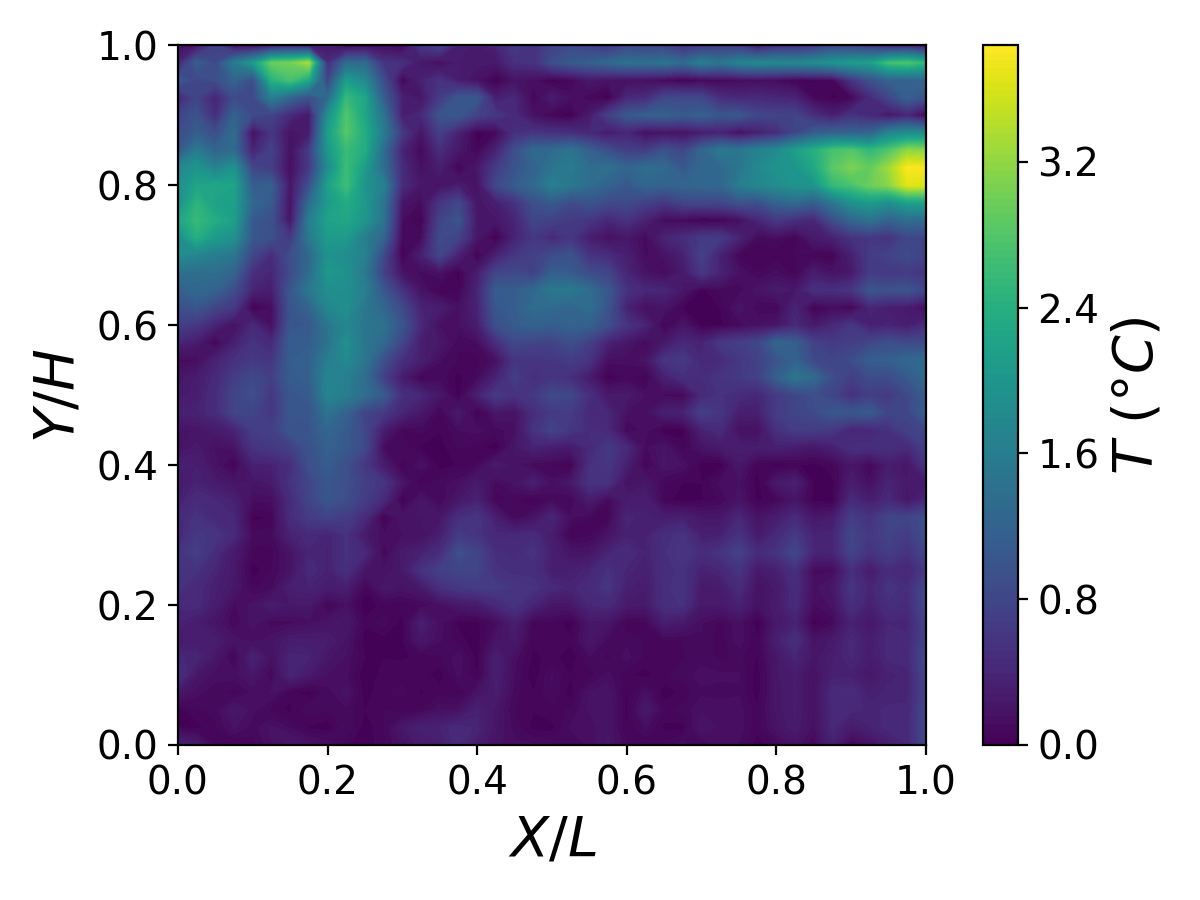}
    \\[\smallskipamount]
    \begin{subfigure}{.24\textwidth}
        \includegraphics[width=\linewidth]{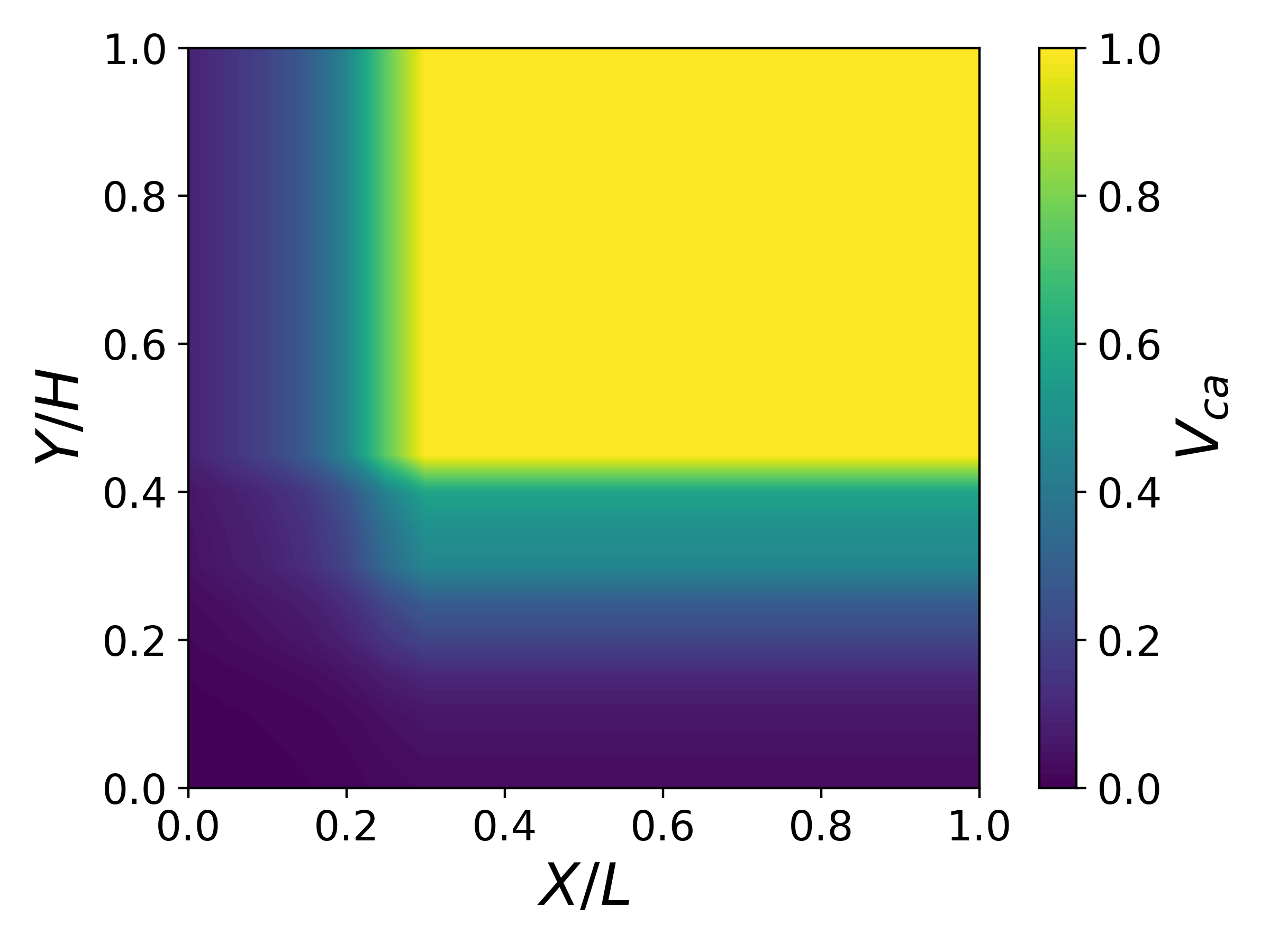}
        \subcaption{}
    \end{subfigure}\hfill
    \begin{subfigure}{.24\textwidth}
        \includegraphics[width=\linewidth]{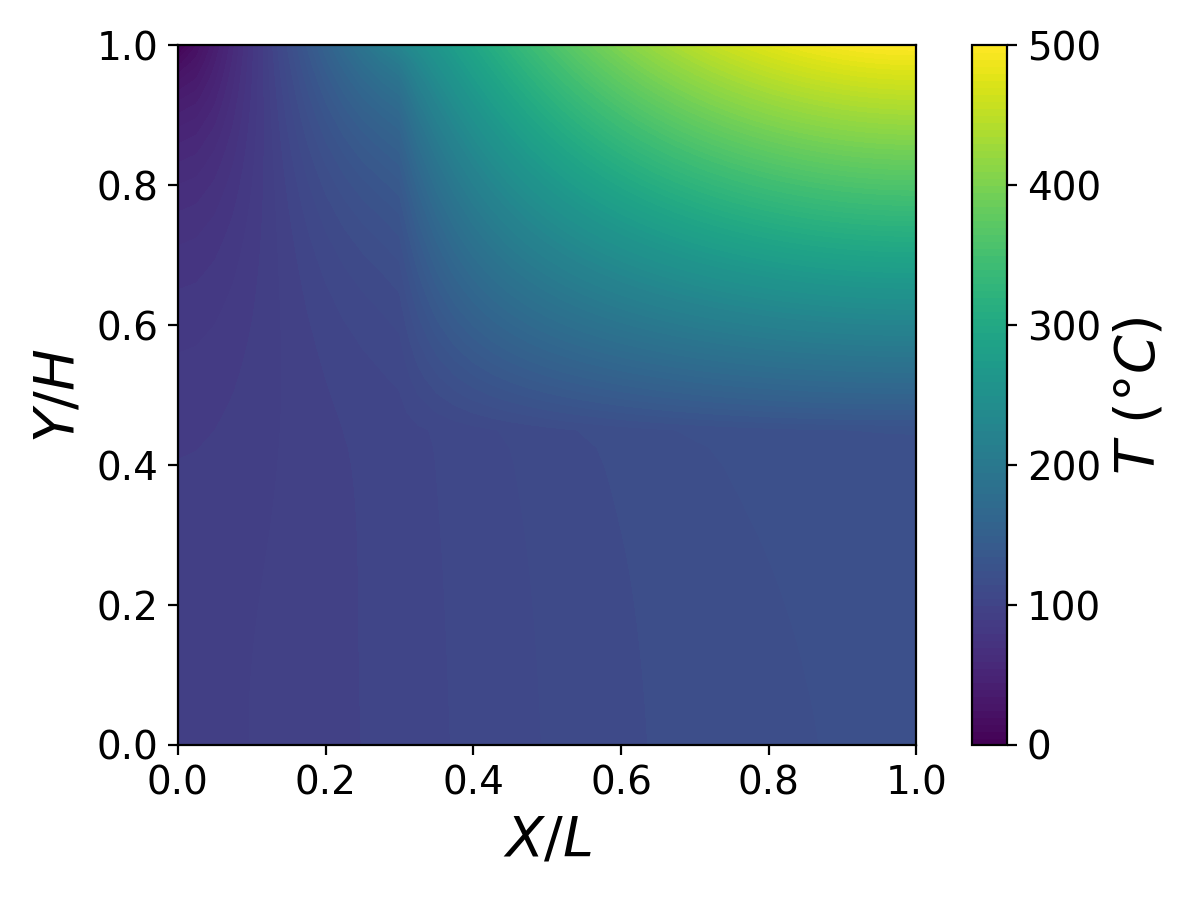}
        \subcaption{}
    \end{subfigure}\hfill
    \begin{subfigure}{.24\textwidth}
        \includegraphics[width=\linewidth]{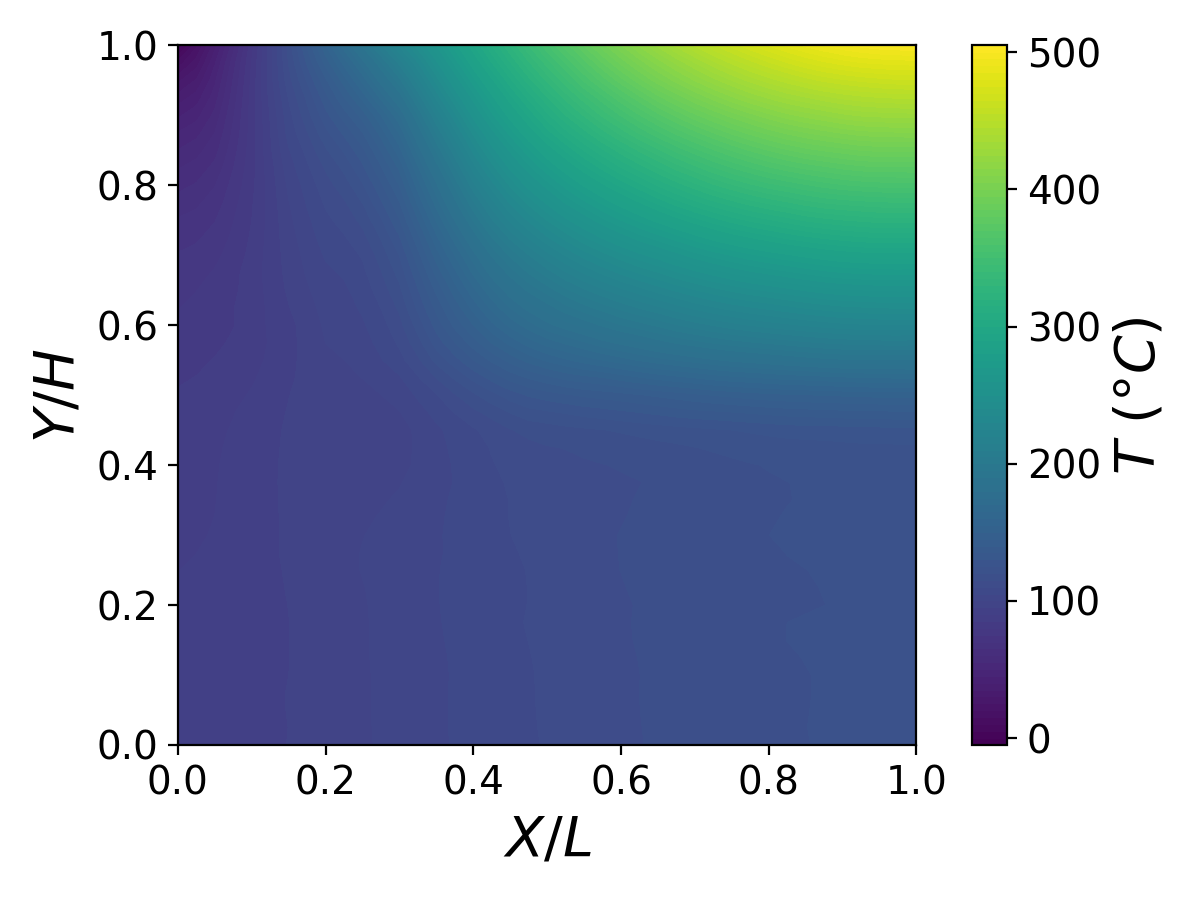}
        \subcaption{}
    \end{subfigure}\hfill
    \begin{subfigure}{.24\textwidth}
        \includegraphics[width=\linewidth]{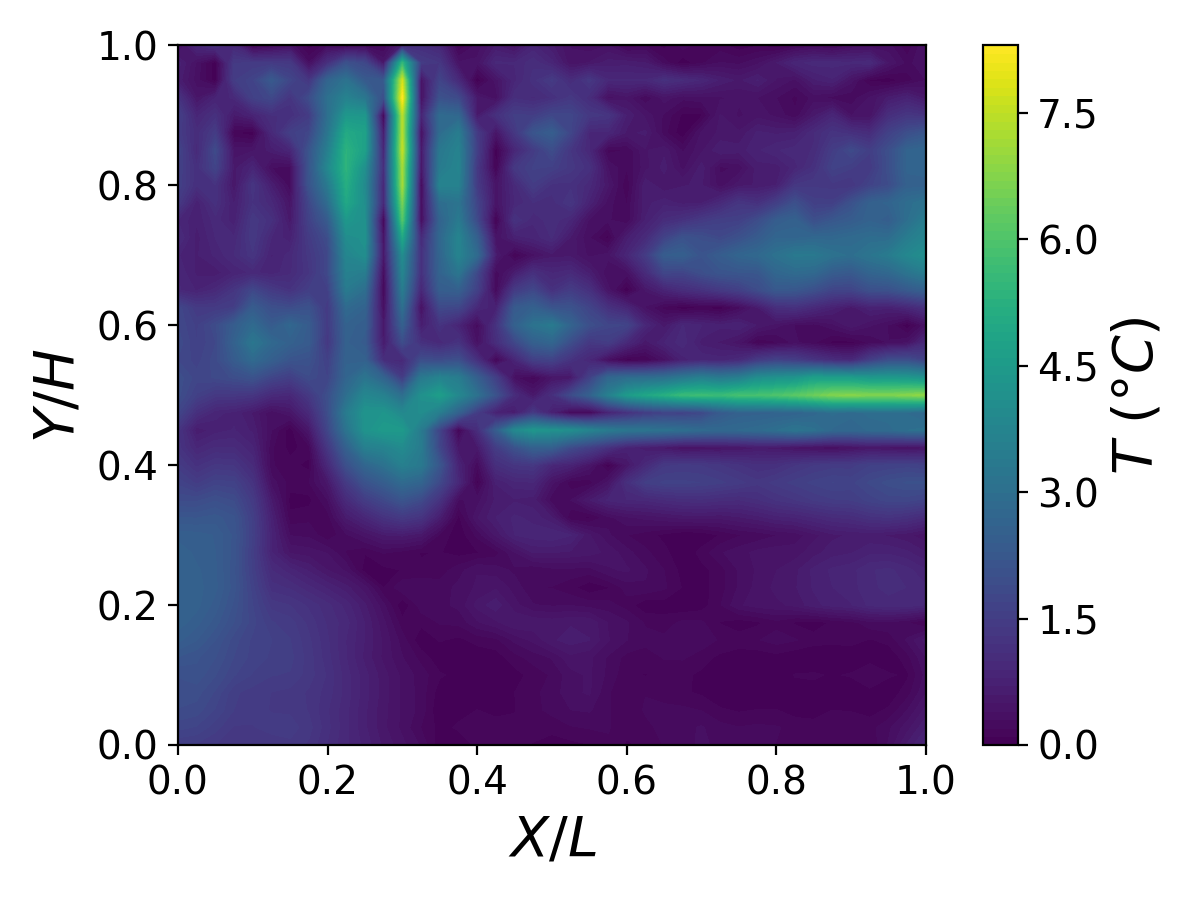}
        \subcaption{}
    \end{subfigure}
    \caption{Temperature field predictions for various profiles of $ \mathrm{Al}/\mathrm{ZrO_2}$ FGM plate (a) FGM profiles, (b) FEM results, (c) DeepONet predictions, (d) absolute error in temperature prediction of DeepONet.}
    \label{deep_prediction}
\end{figure}
%\blindtext
\section{Genetic algorithm}
\label{GA}
Genetic algorithms (GAs) are a family of random search optimization algorithms based on the mechanics of natural selection and processes of evolution.  It takes inspiration from biological evolution and uses selection, crossover, and mutation as operators. 

Before we begin, we define a few key terms used in this work:
\begin{itemize}
    \item Population: A fixed number of FGM  profiles chosen for optimization. 
    \item Chromosome: An individual profile of FGM is referred to as a chromosome.
    \item Genes: The design variables of the FGM  $\phi^{x}_{1}, \phi^{y}_{1}, \boldsymbol{\alpha^{x}}, \boldsymbol{\alpha^{y}}$ are considered as genes.
    \item Generation: A stage in the genetic algorithm where several events (fitness calculation, elite selection, crossover, mutation) occur. In this stage, unfit chromosomes die and are replaced through crossover and mutation operations.
\end{itemize}

\begin{figure}[htbp]
  \centering
  \includegraphics[scale=.25]{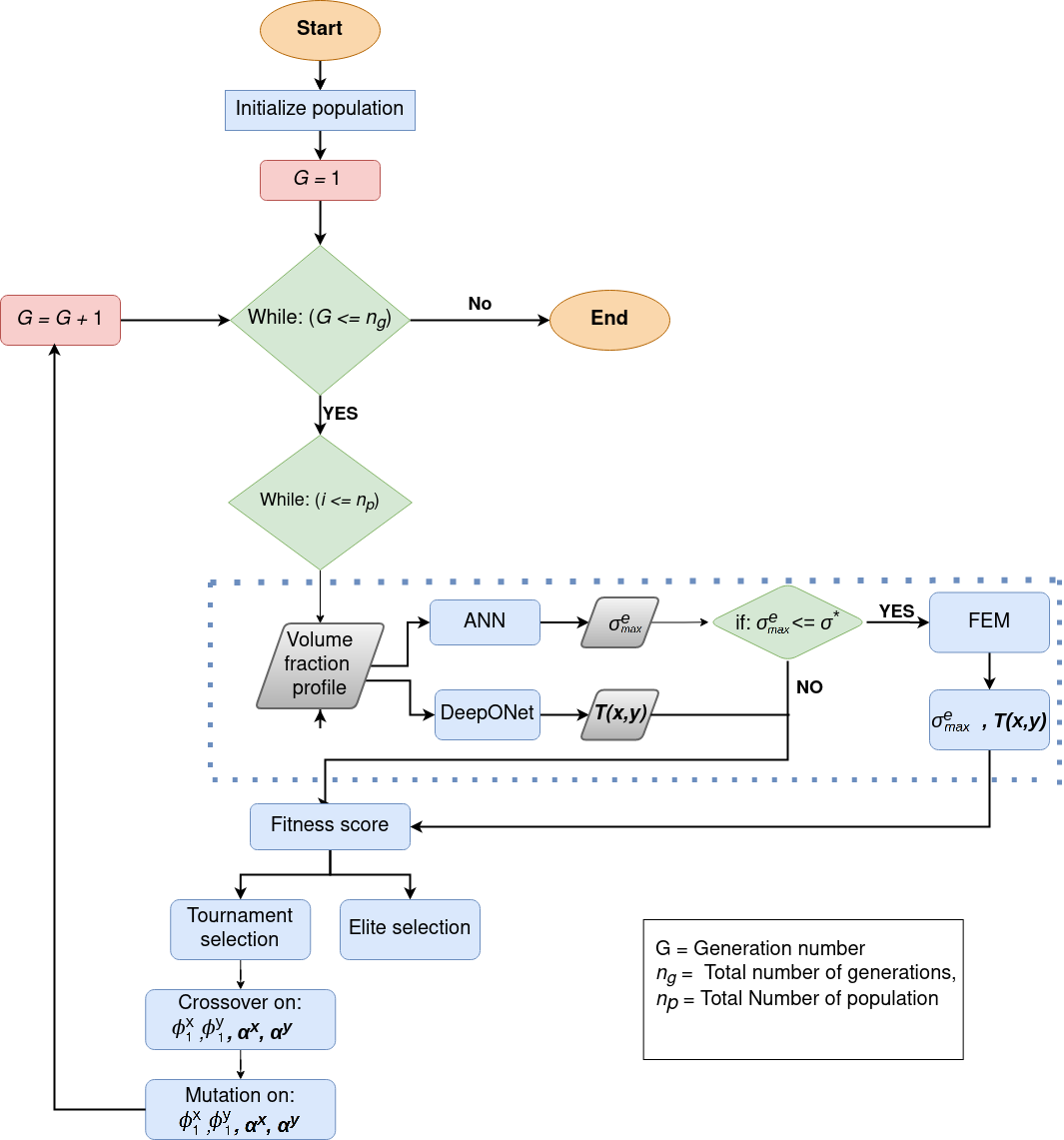}
  \caption{Flowchart of genetic algorithm for FGM optimization.}
  \label{fig:GA_flowchart}
\end{figure}

The implementation of the GA for optimizing the FGM is presented in Fig. \ref{fig:GA_flowchart}. The GA starts with a custom initialization of the population. This population is generated using the proposed profile generation scheme, which is detailed in Section \ref{profile_generation}. The next step in the GA is calculating each FGM profile's fitness function. This calculation is performed using DNN, FEM, and DeepONet models. We have observed that in some cases, even when the overall $\mathrm{R^{2}}$ score for the DNN surrogate model is very high, model predictions are not accurate for limiting cases. This low accuracy on limiting cases can be a significant concern since, in optimization, we are finding these cases only. This is because the overall representation of limiting cases in the dataset may be very low. To overcome this issue, we use a hybrid approach for fitness evaluation. According to this approach, first, we calculate the $\sigma^{e}_{max}$ using the DNN. If this value exceeds the $\sigma^{*}$, the fitness score is calculated using DNN and DeepONet; otherwise, the fitness score is determined using FEM. The $\sigma^{*}$ is defined as the value below which the DNN model is unable to capture the accurate predictions of $\sigma^{e}_{max}$. In problems where the prediction by DNN is correct over the entire range of $\sigma^{e}_{max}$ values, $\sigma^{*}$ can be set equal to zero, eliminating the need for FEM in the fitness calculation. This approach ensures that the fitness calculation is efficient and accurate. 

Further, to evolve the population, we perform tournament selection. In this process, a subset of individuals is randomly chosen from the population. These individuals compete against each other based on their fitness scores. The individual with the highest fitness score in the subset is selected as the winner and chosen as the parent for the crossover. 
Crossover and mutation both operations are performed over the variables $\boldsymbol{\alpha^{x}}$, $\boldsymbol{\alpha^{y}}$, $\phi^{x}_{1}$ and $\phi^{y}_{1}$ under bounded conditions. Here, we are using the Simulated Binary Crossover (SBX) technique, which creates two new offspring ($c_1$ and $c_2$) from two-parent individuals ($p_1$ and $p_2$). The crossover's $\eta_{c}$ parameter regulates how widely the children spread out around their parents. This operation preserves the mean value of the parents while introducing variability. Finally, a polynomial-bounded mutation operator introduces diversity by perturbing a parent (p) to generate a mutated child (c) within set bounds, facilitating the exploration of novel solutions. This is achieved by using parameters such as mutation probability, which determines that the parents gene goes under mutation, and the parameter $\eta_{m}$, helps by controlling the shape of polynomial distribution under the mutation process. These processes can enhance algorithm robustness in optimizing FGMs. The crossover and mutation operator details are given Deb \citep{deb2000efficient}. We also use elitism in the GA to conserve the fittest chromosomes immune from mutation. These steps under the GA are repeated over multiple generations until the convergence criterion is satisfied.

 To handle the inequality constraints in the evolutionary algorithms, various methods are available in the literature, such as penalty function, repair algorithm, decoder, and hybrid methods \citep{deb2001multi}. Penalty-based methods are one of the most commonly used methods in evolutionary algorithms. Detail about the penalty method is given by \citep{coello2002theoretical}. Our work uses a static penalty function to handle inequality constraints.

\section{Numerical examples}

In this section, we present various numerical examples to demonstrate the effectiveness of our proposed methodology, as outlined in Section 2. The entire process includes the design and generation of FGM profiles, FEM analysis for the maximum effective stress and the temperature field determination followed by maximum effective stress prediction using deep neural network, temperature field prediction using DeepONet, and finally, volume fraction optimization using a real coded genetic algorithm. The first problem aims to optimize the material distribution in an FGM to minimize maximum effective stress. The second problem aims to minimize the weight or maximum effective stress subjected to the diverse constraint conditions \citep{goupee2006two}. 

 \subsection{Problem 1: $Ni/Al_2O_3$ FGM plate uniformly cooled from a high temperature}
In the first problem, we aim to minimize the maximum effective stress within a functionally graded material (FGM) plate under plane strain conditions. In this unconstrained optimization problem, we consider one-half of a simply supported three-layered Ni/$\mathrm{Al_2O_3}$ 2D FGM plate, as shown in Fig. \ref{fig: FGM_Prob1}. The properties of these materials are given in Table \ref{tab:1}.

\begin{table}[htbp]
\centering
\renewcommand{\arraystretch}{1.3}
\begin{tabular}{l c r}
% Specifies 3 columns with left, center, and right alignment, enclosed in borders
\hline % Horizontal line at the top
Material Property & Ni  & $\mathrm{Al_2O_3}$ \\ % Header row
\hline % Horizontal line
$E$(GPa)   & 199.5   & 393.0   \\ % First row
$\nu $  & 0.3   & 0.3   \\ % Second row
$\alpha(10^{-6} \mathrm{K}^{-1})$   & 15.4   & 7.4  \\
$\kappa$ (W/mK)   & 60.7   & 30.0   \\
$\rho$ (kg/$\text{m}^{3}$)   & 8880   & 3960   \\ 
\hline % Horizontal line at the bottom
\end{tabular}
\caption{Problem 1: Material properties of nickel and alumina.}
\label{tab:1} % Optional label for referencing
\end{table}

\begin{figure}[htbp]
  \centering
  \includegraphics[width=0.45\textwidth]{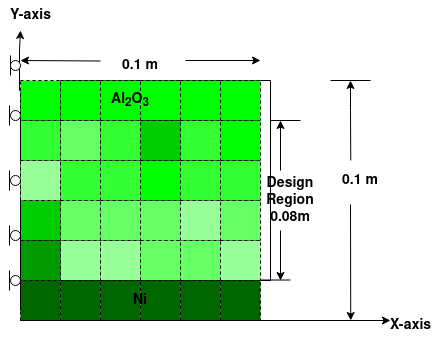}
  \caption{Problem1: Schematic of  a $ \mathrm{Ni} / \mathrm{Al_2O_3} $ FGM plate.}
  \label{fig: FGM_Prob1}
\end{figure}

The $ \mathrm{Ni} / \mathrm{Al_2O_3} $ FGM plate has a length \(L = \SI{0.1}{\meter}\) and a height \(H = \SI{0.1}{\meter}\). The plate is uniformly cooled from an initial temperature of 1000 K to 300 K. The left edge of the plate is constrained by a symmetric boundary condition $u_1 =0$. To design the 2D FGM, we are using the proposed profile generation algorithm described in section \ref{profile_generation}. In the optimization problem, we find the values of the parameters $\boldsymbol{\alpha^{x}}$, $\boldsymbol{\alpha^{y}}$, $\phi^{x}_{1}$, and $\phi^{y}_{1}$ for the optimal FGM design. The optimization problem is given as follows:

\begin{equation}
\begin{aligned}
\textbf{Minimize:}\; & \quad \sigma^{e}_{max}(\phi^{x}_{1}, \phi^{y}_{1}, \boldsymbol{\alpha^{x}}, \boldsymbol{\alpha^{y}}).\\ 
\end{aligned}
\end{equation}

Our proposed profile generation scheme is integrated with the GA to obtain the random designs.
In the profile generation scheme, we set the range for the upper bound $\alpha_{u}$ to be [1, 3] and lower bound $\alpha_{l}$ to be one. The range of $\phi^{x}_{1}$ is obtained using a single bucket of values [0.001, 1], While The range of $\phi^{y}_{1}$ is obtained using two buckets with the values [0.001, 0.01] and [0.01, 0.1].

The fitness score in the optimization process is the $\sigma^{e}_{max}$, obtained using the DNN. We are using thermoelastic FEA for the generation of training and testing data for DNN, as mentioned in section \ref{FEM}. The plate is discretized into 9-node quadrilateral elements in the FEA with a mesh density of $40\times40$ elements. The schematic of DNN architecture is shown in Fig. \ref{fig:DNN}, which can predict the $\sigma^{e}_{max}$ as an output based on the input $\boldsymbol{\phi^{x}}$ and $\boldsymbol{\phi^{y}}$. Initially, for the training of DNN $\sigma^{e}_{max}$ values is normalized by dividing them with $\mathrm{10^{7}}$. The learning rate was adjusted using a stepwise schedule: starting at 0.001 for the first 20 epochs, reducing to 0.0001 for the next 50 epochs, and decreasing to 0.00005 for the final 200 epochs. The overall dataset utilized is 30,000, out of which 80\% is used in the training, and the rest of the data is used in the testing of DNN. The model is compiled using the Adam optimizer with mean squared error (MSE) as the loss function and batch size 32. The effectiveness of the DNN model is given by comparing the actual $\sigma^{e}_{max}$ and predicted $\sigma^{e}_{max}$ values during both training and testing phases, as shown in the parity plots (Fig. \ref{fig:prob1_nn}). The accuracy is measured with $\mathrm{R^{2}}$ score, which has a value of 0.9984 in training and 0.9967 in the testing phases of the model. This high accuracy suggests that the DNN effectively captures the underlying patterns in stress prediction.

\begin{figure}[htbp]
  \centering
  \includegraphics[width=0.6\textwidth]{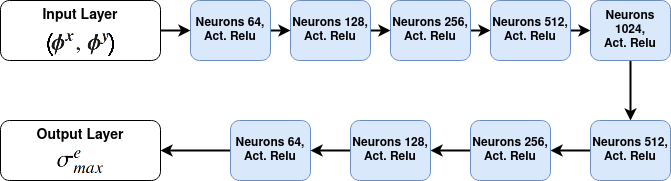}
  \caption{Schematic of the DNN architecture.}
  \label{fig:DNN}
\end{figure}

\begin{figure}[H]
  \centering
  \begin{subfigure}[b]{0.40\textwidth} % Adjust the width
    \centering
    \includegraphics[width=\textwidth]{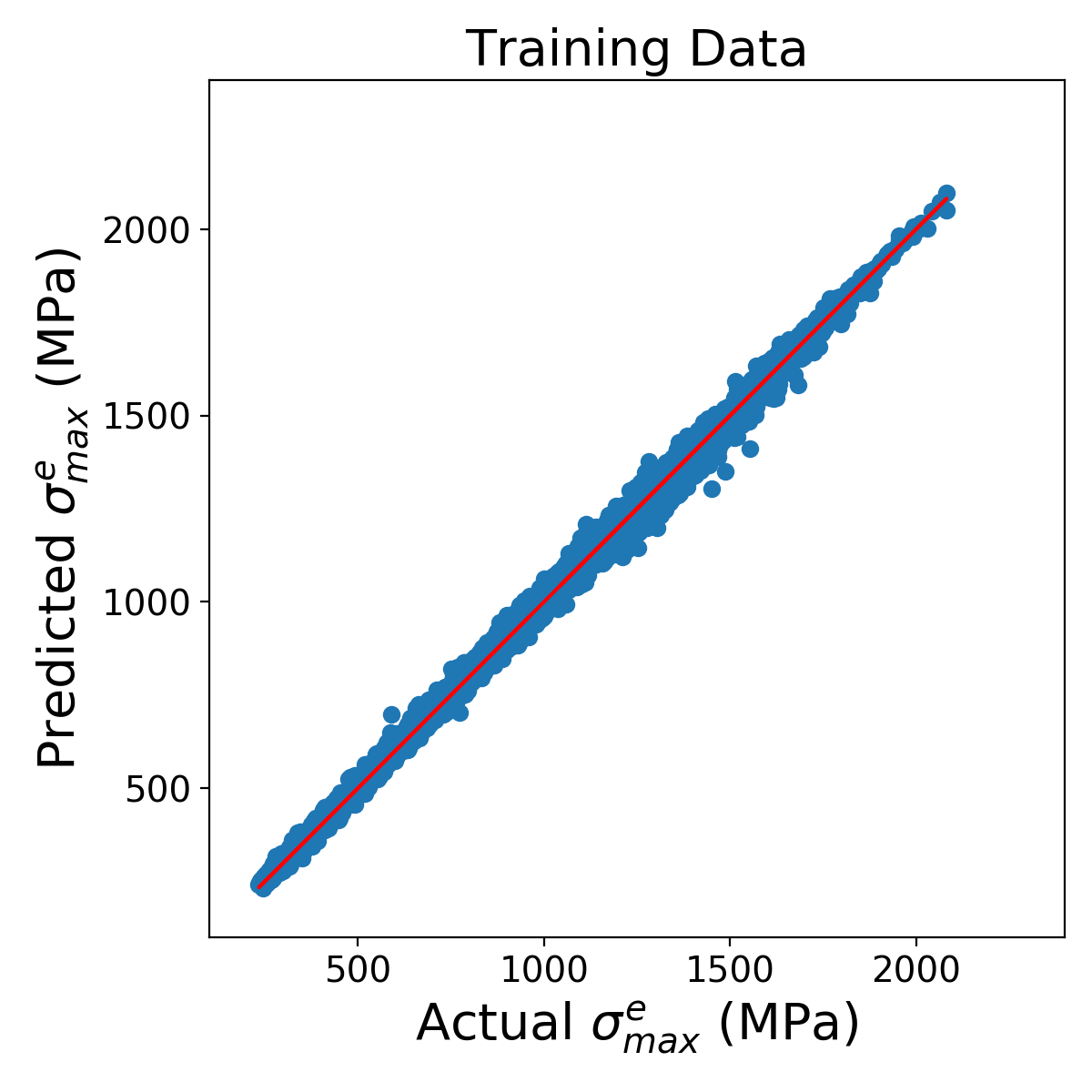}
    \caption{}
    \label{fig:sub1}
  \end{subfigure}
  \hspace{1cm}
%  \hfill
  \begin{subfigure}[b]{0.40\textwidth}
  % Adjust the width
    \centering
    \includegraphics[width=\textwidth]{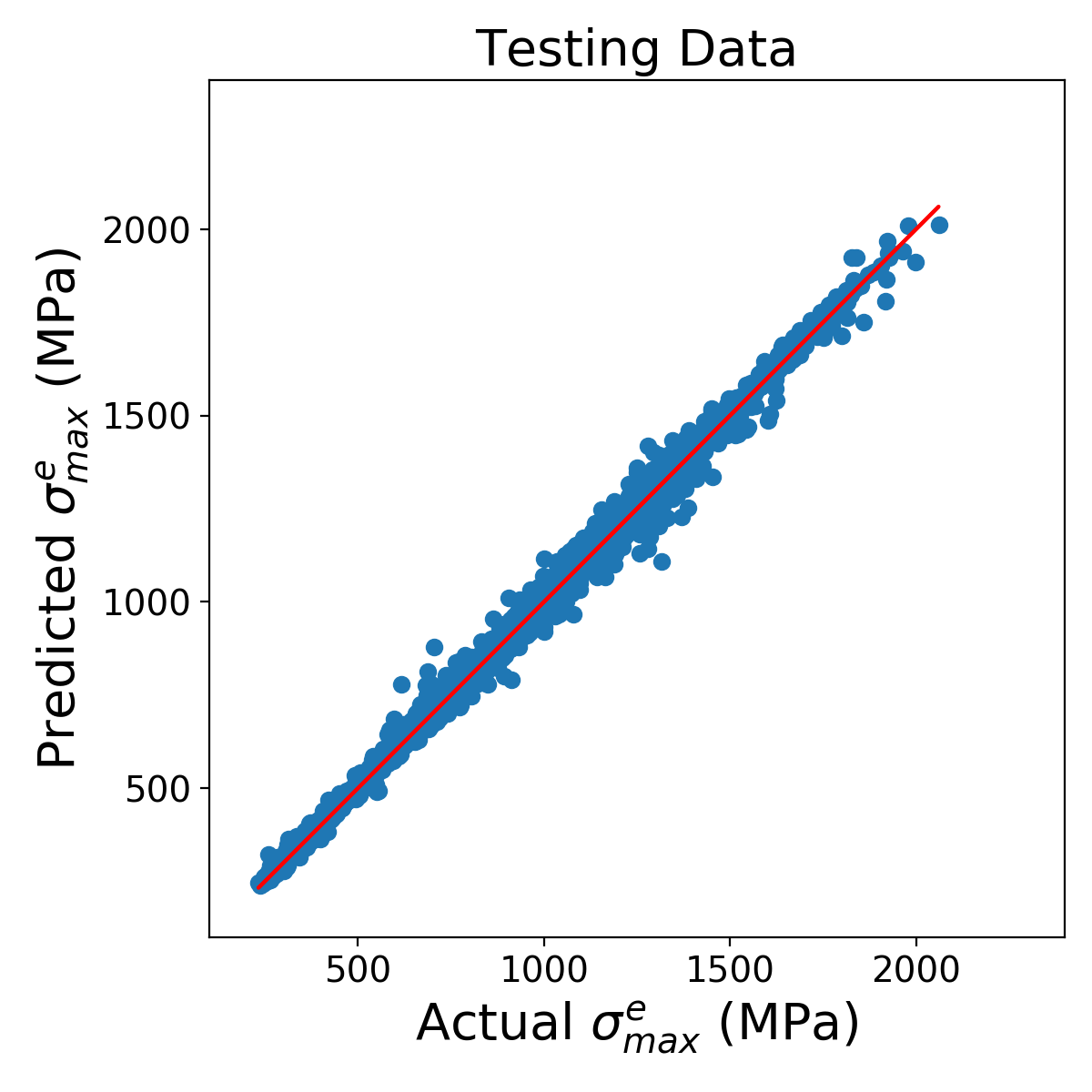}
    \caption{}
    \label{fig:sub2}
  \end{subfigure}
  \caption{Problem 1: Performance of the DNN in (a) training dataset, (b) testing dataset.}
  \label{fig:prob1_nn}
\end{figure}

For this particular problem, the parameters of the GA are detailed in Table \ref{tab:GA_parameters}. The initial population size is 200, while the termination criteria in the GA are as follows: (1) minimum of 50 generations have been completed. (2) the fitness of the best individual has not improved by 0.001 MPa over the last 10 generations.

The $\sigma^{e}_{max}$ equals 225 MPa in the obtained optimal profile. The surface plot Fig. \ref{vf_1} depicting the optimal volume fraction distribution and the convergence of GA with the generations is shown in Fig. \ref{conv_1}. To compare the optimal value, it is noted that the minimum $\sigma^{e}_{max}$ in the training and testing data is 263 MPa. Further, we also compare with various profiles obtained from power law. The power law index of 0.5, 1, 2, and 3 have the value of 410 MPa, 309 MPa, 355 MPa, and 486 MPa. Thus, the proposed optimization framework is able to provide the profile with the reasonable value of the $\sigma^{e}_{max}$ for the unconstrained problem.

\begin{table}[htbp]
    \centering
    \renewcommand{\arraystretch}{1.3}
    \begin{tabular}{l c}
        \hline
        \textbf{Parameter} & \textbf{Value} \\
        \hline
        Tournament size & 4 \\
        Crossover strength parameter, \(\eta_c\) & \(2 \left[1+\frac{1}{2}\left(1-e^{\frac{g}{100}}\right)\right]\) \\
        Mutation strength parameter, \(\eta_m\) & \(10 \left[1+\frac{1}{2}\left(1-e^{\frac{g}{100}}\right)\right]\) \\
        Mutation probability & 0.3 \\
        \hline
    \end{tabular}
    \caption{Problem 1: Parameters of the genetic algorithm.}
    \label{tab:GA_parameters}
\end{table}

\begin{figure}[htbp]
  \centering
  \begin{subfigure}[b]{0.40\textwidth}
    \centering
    \includegraphics[width=\textwidth]{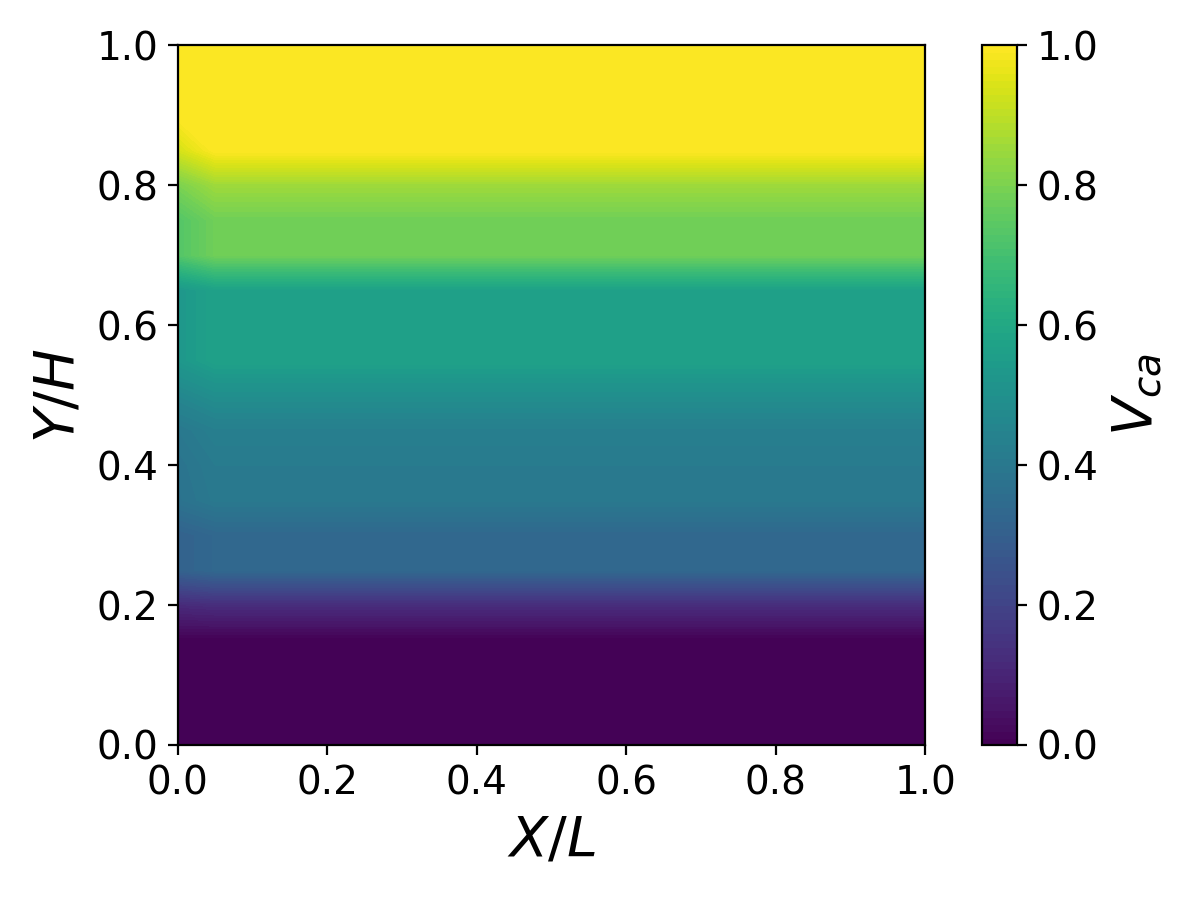}
    \caption{}
    \label{vf_1}
  \end{subfigure}
  \hspace{1cm}
  \begin{subfigure}[b]{0.40\textwidth}
    \centering
    \includegraphics[width=\textwidth]{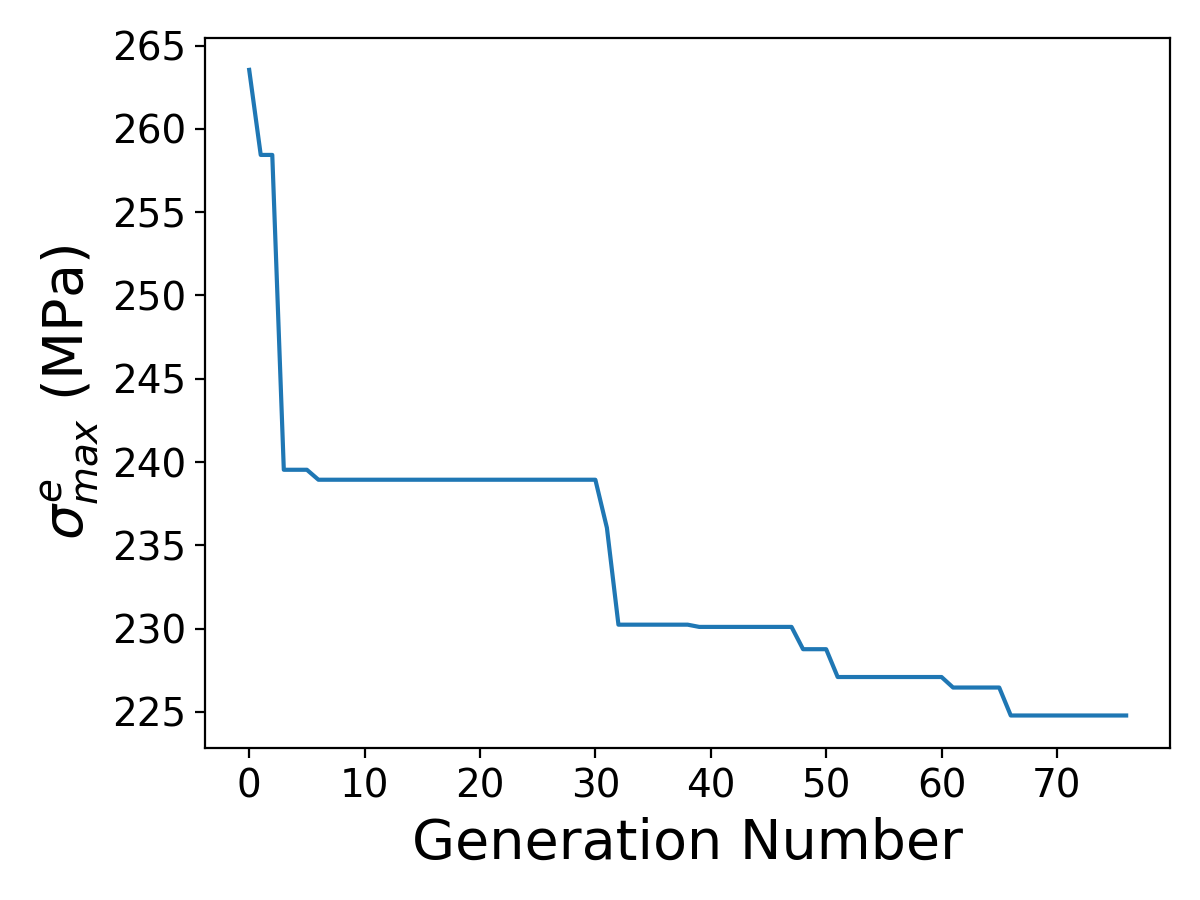}
    \caption{}
    \label{conv_1}
  \end{subfigure}
  \caption{ Optimized $ \mathrm{Ni} / \mathrm{Al_2O_3} $ FGM plate: (a) contour plot of the optimal ceramic volume fraction distribution , (b) $\sigma^{e}_{max}$ values with respect to generations. }
  \label{prb_1}
\end{figure}

\subsection{Problem 2: $Al/ZrO_2$ FGM plate subjected to sinusoidal temperature loading}

In the second problem, our aim is to minimize the maximum effective stress or minimize the average ceramic volume fraction for a 2D FGM plate. This optimization task is conducted under diverse constraint conditions such as temperature, average ceramic volume fraction, and maximum effective stress. The FGM aluminum/zirconia plate of length $L= \SI{0.15}{\meter}$ and the height $H = \SI{0.06}{\meter}$, which is one-half of the plane stress thermoelastic problem as shown in Fig. \ref{prb2_diagram}. The right edge of the plate is subjected to adiabatic ($q=0$) boundary condition with displacement along the X-axis zero ($u_1 =0$). The bottom and left edges are subjected to convective heat transfer and traction-free boundary conditions. The vertical displacement at the plate's lower left corner is zero. Surrounding temperature is specified $\theta_{\infty}$ = \SI{0}{\degreeCelsius} and the convection coefficient $h$ = \SI{50}{\watt\per\meter\squared\per\degreeCelsius}. The top surface of the plate is traction free and subjected to temperature loading given by $\theta$ =  500 sin($\pi$$x$/2L) $^0$C. Properties of aluminum and zirconia are given in Table \ref{prop_2}, while the parameters for the GA are given in Table \ref{tab:GA_parameters2}.\\\\
\begin{figure}[htbp]
  \centering
  \includegraphics[width=0.65\textwidth]{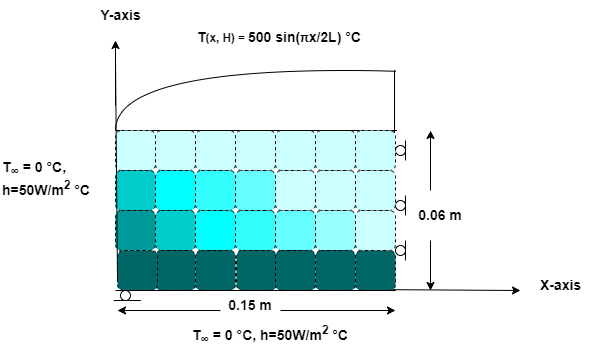}
  \caption{Problem 2: Schematic of $ \mathrm{Al} / \mathrm{ZrO_2} $ FGM plate.}
  \label{prb2_diagram}
\end{figure}

\begin{table}[htbp]
\centering
\renewcommand{\arraystretch}{1.3}
\begin{tabular}{l c r}
% Specifies 3 columns with left, center, and right alignment, enclosed in borders
\hline % Horizontal line at the top
Material Property & Al  & $\mathrm{ZrO_2}$ \\ % Header row
\hline % Horizontal line
$E$(GPa)   & 70.0   & 200.0   \\ % First row
$\nu $  & 0.3   & 0.3   \\ % Second row
$\alpha(10^{-6} \mathrm{K}^{-1})$   & 23.4   & 10.0  \\
$\kappa$ (W/mK)   & 233.0   & 2.2   \\
$\rho$ (kg/$\text{m}^{3}$)   & 2707   & 5700   \\ 
\hline % Horizontal line at the bottom
\end{tabular}
\caption{Problem 2: Material properties of aluminum and zirconia.}
\label{prop_2} % Optional label for referencing
\end{table}

\begin{table}[H]
    \renewcommand{\arraystretch}{1.3}
    \centering
    \begin{tabular}{l c}
        \hline
        \textbf{Parameter} & \textbf{Value} \\
        \hline
        Population size & 200 \\
        Tournament size & 4 \\
        Crossover strength parameter, \(\eta_c\) & \(2 \left[1+\frac{1}{2}\left(1-e^{\frac{g}{100}}\right)\right]\) \\
        Mutation strength parameter, \(\eta_m\) & \(10 \left[1+\frac{1}{2}\left(1-e^{\frac{g}{100}}\right)\right]\) \\
        Mutation probability & 0.4 \\
        \hline
    \end{tabular}
    \caption{Problem 2: Parameters of the genetic algorithm.}
    \label{tab:GA_parameters2}
\end{table}

The parameters in the proposed profile generation scheme, such as the upper bound $\alpha_{u}$, lie within the range of  [1, 3], and the lower bound $\alpha_{l}$ is equal to one. The range of $\phi^{x}_{1}$ is defined using a single bucket of values [0.001, 1]. While the range of $\phi^{y}_{1}$ is obtained using two buckets with the values [0.001, 0.01] and [0.01, 0.1]. The thermoelastic finite element analysis is used to train the computational models DNN and DeepONet. The plate is discretized into uniform 9-node quadrilateral elements in the FEA with a mesh density of 20$\times$20 elements.

The schematic of DNN model is used to predict the $\sigma^{e}_{max}$ as an output for a given input $\boldsymbol{\phi^{x}}$ and $\boldsymbol{\phi^{y}}$, is shown in Fig. \ref{fig:DNN}. For the training purpose of DNN $\sigma^{e}_{max}$ values are normalized by dividing them with $\mathrm{10^{6}}$. We are using a step learning rate by scheduling. We start with a learning rate of 0.001 for the first 20 epochs, followed by a learning rate of 0.0001 for the next 100 epochs, and finally, we use a learning rate of 0.00005 for the last 200 epochs. The DNN model utilized 20,000 datasets, with 80\% used for the training and the rest used for testing. The model is compiled using Adam optimizer with mean squared error (MSE) as the loss function and batch size 32. The parity plots (Fig. \ref{dnn_prb2}) show the comparison between the actual $\sigma^{e}_{max}$ and the predicted $\sigma^{e}_{max}$ for both the training and testing phases of the DNN. The accuracy given by $R^2$ score has a value of 0.997 in training and testing. However, we observe that the accuracy is significantly lower in the region where the $\sigma^{e}_{max}$ predicted by the DNN is less than 50 MPa, compared to the region where the $\sigma^{e}_{max}$ predicted by the DNN exceeds 50 MPa. The frequency of the percentage profiles with the percentage error in both regions is demonstrated in the histogram in Fig. \ref{histogram_prb2}. The mean percentage error in the region of $\sigma^{e}_{max}$ prediction by DNN less than 50 MPa is 7.1\%, while other regions contain the mean percentage error of 2.2\%. The higher inaccuracy in predicting $\sigma^{e}_{max}$ in this specific region is the low data density in that region. To remove inaccuracy in the results and maintain the efficacy of the proposed optimization framework, we are using the hybrid approach for the $\sigma^{e}_{max}$ prediction in the GA. This approach holds the condition that if, for any profile, the prediction of $\sigma^{e}_{max}$ by DNN is lesser than 50 MPa, we proceed with the FEA. Otherwise, we will use DNN for the whole optimization process. This approach helps us to achieve accurate and efficient results across the full range of $\sigma^{e}_{max}$ values by utilizing the effectiveness of both models.

\begin{figure}[htbp]
  \centering
  \begin{subfigure}[b]{0.40\textwidth} % Adjust the width
    \centering
    \includegraphics[width=\textwidth]{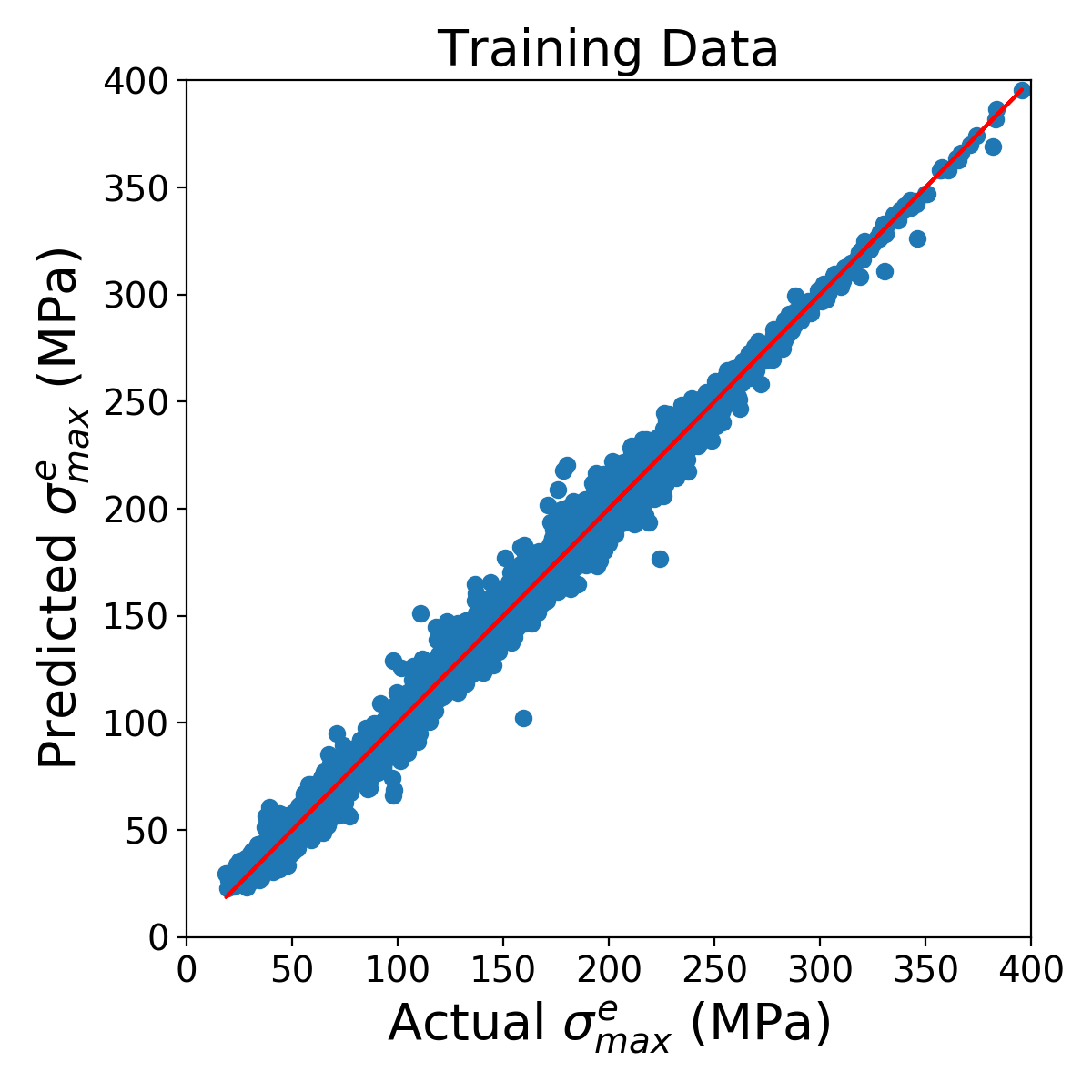}
    \caption{}
    \label{fig:sub3}
  \end{subfigure}
  \hspace{1cm}
  \begin{subfigure}[b]{0.40\textwidth} % Adjust the width
    \centering
    \includegraphics[width=\textwidth]{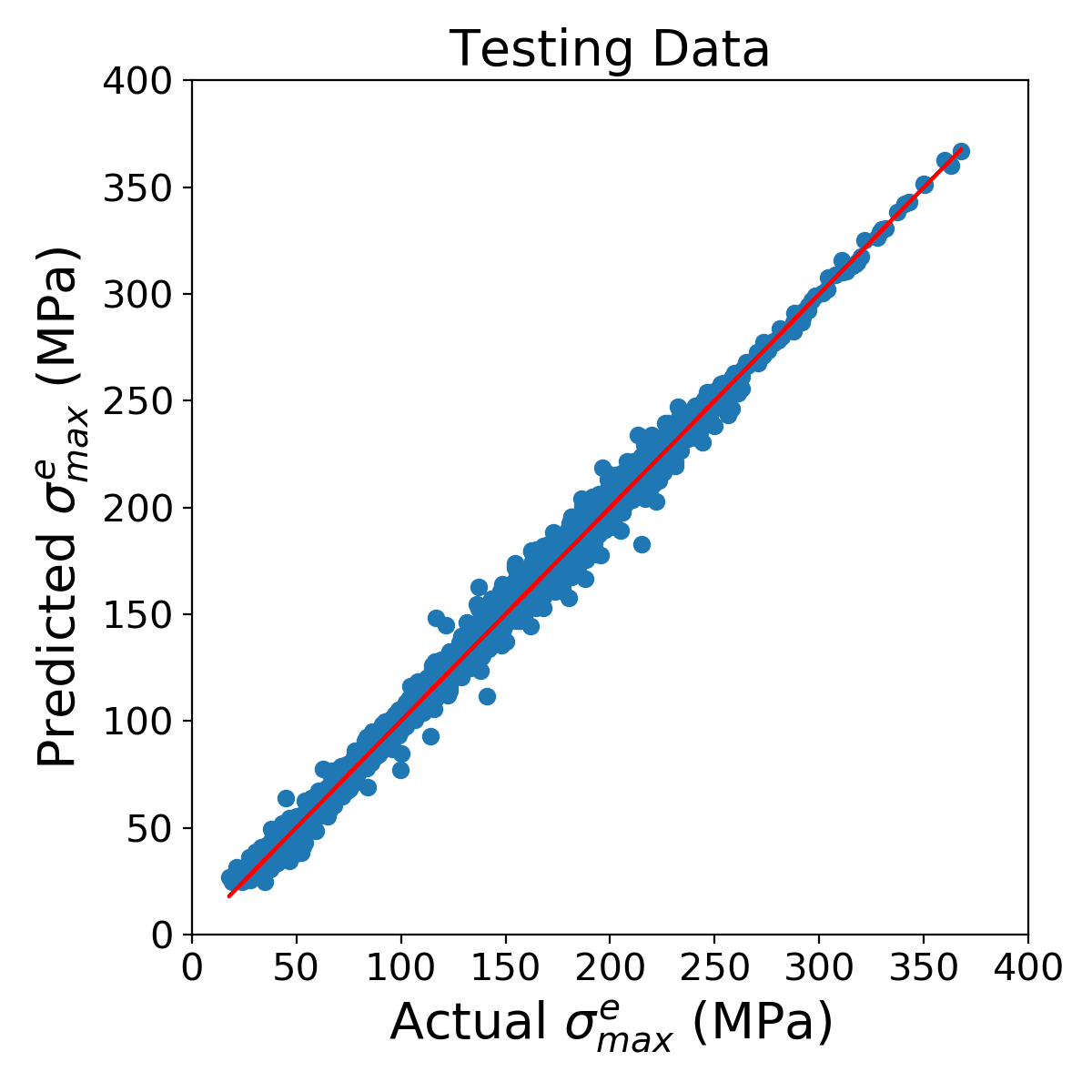}
    \caption{}
    \label{}
  \end{subfigure}
  \caption{Problem 2 : Performance of the DNN in (a) training dataset, (b) testing dataset.}
  \label{dnn_prb2}
\end{figure}

\begin{figure}[htbp]
  \centering
  \begin{subfigure}[b]{0.40\textwidth} % Adjust the width
    \centering
    \includegraphics[height=6cm,width=\textwidth]{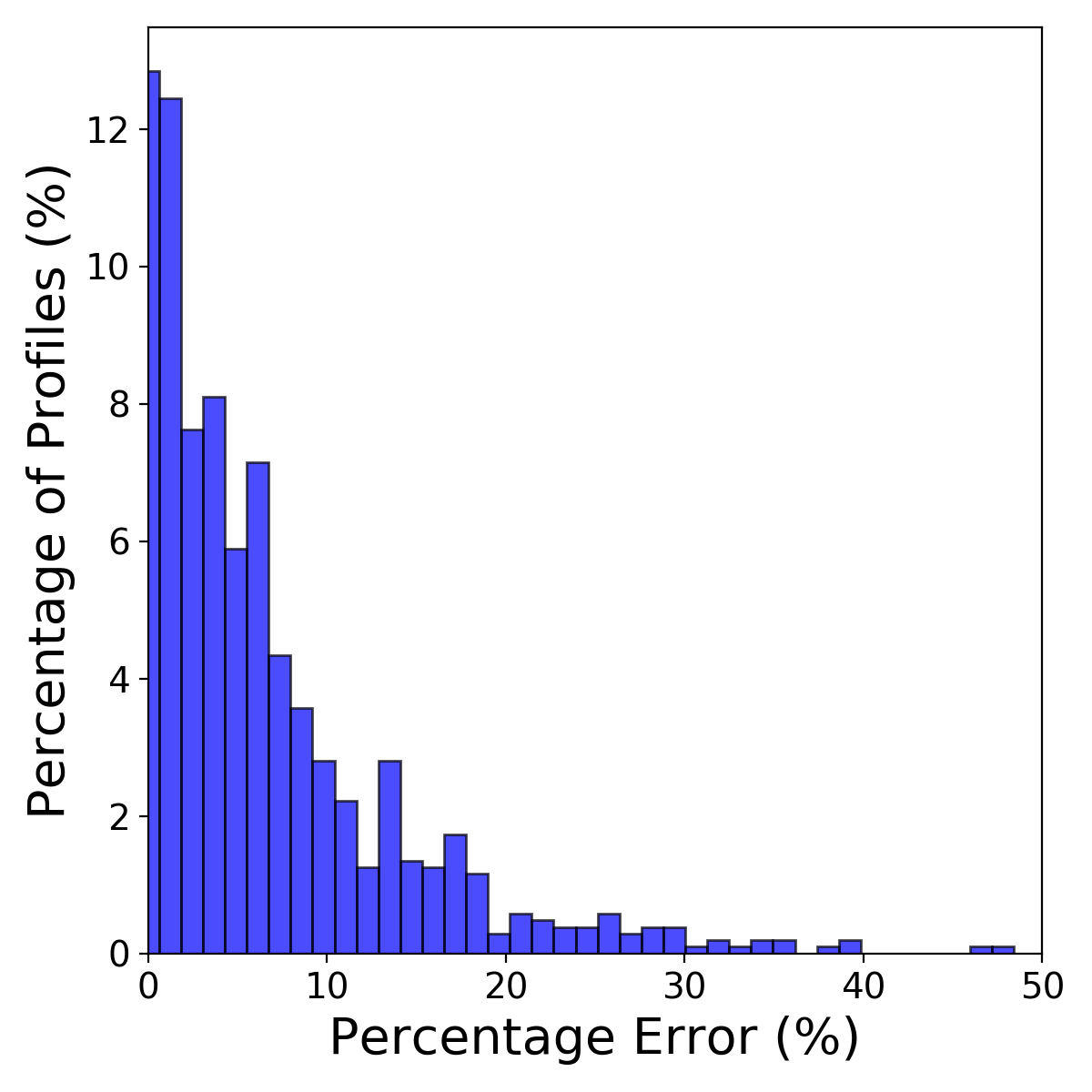}
    \caption{}
    \label{fig:sub5}
  \end{subfigure}
  \hspace{1cm}
  \begin{subfigure}[b]{0.40\textwidth} % Adjust the width
    \centering
    \includegraphics[height=6cm,width=\textwidth]{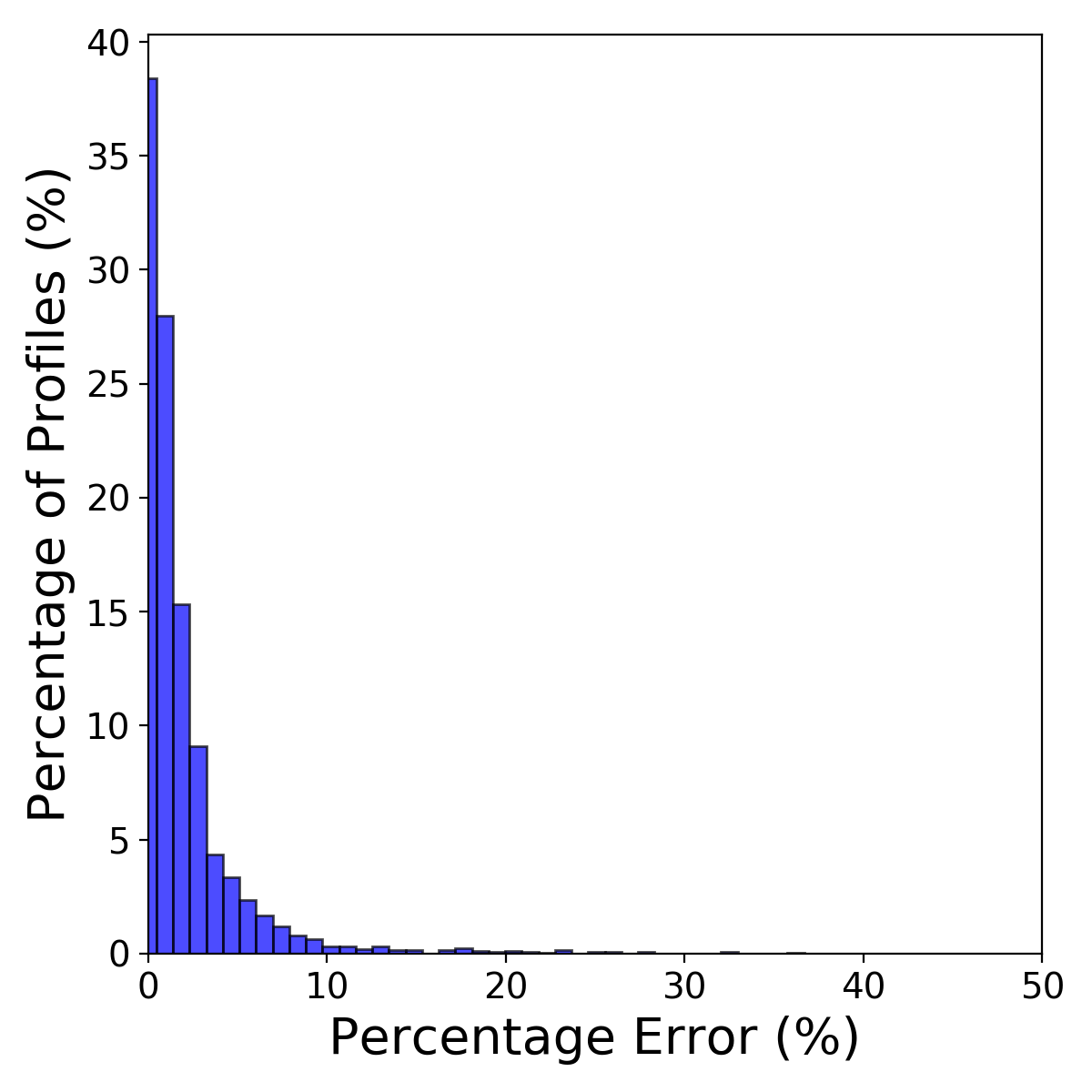}
    \caption{}
    \label{fig:sub6}
  \end{subfigure}
  \caption{Problem 2: Percentage error in the result of DNN for different range of maximum effective stress ( $\sigma^{e}_{max}$): (a) predicted  $\sigma^{e}_{max} < 50$MPa (b) predicted $\sigma^{e}_{max} \geq 50$MPa.}
  \label{histogram_prb2}
\end{figure}

The temperature field is determined using either FEM or DeepONet, depending on the $\sigma^{e}_{max}$ prediction. DeepONet is applied for the profiles where the $\sigma^{e}_{max}$ predicted by DNN, whereas for profiles where $\sigma^{e}_{max}$ is calculated using FEM, the temperature field is also obtained through FEM. The inputs of the DeepONet model are  spatial coordinates $(x,y)$ and volume fraction ($\boldsymbol{\phi^{x}}$, $\boldsymbol{\phi^{y}}$) while the output is the temperature field ($T(\boldsymbol{\phi^{x}},\boldsymbol{\phi^{y}})(x,y)$). The features utilized for the DeepONet are detailed in Section \ref{Deeponet}. The accuracy is evaluated using the $\mathrm{\mathrm{R^{2}}}$ score, which achieved a value of 0.9998 during the training phase and 0.9994 during the testing phase. The histogram in Fig. \ref{histo_deep} illustrates the frequency distribution of the maximum percentage error across profiles, further demonstrating the high accuracy of the DeepONet-based surrogate model in the temperature field prediction. The effectiveness of the proposed optimization framework for this problem is demonstrated through various cases, which are detailed in the following paragraph.\\

In the various cases, the objective is to minimize the maximum effective stress or to minimize the average ceramic volume fraction. Further, we consider various constraints, i.e., thermal, structural, and volume fractions. Termination criteria for the GA are as follows: (1) the minimum number of 50 generations in the optimization algorithm, (2) the fitness of the best individual in the population failed to improve by more than 0.001 MPa of maximum effective stress or 1\% average ceramic volume fraction in the latest 10 generations.

\begin{figure}[htbp]
  \centering
  \includegraphics[width=0.5\textwidth]{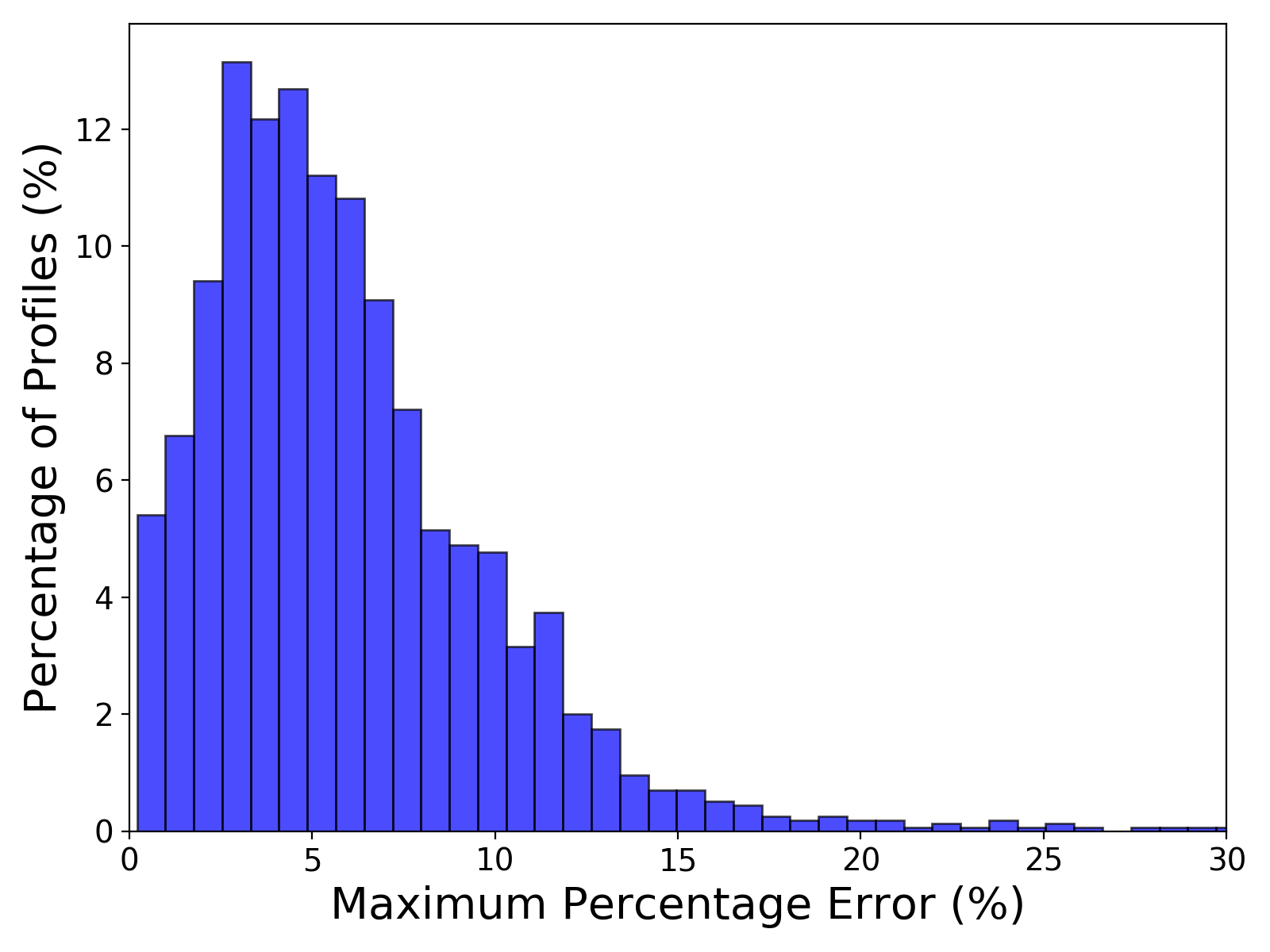}
  \caption{Problem 2: Maximum percentage error in the temperature predictions made by DeepONet.}
  \label{histo_deep}
\end{figure}

\subsubsection{Case 1}\noindent
In the first case, we start with the unconstrained optimization, where the objective is to minimize the $\sigma^{e}_{max}$. The optimization problem is given as:
\begin{equation}
\begin{aligned}
\textbf{Minimize:}\; & \quad\sigma^{e}_{max}(\phi^{x}_{1}, \phi^{y}_{1}, \boldsymbol{\alpha^{x}}, \boldsymbol{\alpha^{y}}).
\nonumber
\\ 
\end{aligned}
\end{equation}

The optimum volume fraction distribution for the FGM plate is given in Fig. \ref{vf_21a}.  The profile of FGM shows a smooth transition from the metal to the ceramic region, which is expected in the minimization of $\sigma^{e}_{max}$. 
The convergence of the GA is illustrated in Fig. \ref{conv_21a}. The $\sigma^{e}$ distribution in the optimal FGM plate is illustrated in Fig. \ref{minimum_stress}, where the maximum value equals 15.95 MPa. Note that this stress value is significantly lower than the maximum stress of 80 MPa observed in the profile with a linear volume fraction gradation along the Y-axis, as shown in Fig. \ref{1d_plot}. Additionally, the $\sigma^{e}_{max}$ in the optimized profile is also lower than in the profile with a 2D bilinear volume fraction gradation. This 2D bilinear profile has $\sigma^{e}_{max}$ equal to 97.2 MPa as shown in the Fig. \ref{2d_plot}.

\begin{figure}[htbp]
  \centering
  \begin{subfigure}[b]{0.40\textwidth} % Adjust the width
    \centering
    \includegraphics[width=\textwidth]{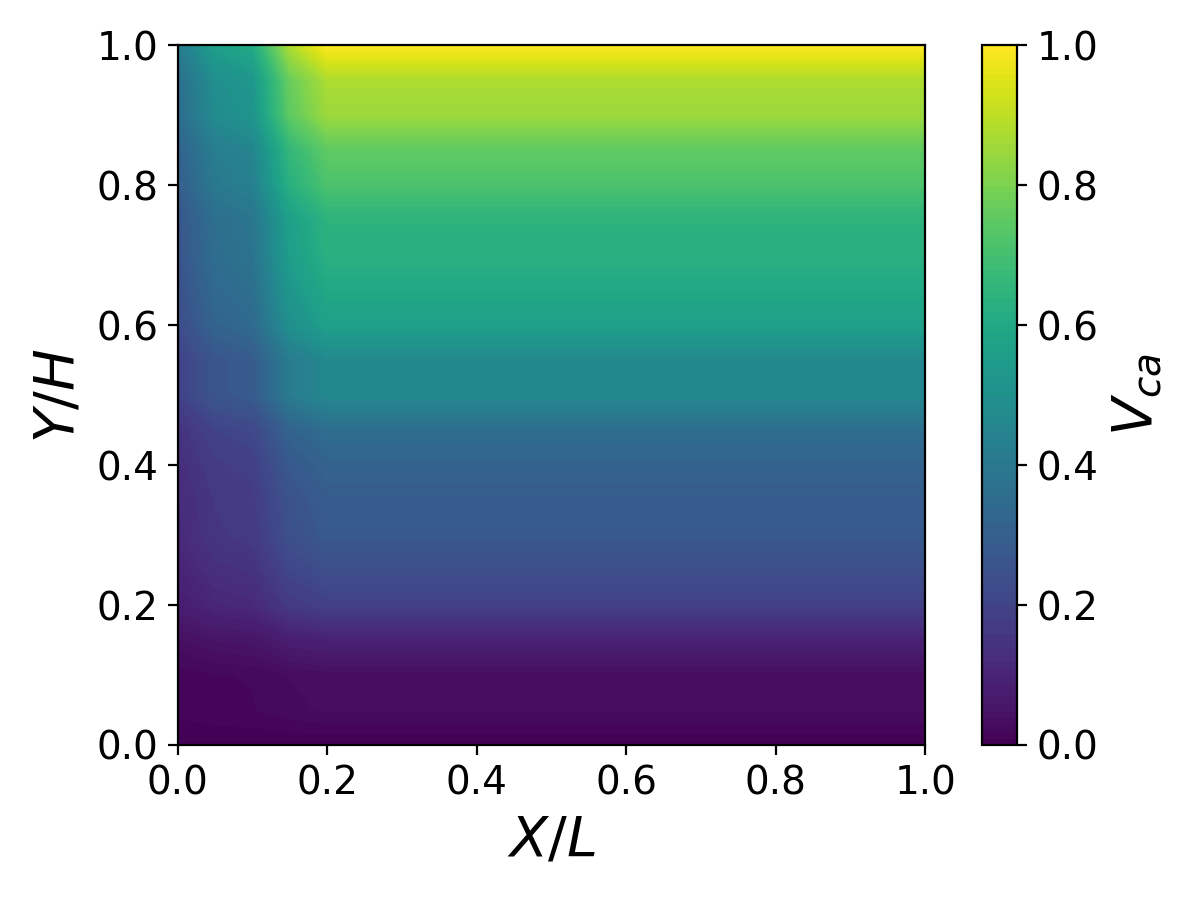}
    \caption{}
    \label{vf_21a}
  \end{subfigure}
  \hspace{1cm}
  \begin{subfigure}[b]{0.40\textwidth} % Adjust the width
    \centering
    \includegraphics[width=\textwidth]{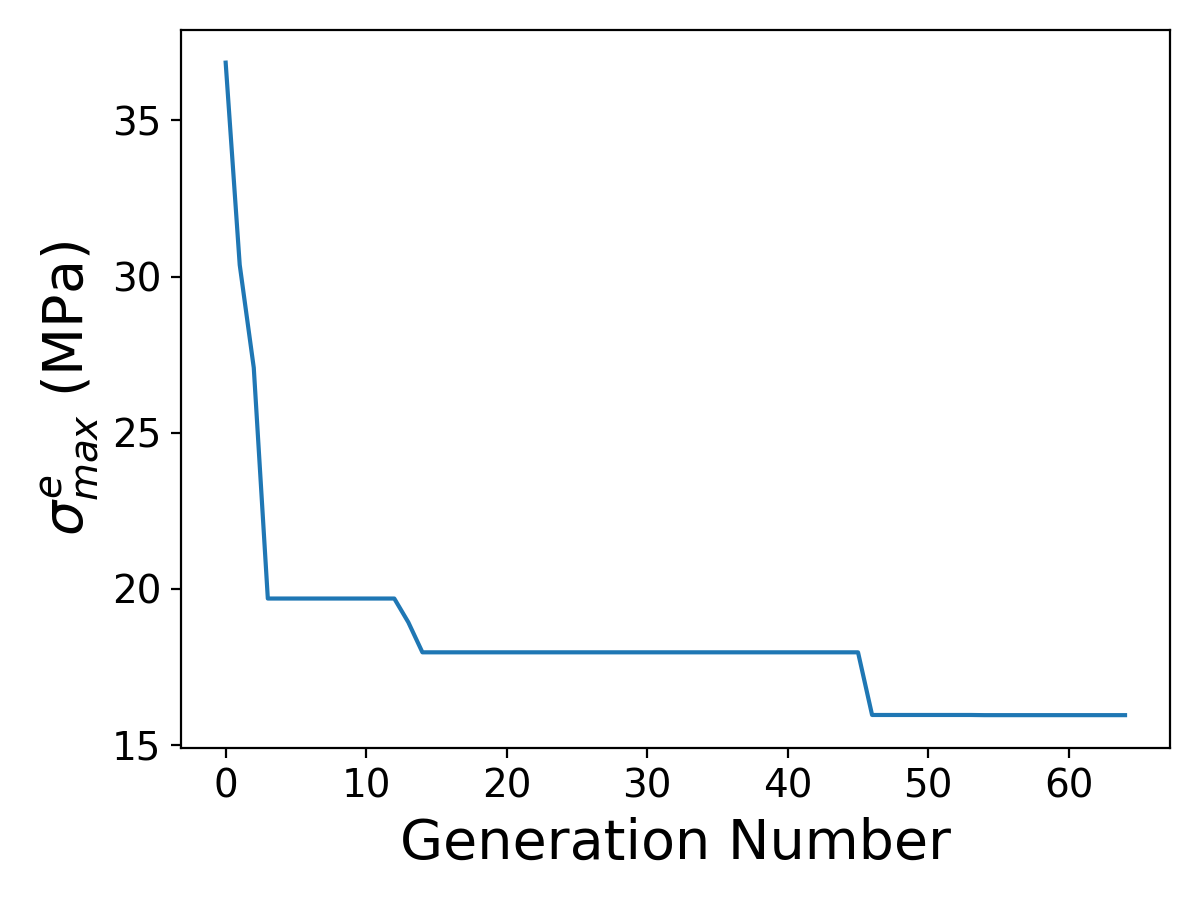}
    \caption{}
    \label{conv_21a}
  \end{subfigure}

  \caption{ Unconstrained optimization of FGM plate: (a) contour plot of the optimal ceramic volume fraction distribution, (b) $\sigma^{e}_{max}$ values with respect to generations.}
  \label{prb_21a}
\end{figure}

\begin{figure}[htbp]
    \centering
    \begin{subfigure}[b]{0.3\textwidth}
        \centering
        \includegraphics[width=\textwidth]{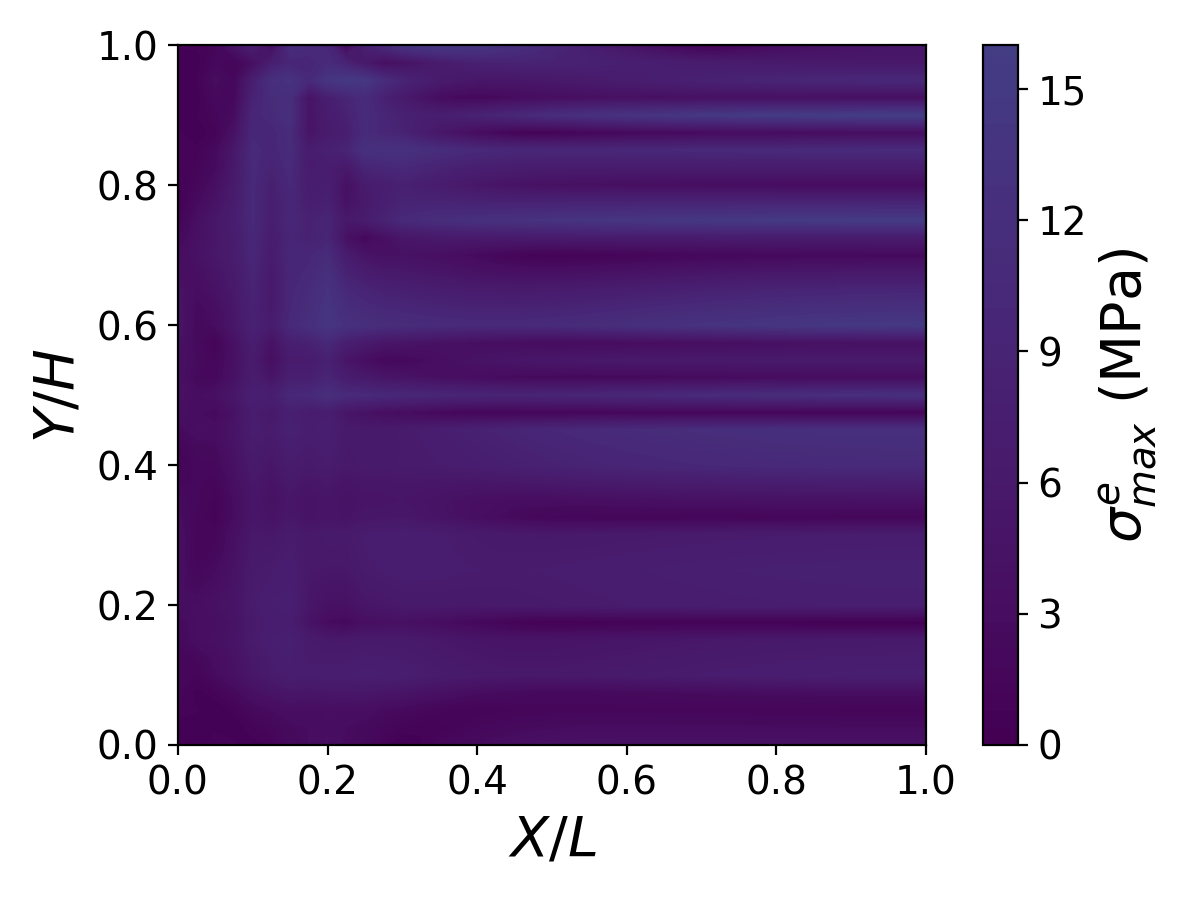}
        \caption{}
        \label{minimum_stress}
    \end{subfigure}
    \begin{subfigure}[b]{0.3\textwidth}
        \centering
        \includegraphics[width=\textwidth]{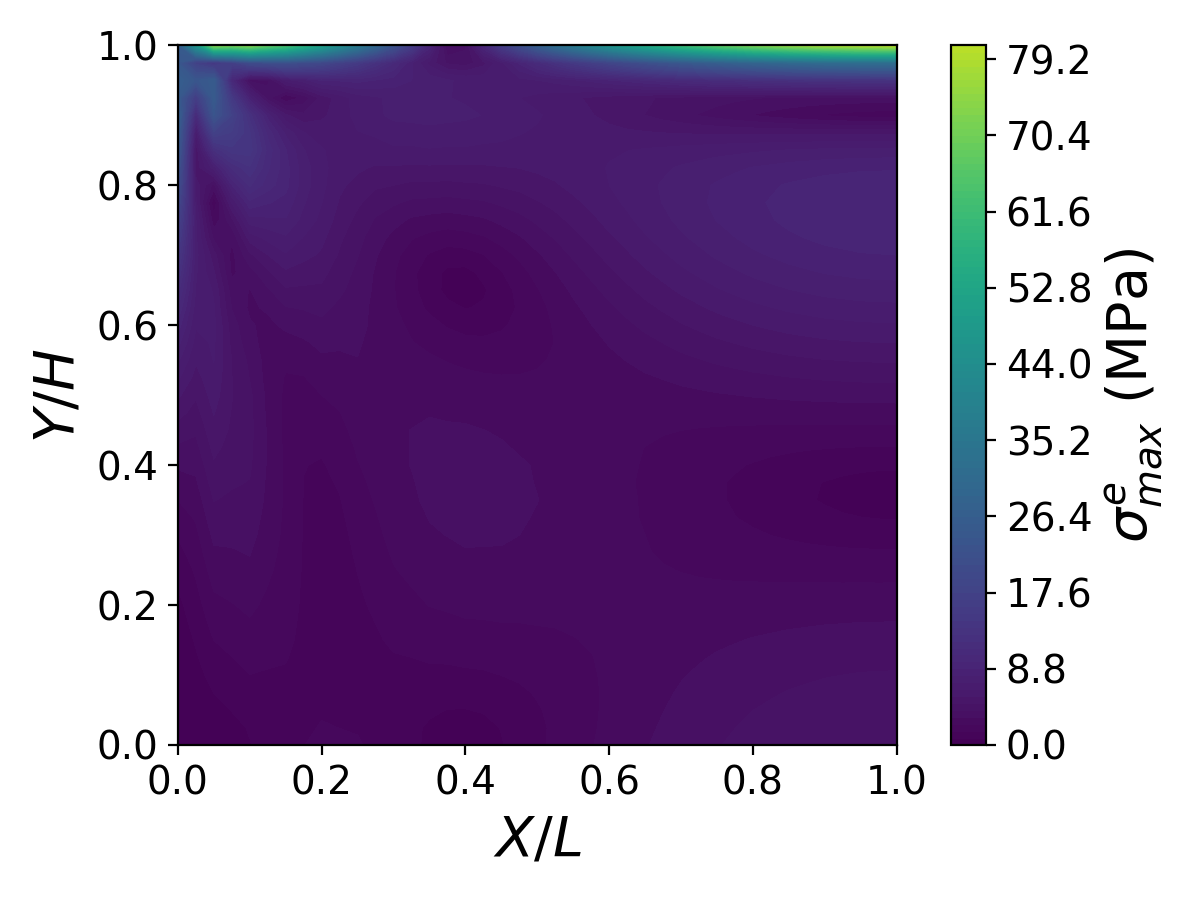}
        \caption{}
        \label{1d_plot}
    \end{subfigure}
    \begin{subfigure}[b]{0.3\textwidth}
        \centering
        \includegraphics[width=\textwidth]{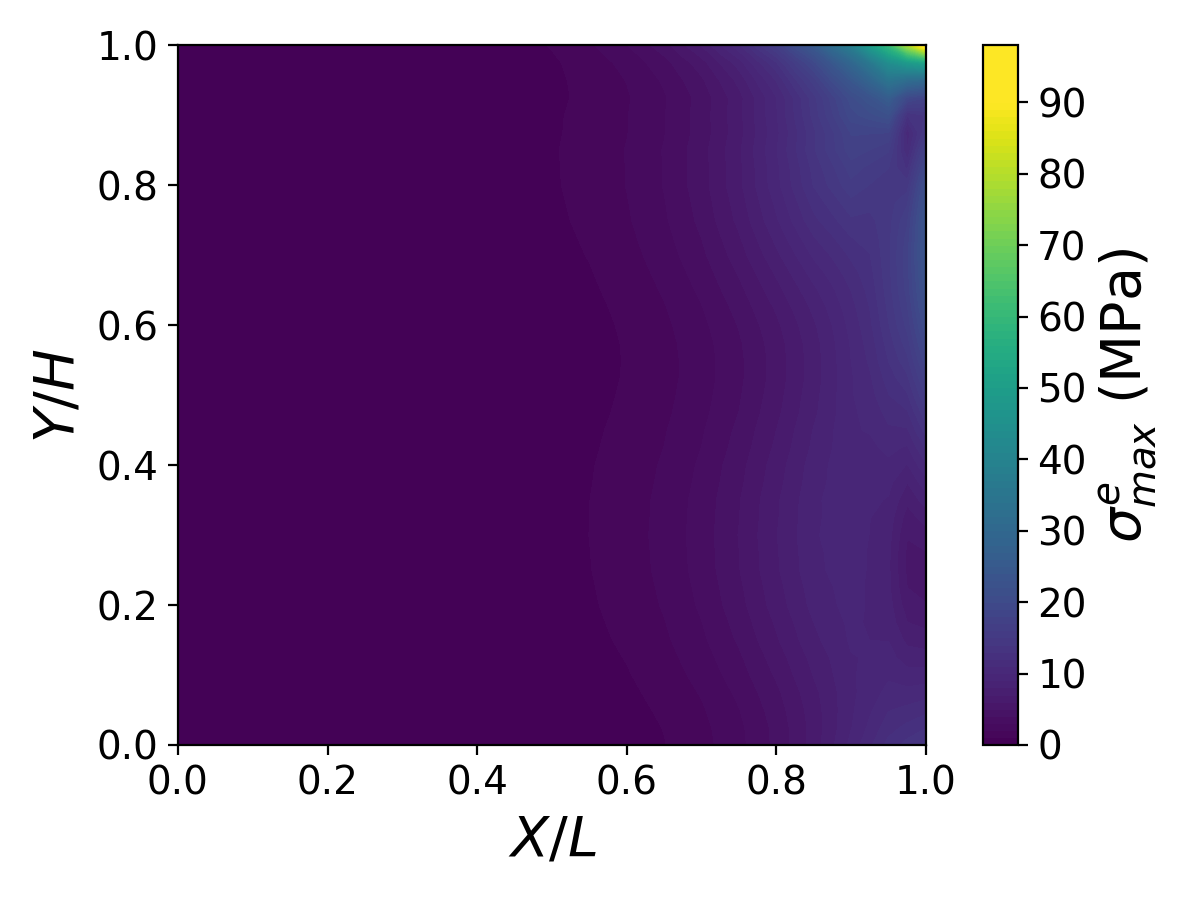}
        \caption{}
        \label{2d_plot}
    \end{subfigure}
    \caption{Contour plot shows the comparison of effective stress profile for different designs of FGM in (a) the optimized profile, (b) the plate with the linear volume fraction gradation along the Y-axis, (c) the plate with 2D bilinear volume fraction gradation.}
    \label{stress_21a}
\end{figure}

%% case(b) minimze stress, volume <=0.15
\subsubsection{Case 2}
In this case, we increase the complexity of the optimization problem by imposing a volume constraint. In particular, the optimization problem is to minimize $\sigma^{e}_{max}$ subjected to the condition that the average ceramic volume fraction should not exceed a specific value. The optimization problem is given as:

\begin{equation}
\begin{aligned}
\textbf{Minimize:}\; & \quad\sigma^{e}_{max}(\phi^{x}_{1}, \phi^{y}_{1}, \boldsymbol{\alpha^{x}}, \boldsymbol{\alpha^{y}}),\\ 
\textbf{Subject to:}\; & \quad V_{ca} \leq V^{*}.
\nonumber
\end{aligned}
\end{equation}
In this example, the value of \(V^{*}\) is taken as 0.15. The obtained optimized profile is shown in Fig. \ref{vf_21b}, and the convergence of GA is given by Fig. \ref{conv_21b}. The optimized profile has the average ceramic volume fraction of 0.149, thus satisfying the permissible limit of volume faction. In the optimum volume fraction profile, we find that the change in volume fraction is minimal in the region $0.0 \leq Y/H \leq 0.6$, while it increases significantly in the region $0.6 \leq Y/H \leq 1.0$. Note that this variation significantly differs from the unconstrained optimization presented in case 1. The large constant region is present due to the large metal volume fraction requirement. The value of the \(\sigma^{e}_{max}\) for the optimum profile is equal to 45.25 MPa. As expected, the value of the minimum \(\sigma^{e}_{max}\) is higher than the stress value obtained in case-1.

\begin{figure}[htbp]
  \centering
  \begin{subfigure}[b]{0.40\textwidth} % Adjust the width
    \centering
    \includegraphics[width=\textwidth]{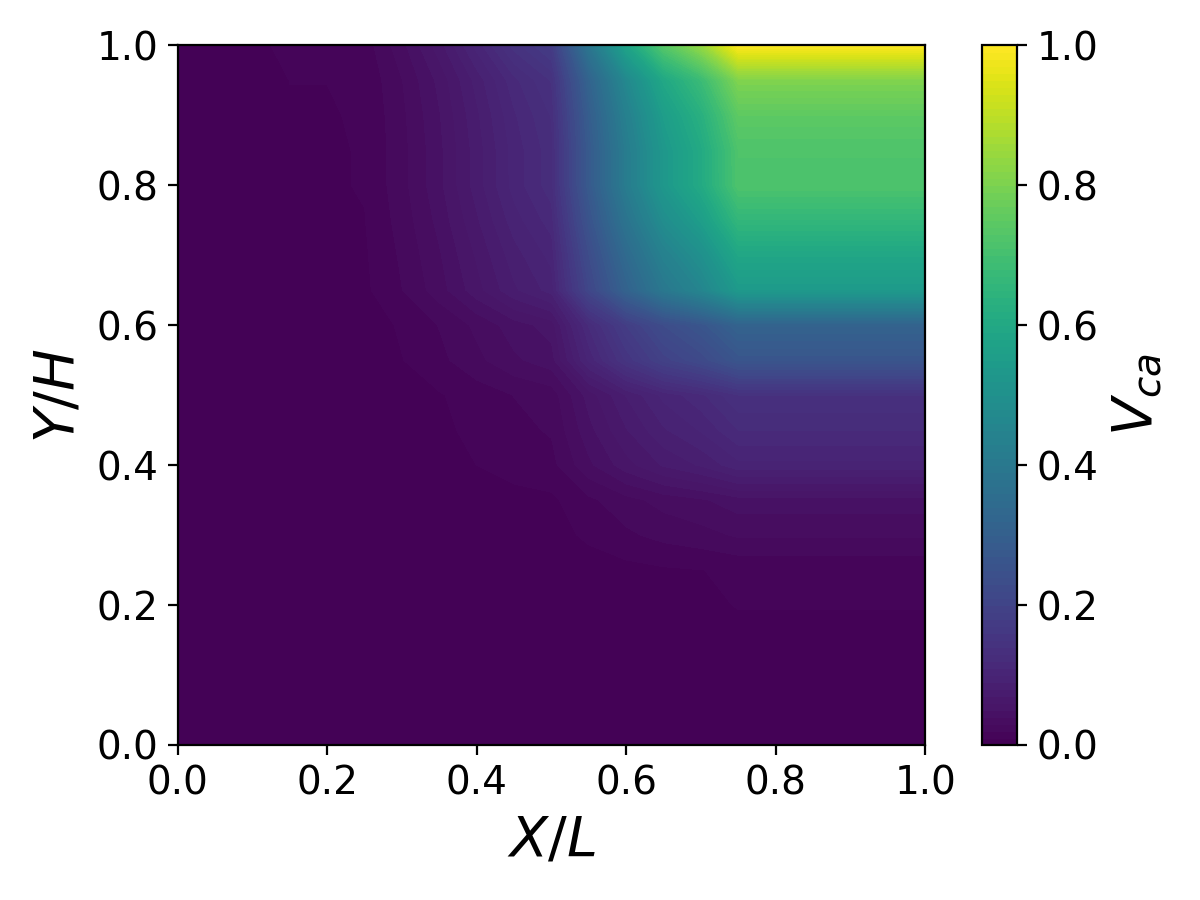}
    \caption{}
    \label{vf_21b}
  \end{subfigure}
  \hspace{1cm}
  \begin{subfigure}[b]{0.40\textwidth} % Adjust the width
    \centering
    \includegraphics[width=\textwidth]{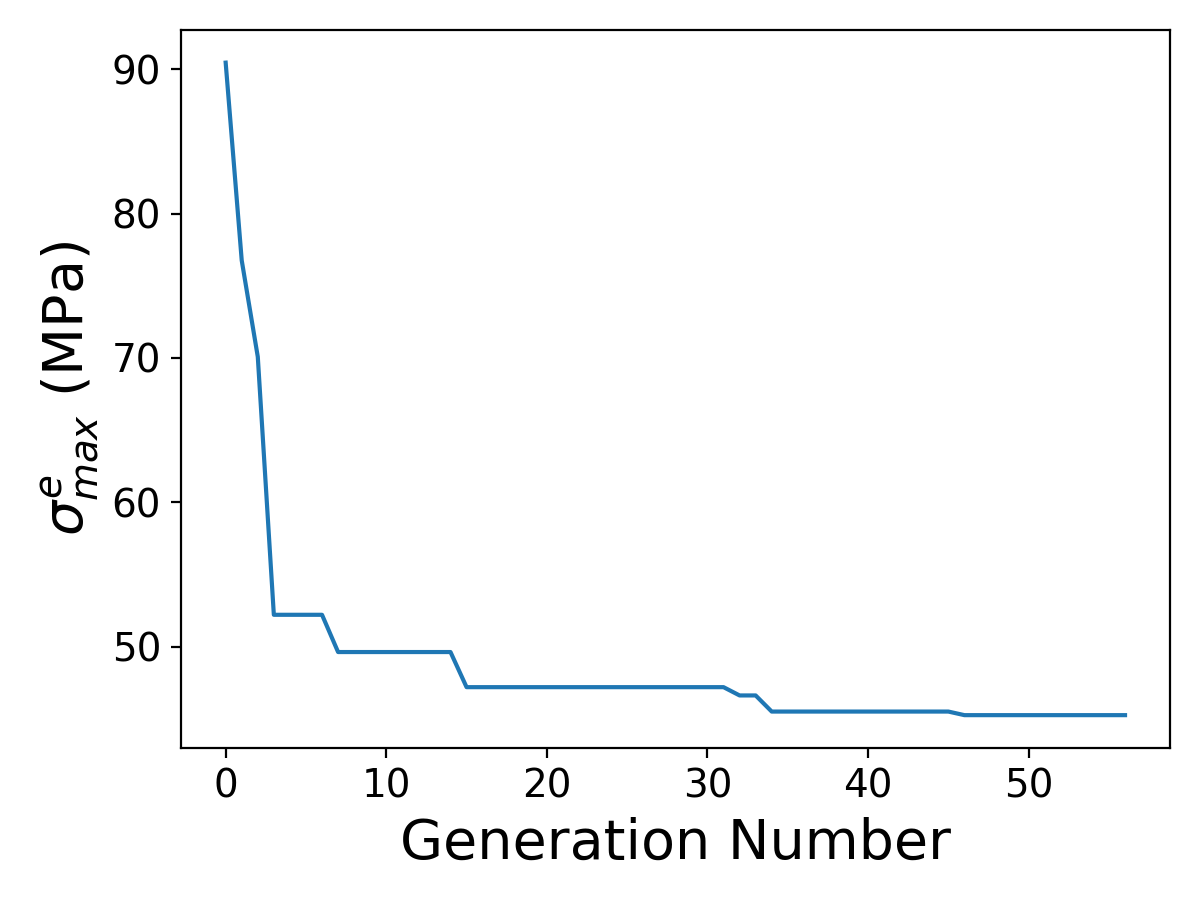}
    \caption{}
    \label{conv_21b}
  \end{subfigure}
      \caption{ Optimization of FGM plate under the constraint of average ceramic volume fraction: (a) contour plot of the optimal ceramic volume fraction distribution, (b) $\sigma^{e}_{max}$ values with respect to generations.}
  \label{prb_211b}
\end{figure}

%% case(c) minimze stress, temperature <= 275

\subsubsection{Case 3}

We now consider a more practical case, where the \(\sigma^{e}_{max}\) is minimized subjected to a thermal constraint. As a thermal constraint, we impose the condition that the metallic part in the FGM should not be subjected to temperature exceeding a specified value. In mathematical terms, the optimization problem is given as:
\begin{equation}
\begin{aligned}
\textbf{Minimize:}\; & \quad\sigma^{e}_{max}(\phi^{x}_{1}, \phi^{y}_{1}, \boldsymbol{\alpha^{x}}, \boldsymbol{\alpha^{y}}),\\ 
\textbf{Subject to:}\; & \quad \frac{1}{\theta_{max}}( \textbf{max} \;\overline{\theta}) \leq 1, \qquad\text { on } \mathit{\Omega_{m}}
\nonumber
\end{aligned}
\end{equation}
where, $\mathit{\Omega_{m}} = (x, y \quad ;\; \phi(x,y) < 1)$.\\

The value of $\theta_{max}$ is chosen as \SI{275}{\degreeCelsius}. The volume fraction distribution and the convergence of GA are shown in Figs. \ref{vf_21c} and \ref{conv_21c}, respectively. In the optimized design, the value of $\sigma^{e}_{max}$ is equal to \SI{71.30}{\mega\pascal}. The minimization of  $\sigma^{e}_{max}$, along with the thermal constraint, resulted in the FGM profile having overall low metal contents compared to the previous cases. The temperature distribution obtained from DeepONet for the optimum FGM profile is compared against FEM predictions in Fig. \ref{temp_21c}. This comparison confirms that the optimum FGM profile satisfies the specified thermal constraint.

\begin{figure}[htbp]
  \centering
  \begin{subfigure}[b]{0.40\textwidth} % Adjust the width
    \centering
    \includegraphics[width=\textwidth]{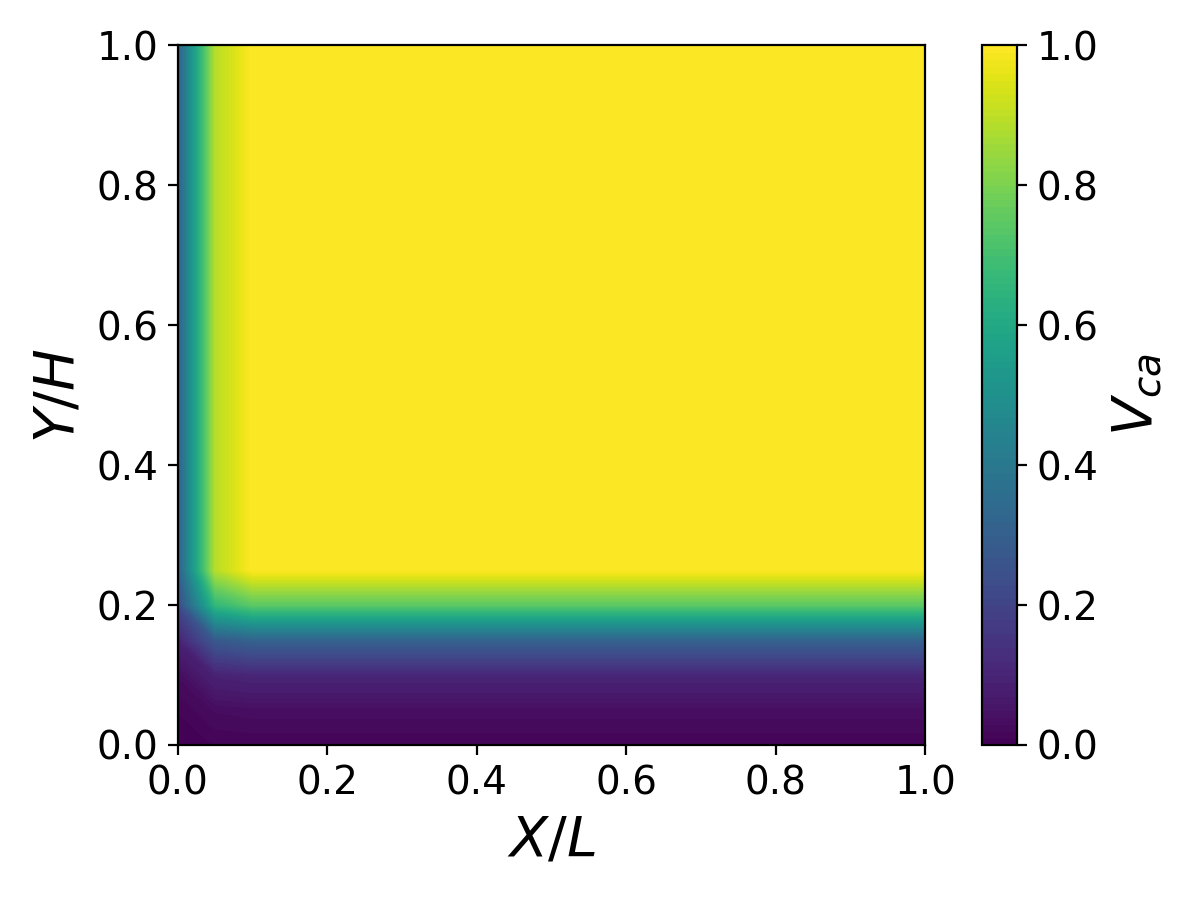}
    \caption{}
    \label{vf_21c}
  \end{subfigure}
  \hspace{1cm}
  \begin{subfigure}[b]{0.40\textwidth} % Adjust the width
    \centering
    \includegraphics[width=\textwidth]{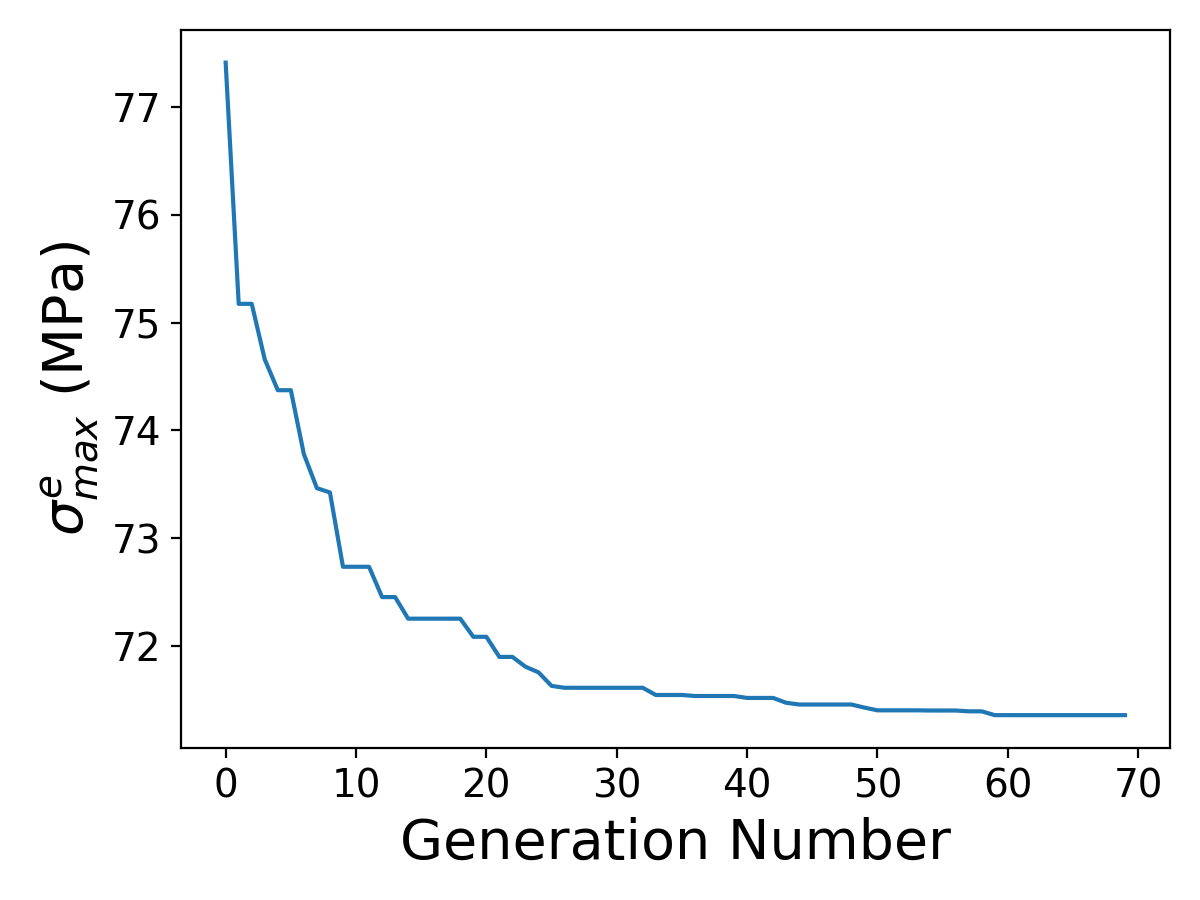}
    \caption{}
    \label{conv_21c}
  \end{subfigure}

  \caption{ Optimization of FGM plate under the constraint of $\theta_{max}$ experienced by metallic phase of the FGM: (a) contour plot of the ceramic volume fraction optimal distribution, (b) $\sigma^{e}_{max}$ value with respect to generations.}
  \label{prb_21c}
\end{figure}

\begin{figure}[htbp]
    \centering
    \begin{subfigure}[b]{0.3\textwidth}
        \centering
        \includegraphics[width=\textwidth]{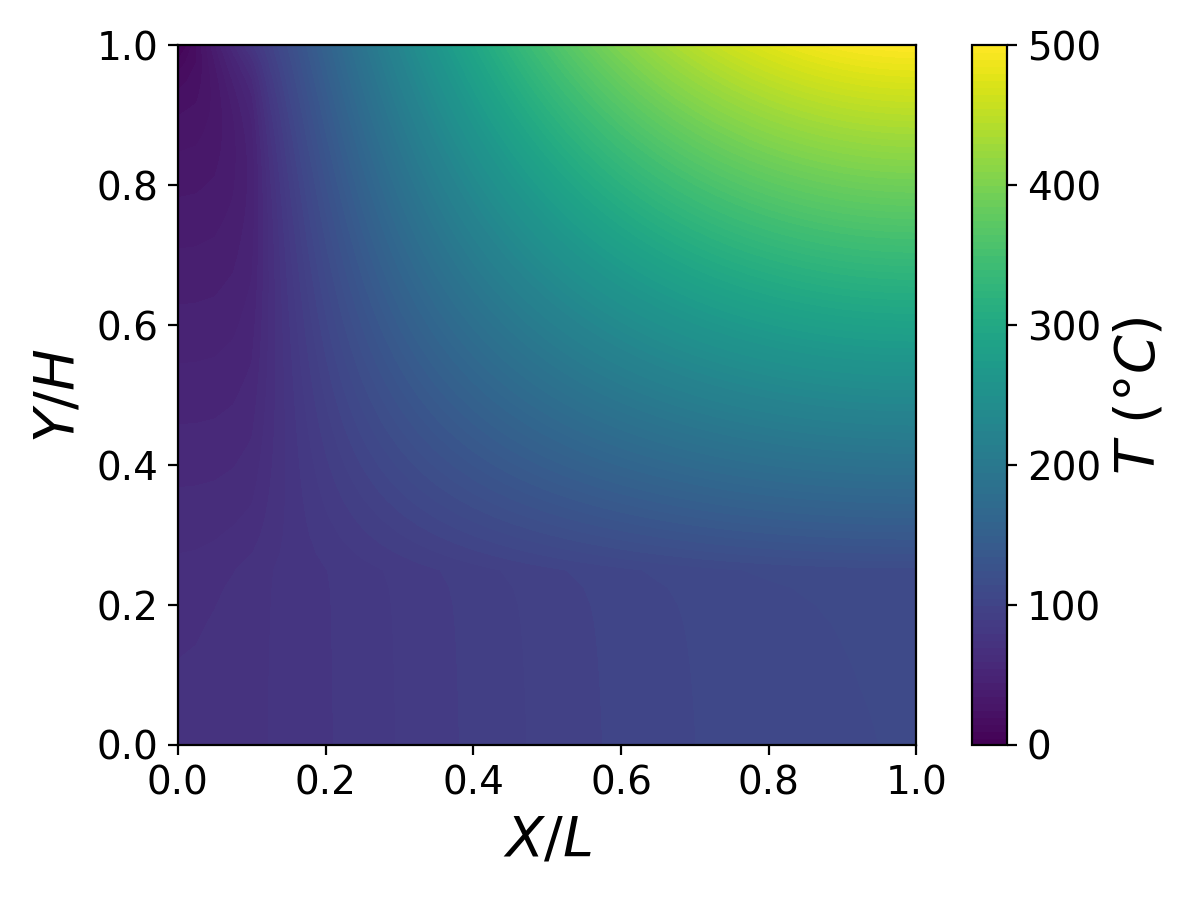}
        \caption{}
        \label{temp_fem_21c}
    \end{subfigure}
    \begin{subfigure}[b]{0.3\textwidth}
        \centering
        \includegraphics[width=\textwidth]{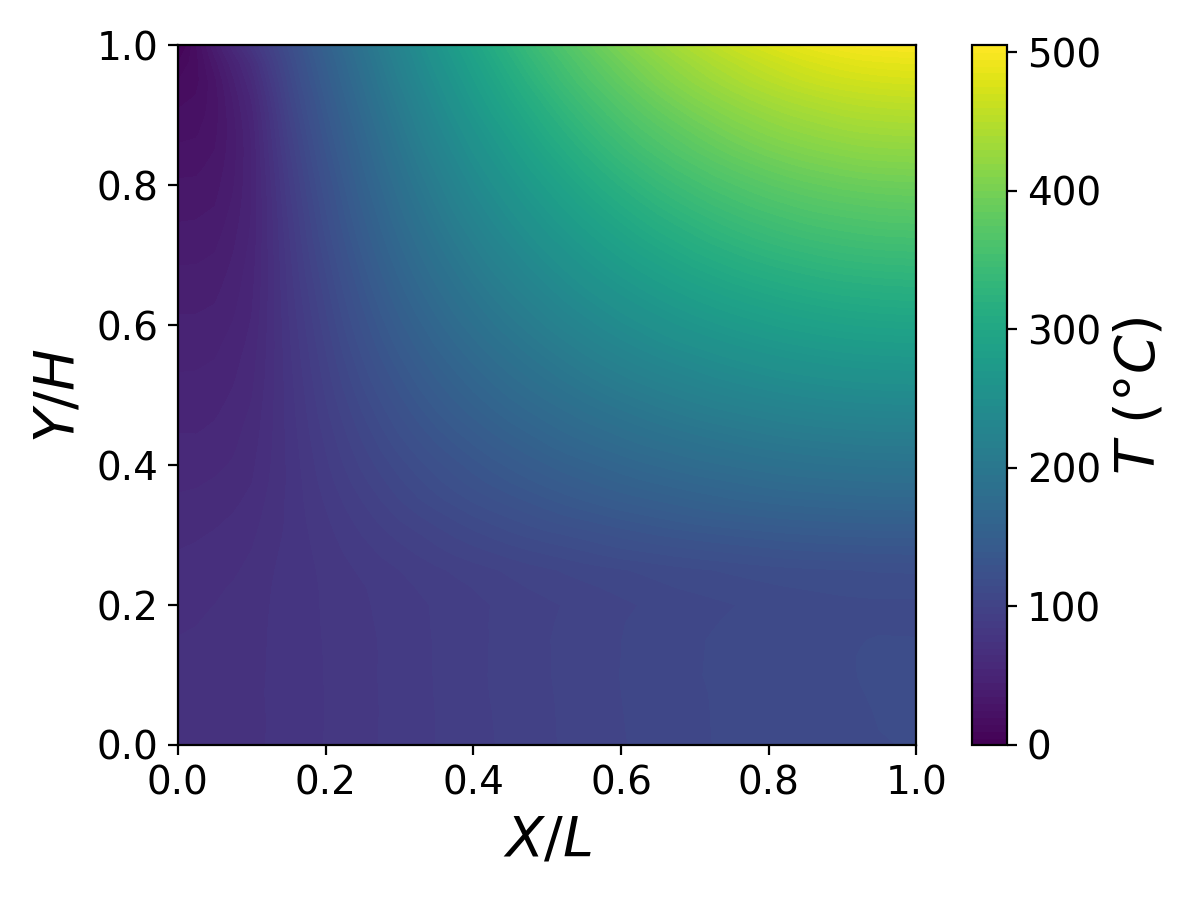}
        \caption{}
        \label{temp_deep_21c}
    \end{subfigure}
    \begin{subfigure}[b]{0.3\textwidth}
        \centering
        \includegraphics[width=\textwidth]{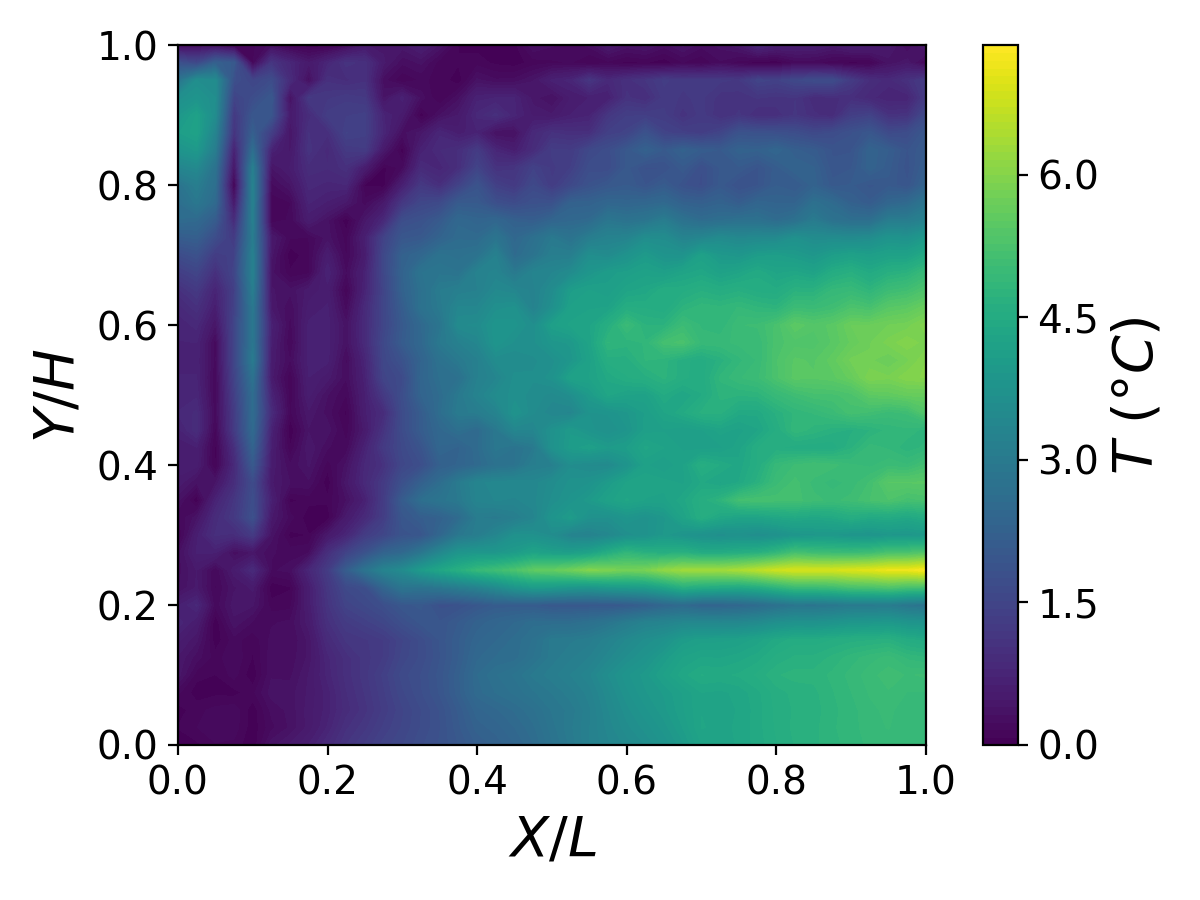}
        \caption{}
        \label{temp_error_21c}
    \end{subfigure}
    \caption{Temperature field in the optimal FGM profile: (a) obtained by FEM, (b) obtained by DeepONet, (c) absolute error in DeepONet prediction.}
    \label{temp_21c}
\end{figure}

\subsubsection{Case 4} 

In this case, we impose both the thermal and structural constraints. The structural constraint is that  $\sigma^{e}_{max}$ should be below a specified value, while the thermal constraint is same as of case 3. The objective function is here to minimize the overall ceramic contents in the FGM. The optimization problem is stated as:
\begin{equation}
\begin{aligned}
\textbf{Minimize:}\; & \quad V_{ca}(\phi^{x}_{1}, \phi^{y}_{1}, \boldsymbol{\alpha^{x}}, \boldsymbol{\alpha^{y}}),\\ 
\textbf{Subject to:}\; &(1) \quad \sigma^{\text{e}}_{\text{max}} \leq \sigma_{\text{a}},\\
&(2)\; \frac{1}{\theta_{max}}  ( \textbf{max} \;\overline{\theta}) \leq 1, \qquad\text { on } \mathit{\Omega_{m}}
\nonumber
\end{aligned}
\end{equation}
where, $\mathit{\Omega_{m}} = (x, y \quad ;\; \phi(x,y) < 1)$.

 Here,  the value of $\theta_{max}$ is taken \SI{275}{\degreeCelsius} and the $\sigma^{e}_{max}$ in the FGM should be equal or lesser than the allowable stress ($\sigma_{\text{a}}$). To demonstrate the effectiveness of our proposed optimization framework across different ranges of $\sigma^{e}_{max}$, we selected three distinct $\sigma_{\text{a}}$ values: 100 MPa, 125 MPa, and 150 MPa. The volume fraction distribution of the optimized profile for the different values of the $\sigma_{\text{a}}$ is shown in Fig. \ref{vf_22b}. We observed that by relaxing the $\sigma_{a}$ condition, as expected, we can achieve the better value of the objective function. Further, to provide an intuitive understanding of the result, we present the overall effective stress distribution from FEM in the accuracy of DNN in the prediction of $\sigma^{e}_{max}$ Fig. \ref{stress_22b}. These plots also verify that the maximum stress value in the optimized profile is within the permissible limit. 
 
 We also present the temperature distribution obtained from DeepOnet and FEM for the optimum profiles in Figs. \ref{temp100}, \ref{temp125}, \ref{temp150}, respectively. From these plots, we can see that DeepOnet and FEM predictions are very close to each other for the optimized profile. This further validates that the temperature predictions from DeepOnet are accurate and can be used in FGM optimization to obtain the optimum profile with the thermal constraint.

\begin{figure}[htbp]
    \centering
    \begin{subfigure}[b]{0.3\textwidth}
        \centering
        \includegraphics[width=\textwidth]{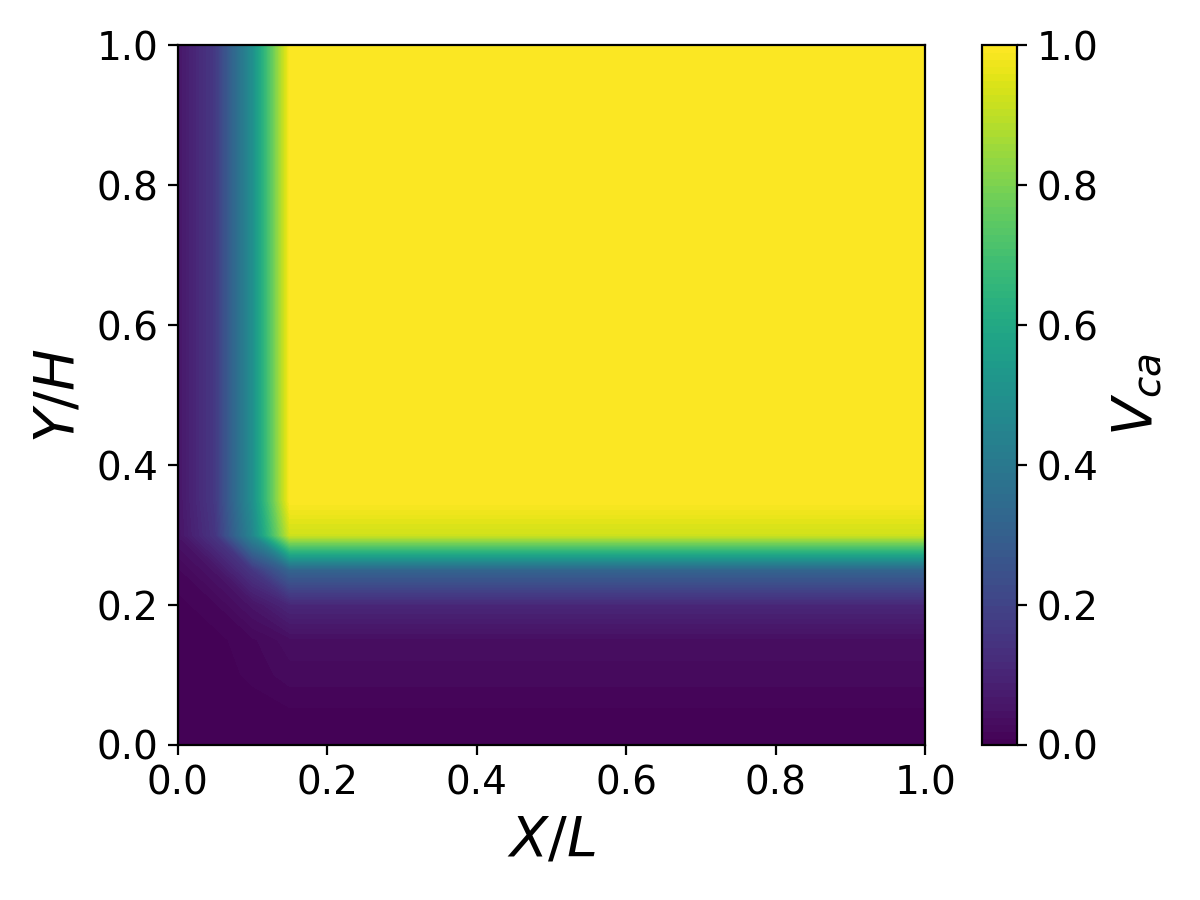}
        \caption{}
        \label{vf_100}
    \end{subfigure}
    \begin{subfigure}[b]{0.3\textwidth}
        \centering
        \includegraphics[width=\textwidth]{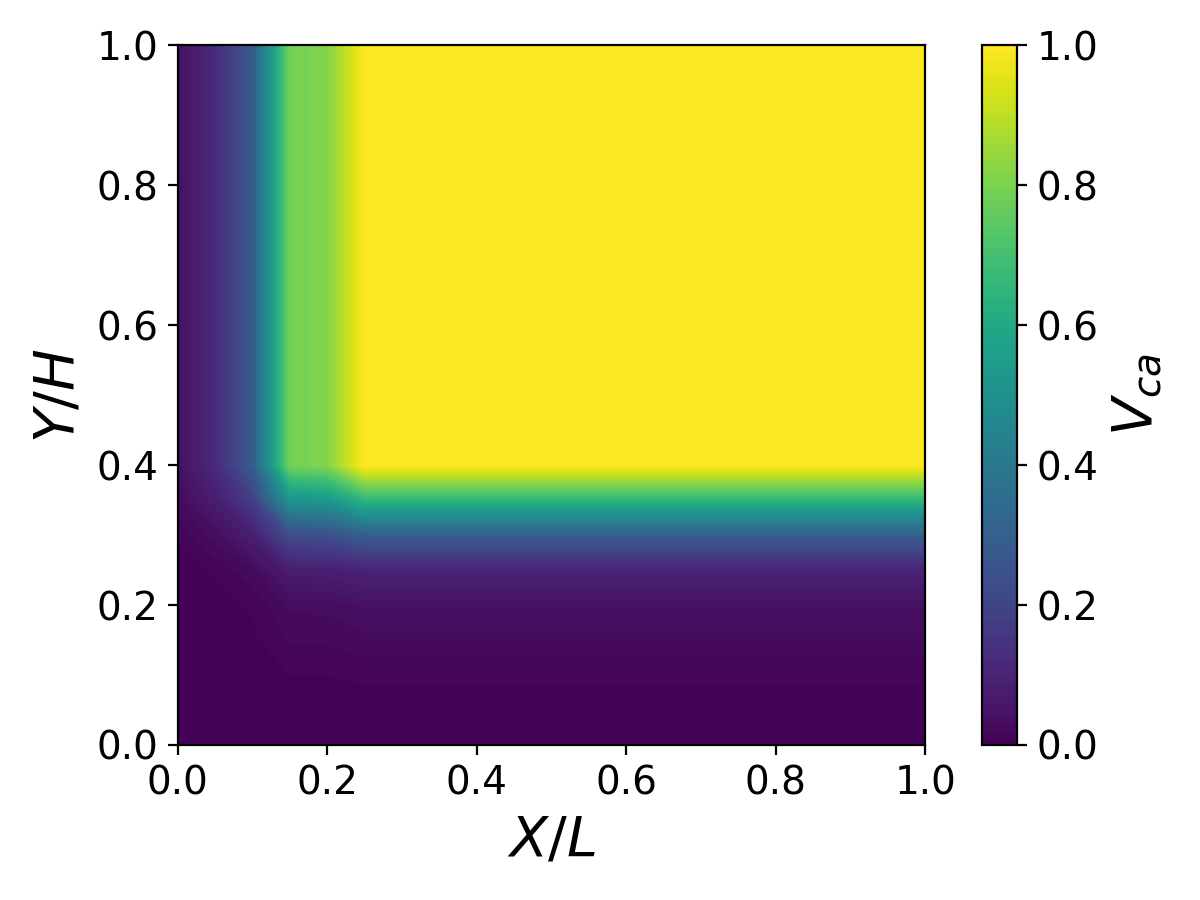}
        \caption{}
        \label{vf_125}
    \end{subfigure}
    \begin{subfigure}[b]{0.3\textwidth}
        \centering
        \includegraphics[width=\textwidth]{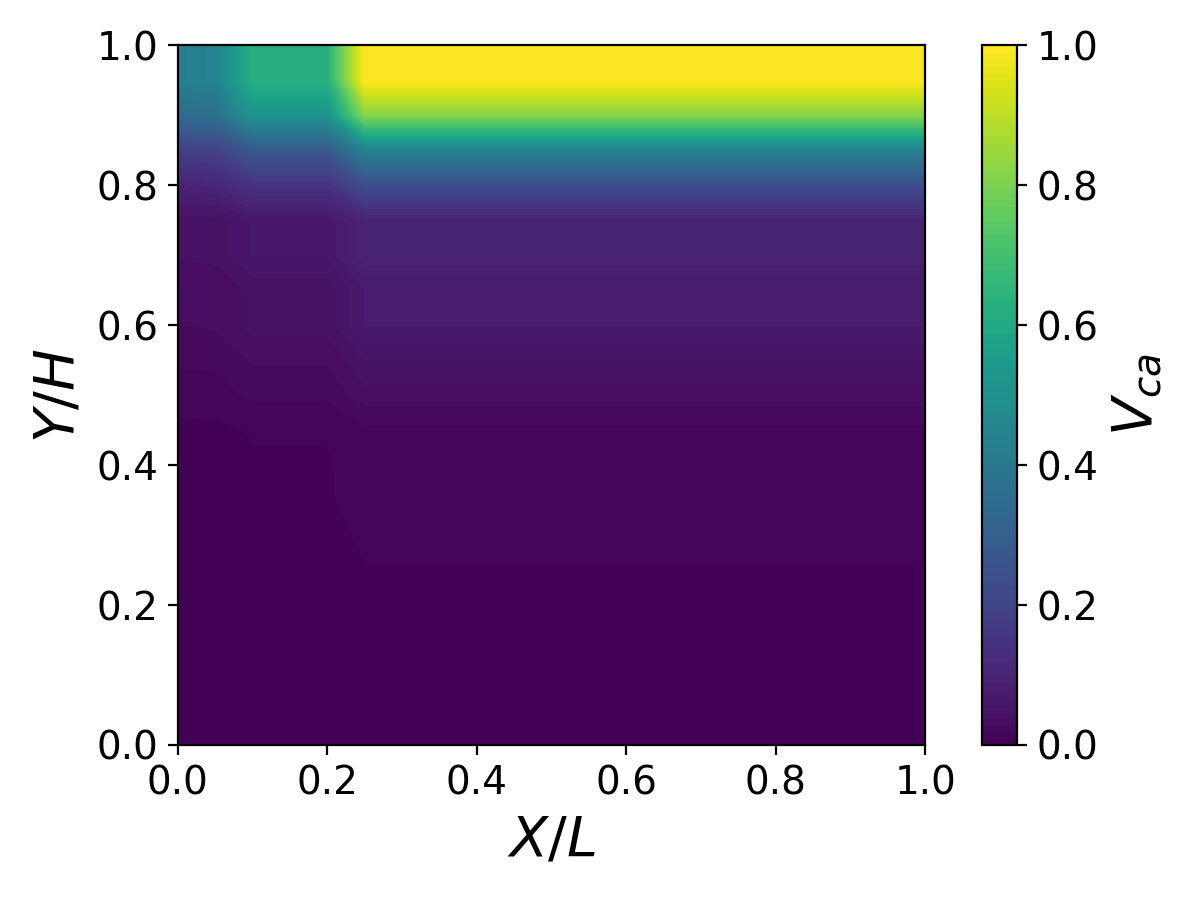}
        \caption{}
        \label{vf_150}
    \end{subfigure}
    \caption{The volume fraction distribution in the optimized FGM profiles: (a) for $\sigma_{\mathrm{a}} \leq$ 100 MPa, $V_{ca}$ = 65\%, (b) for $\sigma_{\mathrm{a}} \leq$ 125 MPa, $V_{ca}$ = 57\%, (c) for $\sigma_{\mathrm{a}} \leq$ 150 MPa, $V_{ca}$ = 17\%. }
    \label{vf_22b}
\end{figure}

\begin{figure}[htbp]
    \centering
    \begin{subfigure}[b]{0.3\textwidth}
        \centering
        \includegraphics[width=\textwidth]{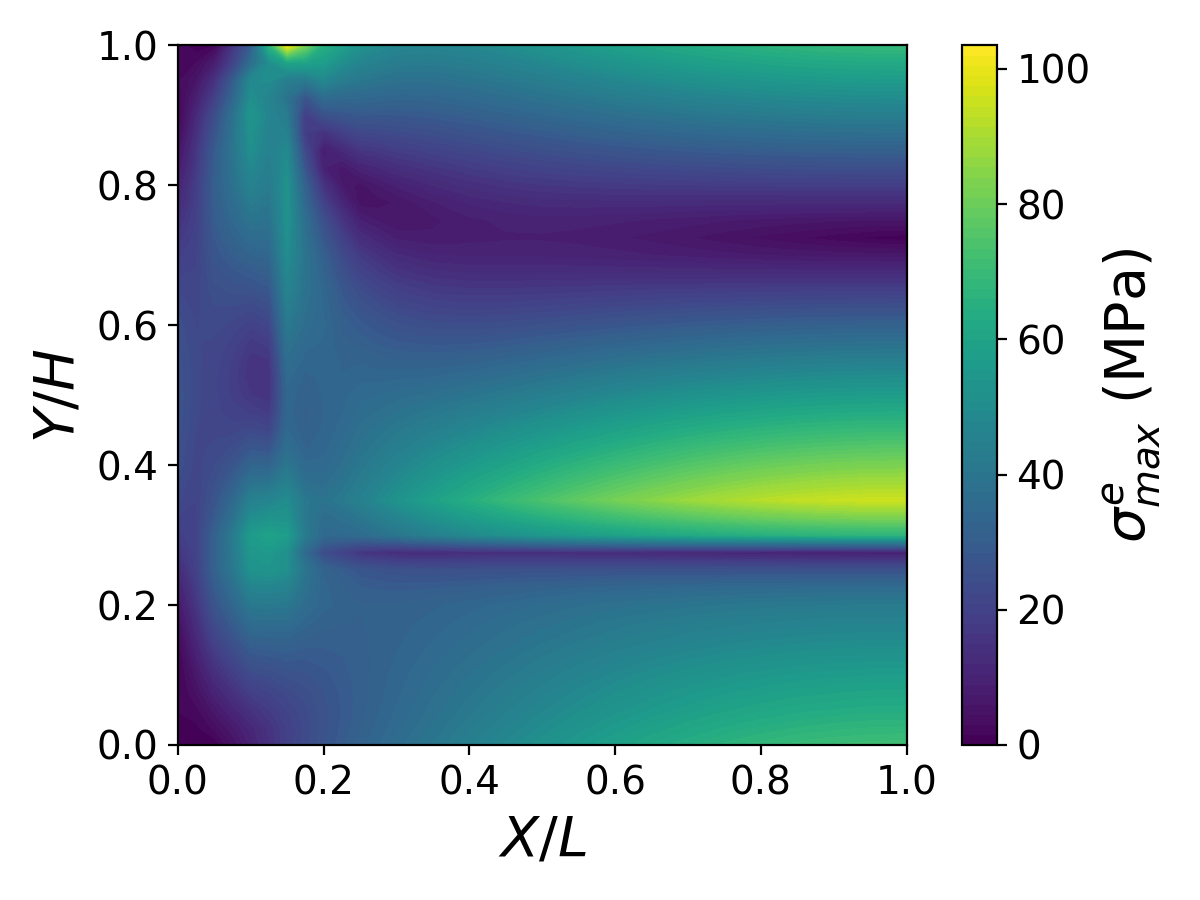}
        \caption{}
        \label{stress_100}
    \end{subfigure}
    \begin{subfigure}[b]{0.3\textwidth}
        \centering
        \includegraphics[width=\textwidth]{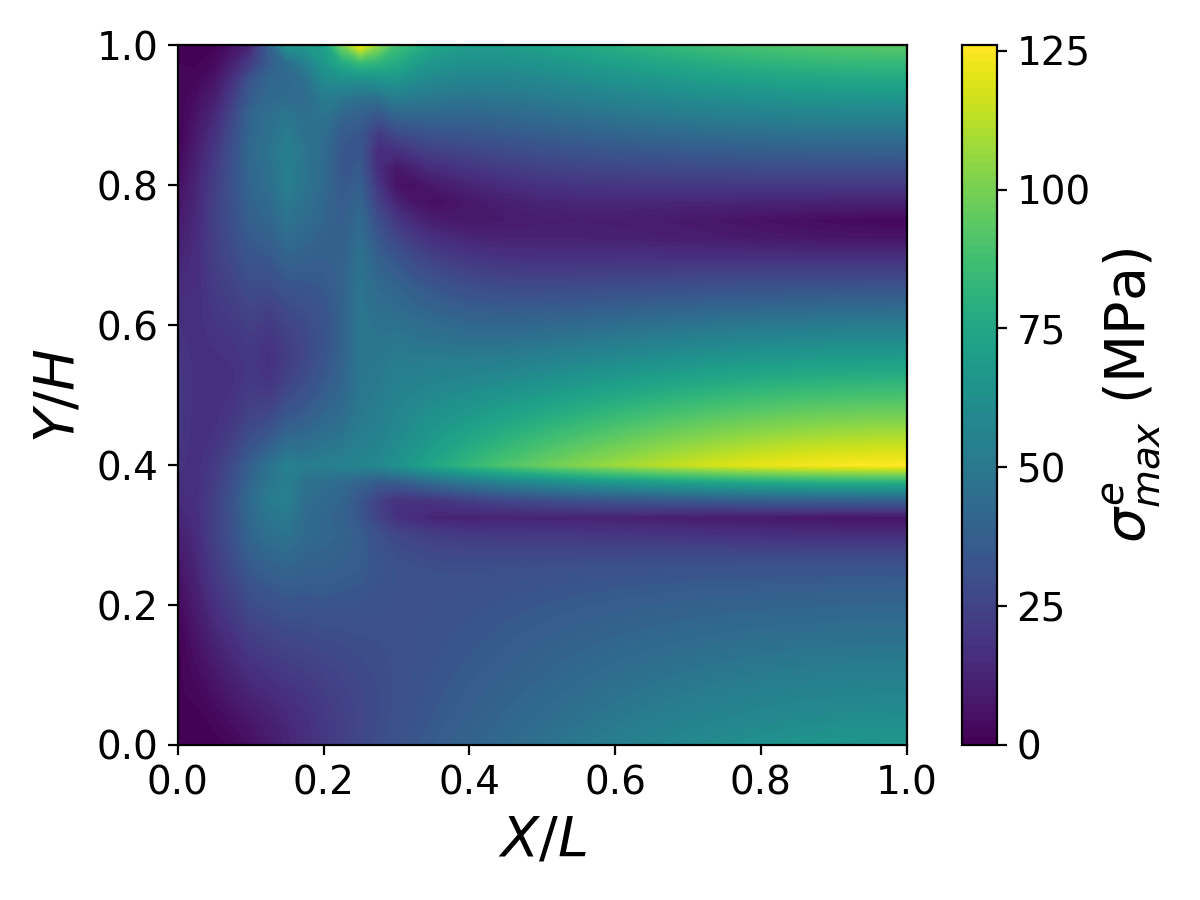}
        \caption{}
        \label{stress_125}
    \end{subfigure}
    \begin{subfigure}[b]{0.3\textwidth}
        \centering
        \includegraphics[width=\textwidth]{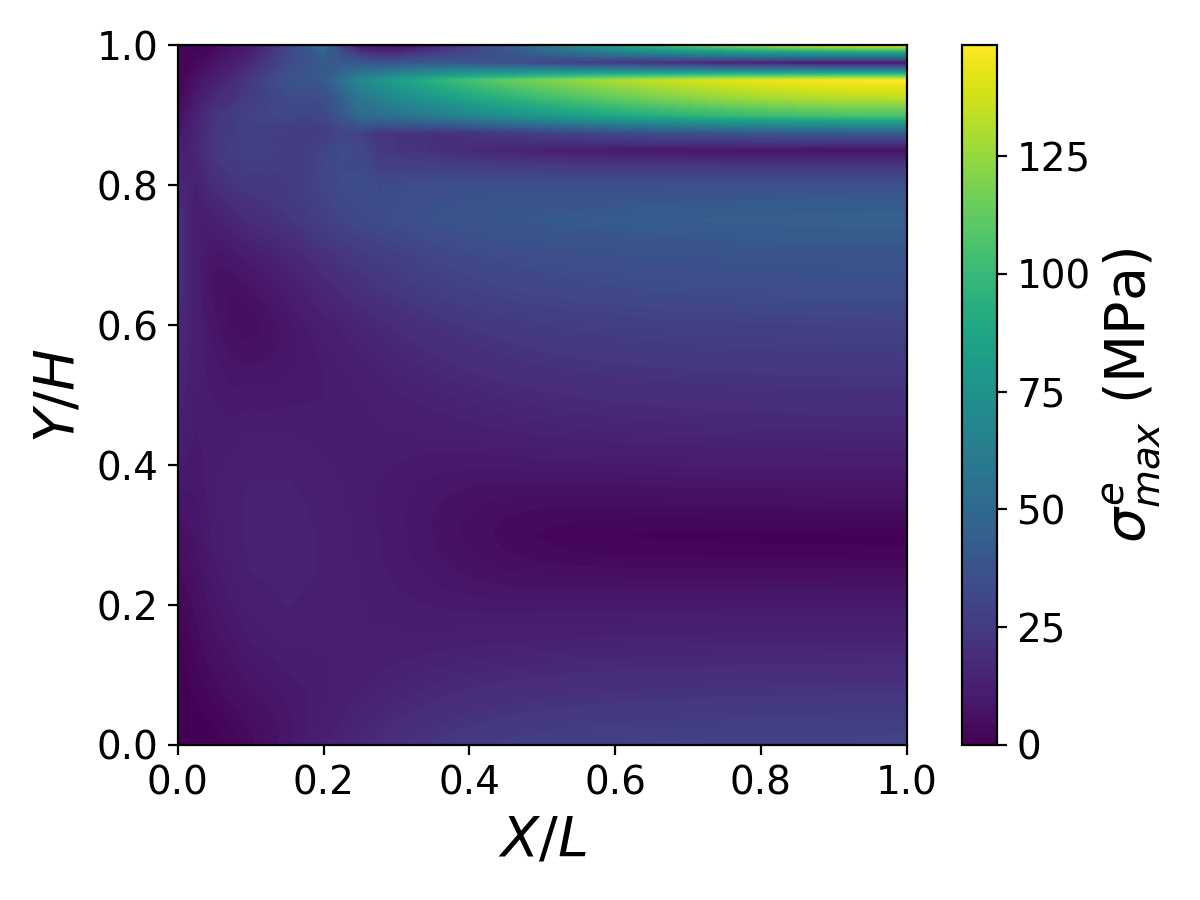}
        \caption{}
        \label{stress_150}
    \end{subfigure}
    \caption{Effective stress from FEM for the optimized FGM profiles: (a) $\sigma_{\mathrm{a}} \leq$ 100 MPa, (b) $\sigma_{\mathrm{a}} \leq$ 125 MPa, (c) $\sigma_{\mathrm{a}} \leq$ 150 MPa.}
    \label{stress_22b}
\end{figure}

% stress =100 MPa

\begin{figure}[htbp]
    \centering
    \begin{subfigure}[b]{0.3\textwidth}
        \centering
        \includegraphics[width=\textwidth]{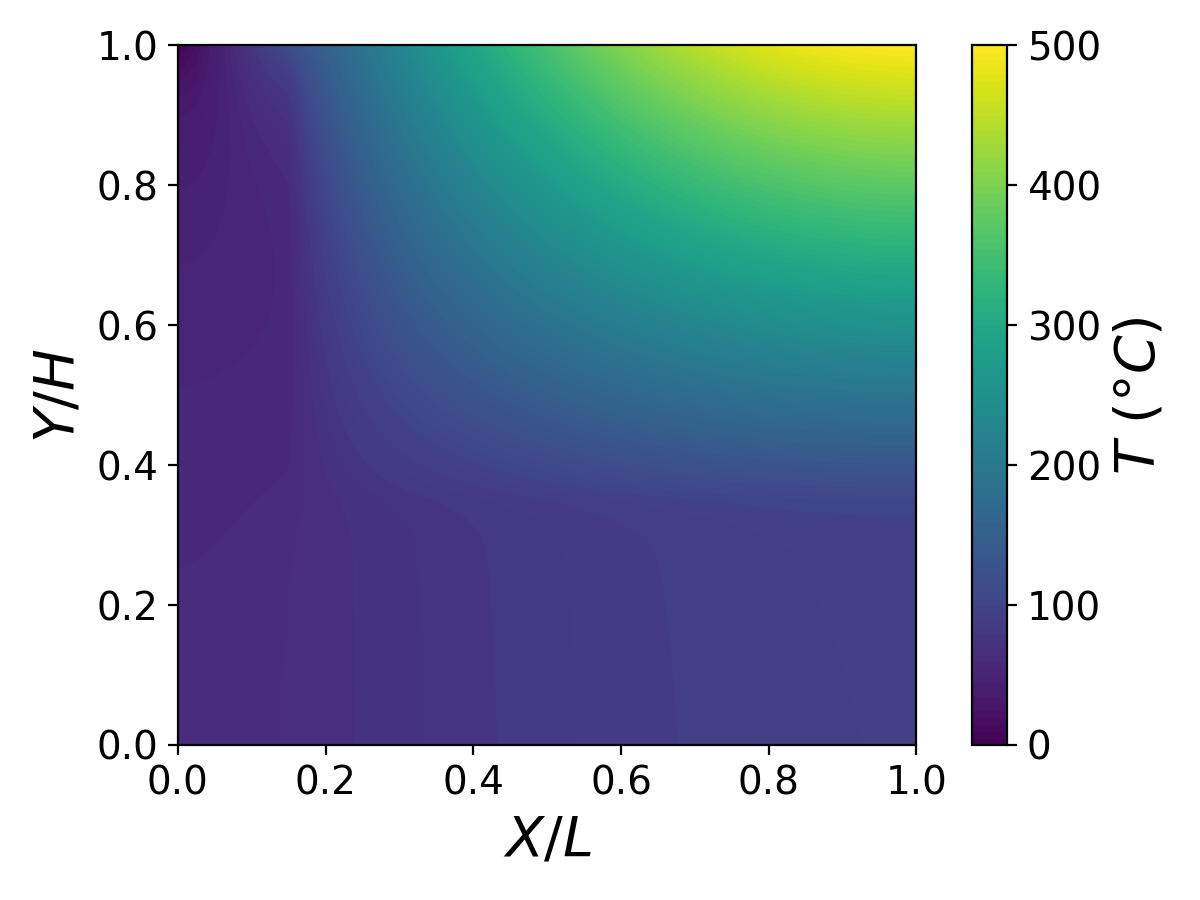}
        \caption{}
        \label{}
    \end{subfigure}
    \begin{subfigure}[b]{0.3\textwidth}
        \centering
        \includegraphics[width=\textwidth]{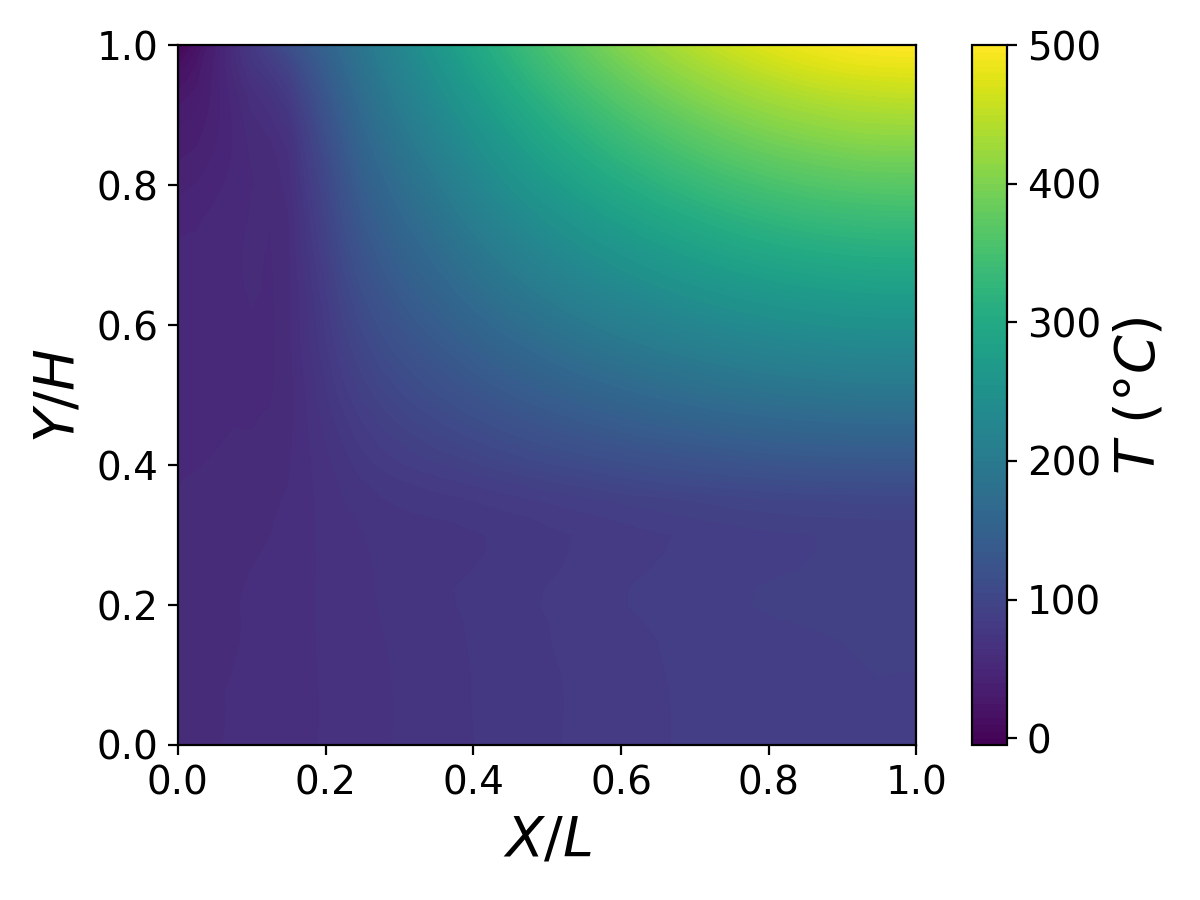}
        \caption{}
        \label{}
    \end{subfigure}
    \begin{subfigure}[b]{0.3\textwidth}
        \centering
        \includegraphics[width=\textwidth]{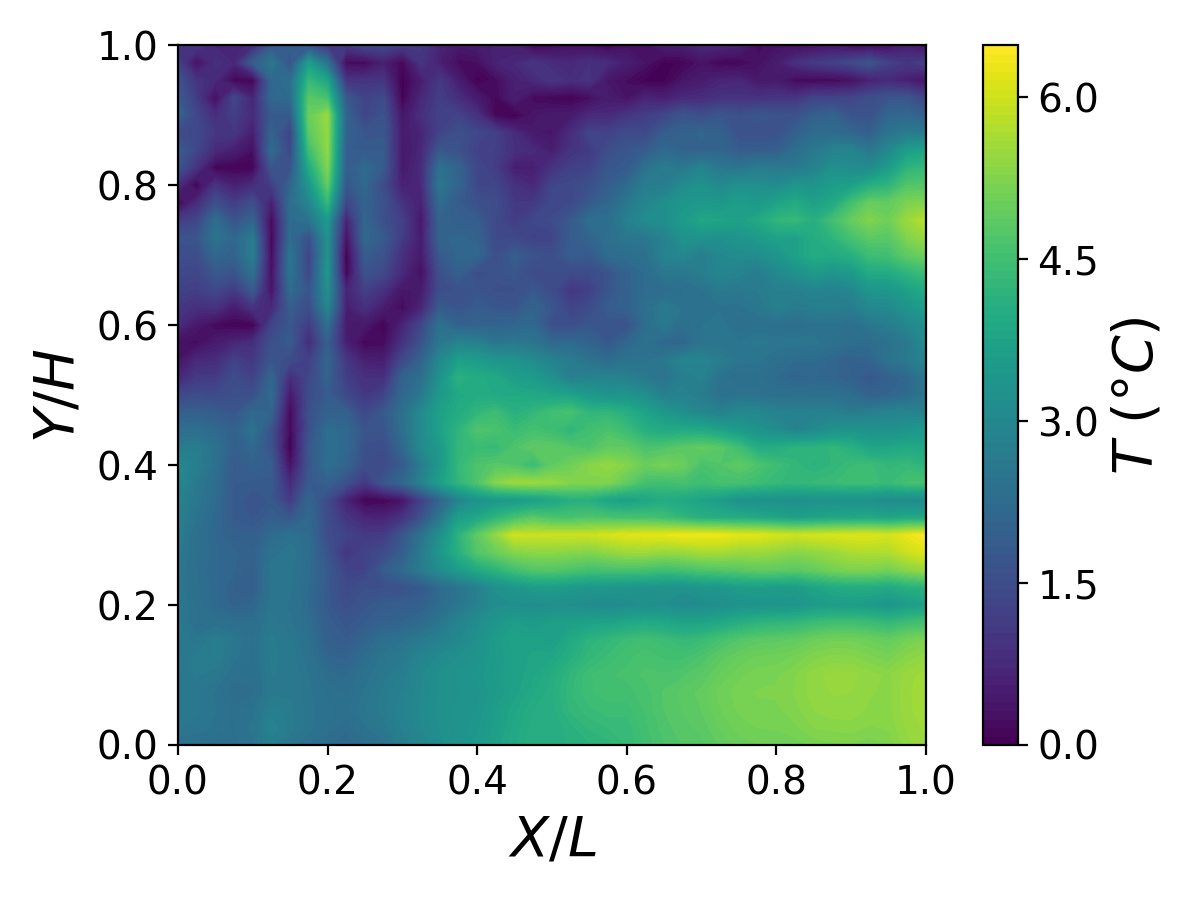}
        \caption{}
        \label{}
    \end{subfigure}
    \caption{Temperature field in the optimal FGM profile with $\sigma_{\mathrm{a}} \leq$ 100 MPa: (a)  obtained by FEM, (b) obtained by DeepONet, (c) absolute error in DeepONet prediction.}
    \label{temp100}
\end{figure}

% stress =125 MPa

\begin{figure}[htbp]
    \centering
    \begin{subfigure}[b]{0.3\textwidth}
        \centering
        \includegraphics[width=\textwidth]{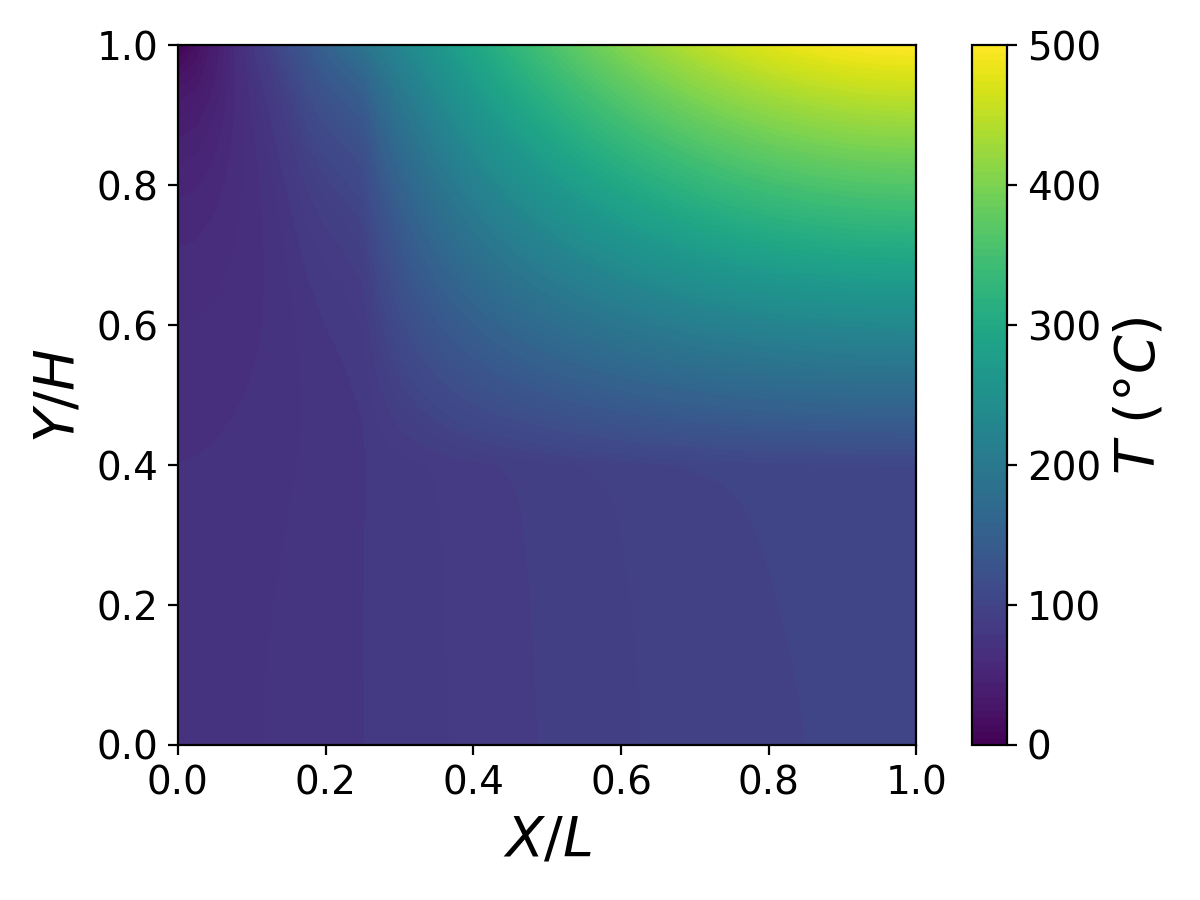}
        \caption{}
        \label{}
    \end{subfigure}
    \begin{subfigure}[b]{0.3\textwidth}
        \centering
        \includegraphics[width=\textwidth]{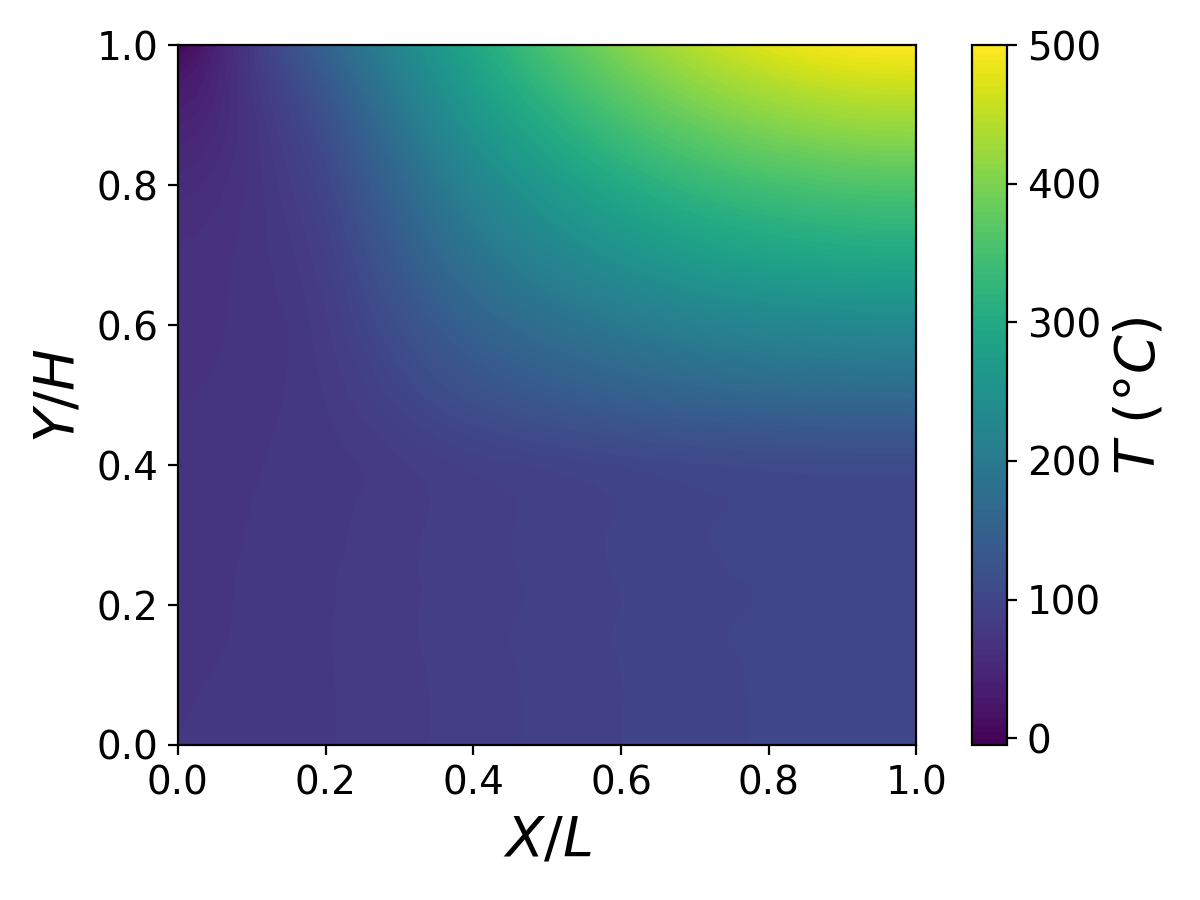}
        \caption{}
        \label{}
    \end{subfigure}
    \begin{subfigure}[b]{0.3\textwidth}
        \centering
        \includegraphics[width=\textwidth]{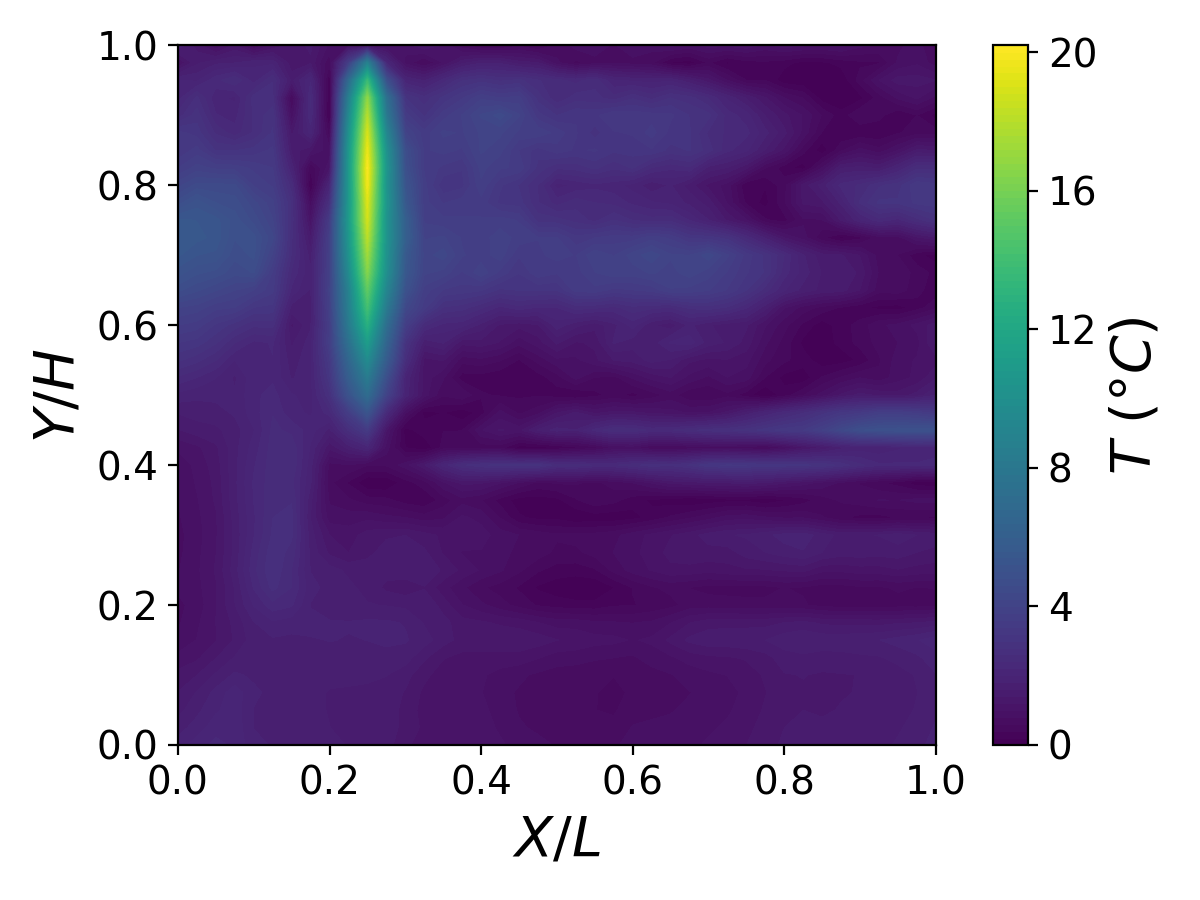}
        \caption{}
        \label{}
    \end{subfigure}
    \caption{Temperature field in the optimal FGM profile with $\sigma_{\mathrm{a}} \leq$ 125 MPa: (a) obtained by FEM, (b) obtained by DeepONet, (c) absolute error in DeepONet prediction.}
    \label{temp125}
\end{figure}

% stress = 150 MPA

\begin{figure}[htbp]
    \centering
    \begin{subfigure}[b]{0.3\textwidth}
        \centering
        \includegraphics[width=\textwidth]{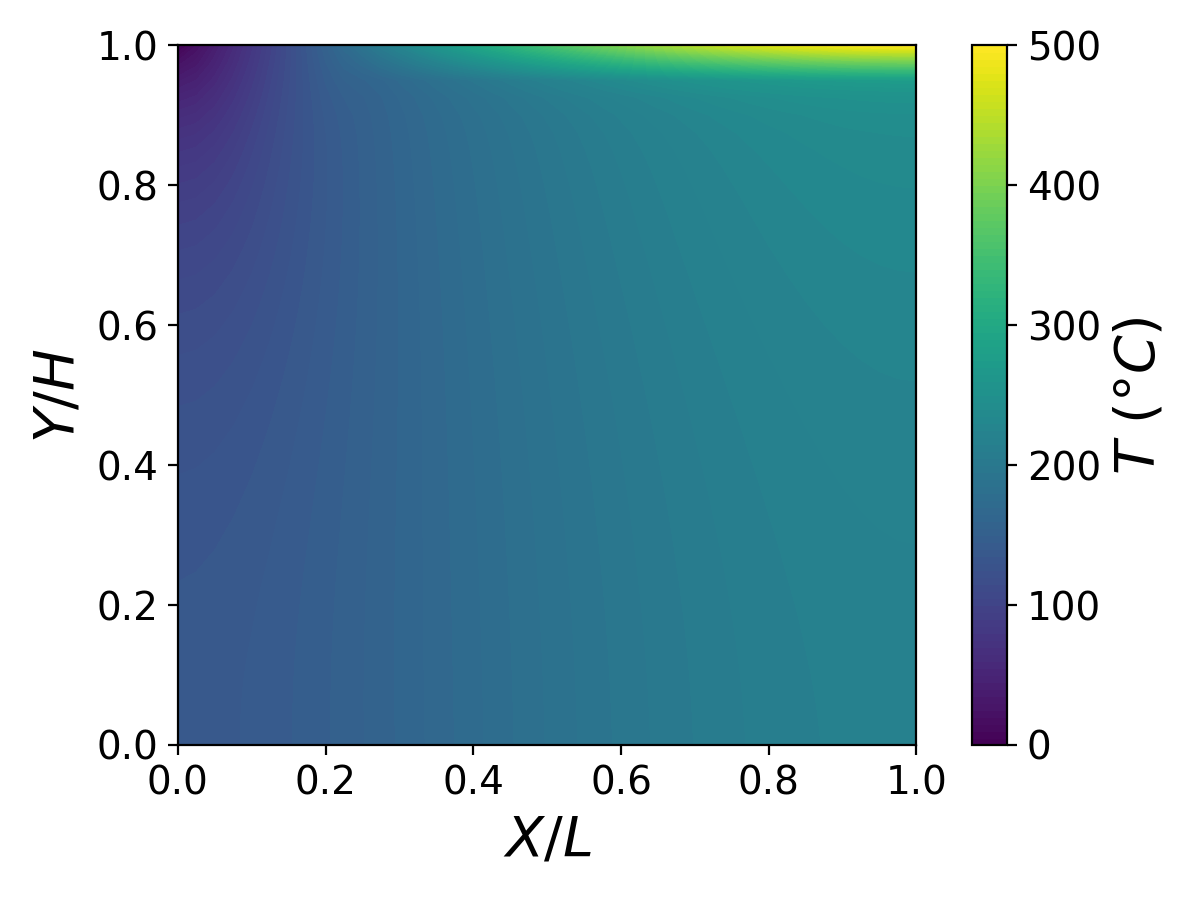}
        \caption{}
        \label{}
    \end{subfigure}
    \begin{subfigure}[b]{0.3\textwidth}
        \centering
        \includegraphics[width=\textwidth]{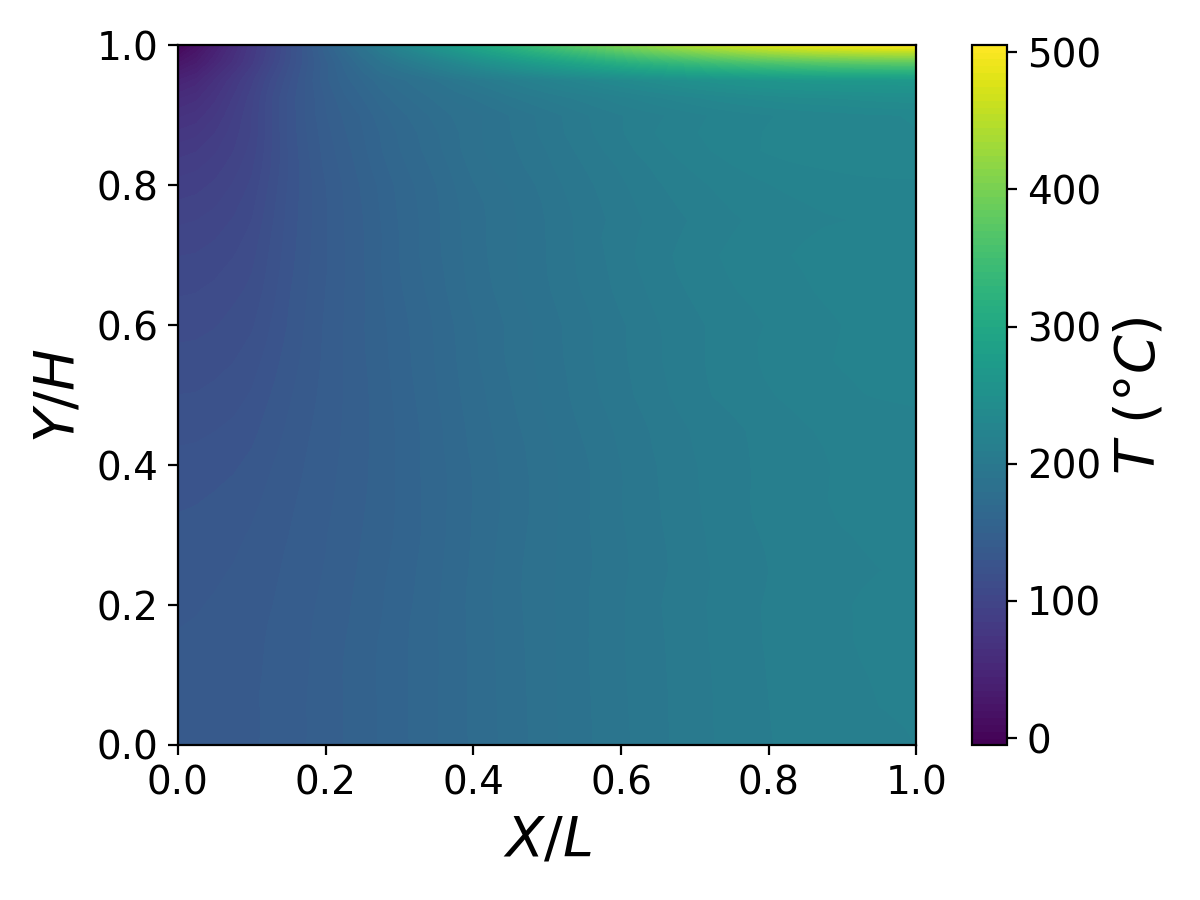}
        \caption{}
        \label{}
    \end{subfigure}
    \begin{subfigure}[b]{0.3\textwidth}
        \centering
        \includegraphics[width=\textwidth]{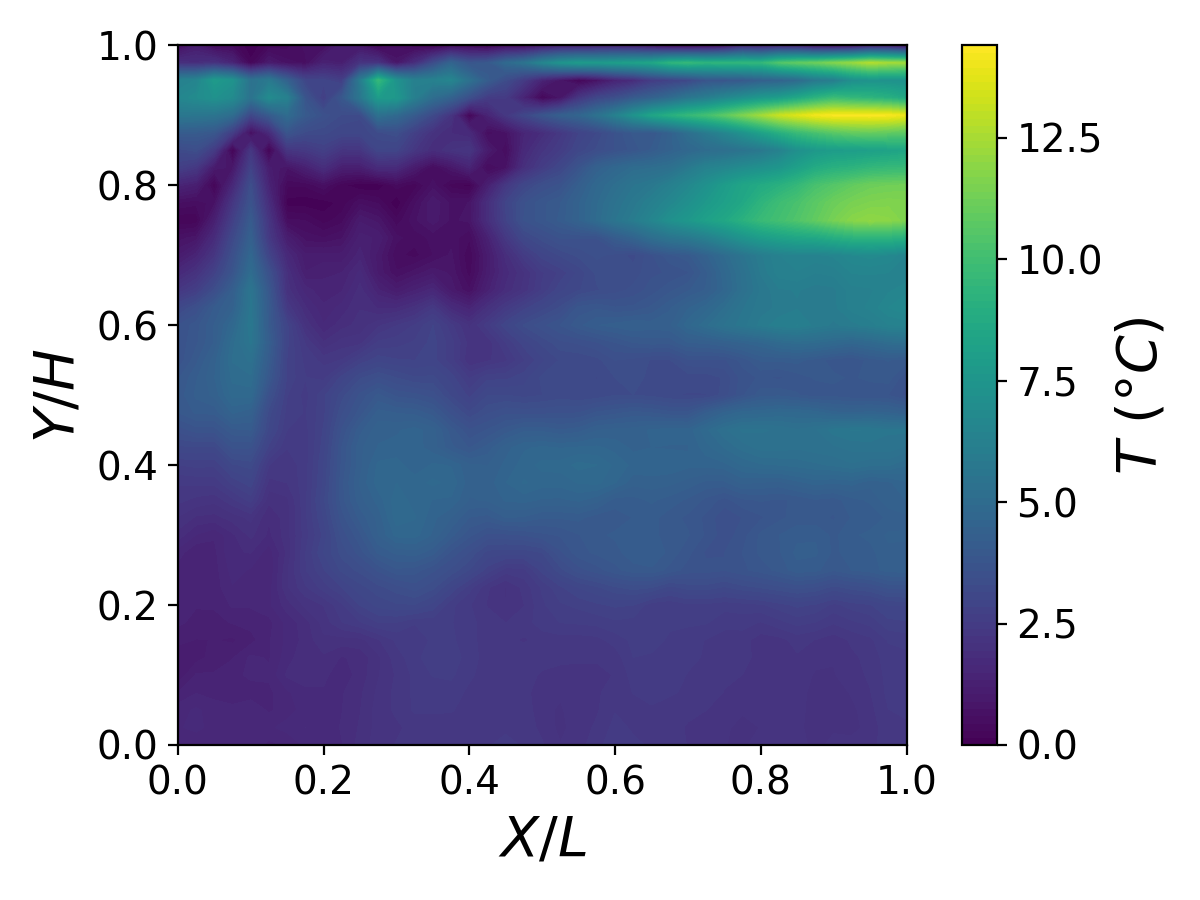}
        \caption{}
        \label{}
    \end{subfigure}
    \caption{Temperature field in the optimal FGM profile with $\sigma_{\mathrm{a}} \leq$ 150 MPa: (a) obtained by FEM, (b) obtained by DeepONet, (c) absolute error in DeepONet prediction.}
    \label{temp150}
\end{figure}

\section{Conclusion}
We have proposed an optimization framework that consists of a novel profile generation scheme, deep learning-based surrogate models, and a genetic algorithm. The proposed profile generation scheme ensures that only smooth profiles exist in the design space. Additionally, it has been demonstrated that this scheme can generate a wide variety of profiles. We show that the deep learning-based surrogate models are not only effective for predicting real-valued metrics such as maximum effective stress but can also be used for predicting the overall field. By deploying the deep operator-based network ‘DeepONet’ for thermal field prediction, point-wise thermal constraints are implemented in FGM optimization in a computationally efficient manner. An interesting observation from this manuscript is that even with high R² scores, surrogate models might fail to predict limiting cases accurately. This can be of concern, since we are mainly interested in the limiting cases in optimization. In this work, we address this issue by employing a hybrid strategy based on FEM and deep learning-based surrogate models in the fitness evaluation.

\section*{Acknowledgments}
We acknowledge the support of SERB, DST (Department of Science and Technology),
Government of India under Project No. SRG/2021/001594.

\bibliographystyle{elsarticle-num}
\bibliography{library_edited}

\end{document}